\documentclass[american]{article}
\usepackage[latin9]{inputenc}
\usepackage{geometry}
\usepackage{booktabs}
\usepackage{graphicx}
\usepackage[font=normalsize]{caption}
\usepackage[font=large]{subcaption}
\PassOptionsToPackage{normalem}{ulem}
\usepackage{ulem}
\usepackage[capposition=top]{floatrow}
\usepackage{changepage} 
\makeatletter
\providecommand{\tabularnewline}{\\}
\makeatother
\usepackage{fancyhdr}
\usepackage{sgame}
\usepackage{color}
\usepackage{amsmath}
\usepackage{amsfonts}
\usepackage{graphicx}
\usepackage{setspace}
\usepackage{verbatim}
\usepackage{amssymb}
\usepackage{commath}
\usepackage{enumerate}
\usepackage{epstopdf}
\usepackage{pdflscape}
\usepackage{array}
\usepackage[flushleft]{threeparttable}
\usepackage[section]{placeins}
\usepackage{multirow}
\usepackage[multiple]{footmisc}
\usepackage{afterpage}
\usepackage{enumitem}
\usepackage{microtype}
\usepackage[%  
    colorlinks=true,
    pdfborder={0 0 0},
    linkcolor=blue,
    backref=false
]{hyperref}
\hypersetup{
    colorlinks,
    citecolor=blue,
    filecolor=black,
    linkcolor=black,
    urlcolor=blue
}
\usepackage[round]{natbib}
\setcitestyle{authoryear,open={(},close={)}}

\usepackage{titlesec}
\titlespacing\section{0pt}{10pt}{10pt}
\titlespacing\subsection{0pt}{10pt}{10pt}

\titleformat{\section}
  {\normalfont\fontsize{12}{15}\bfseries}{\thesection}{1em}{}
  
\titleformat{\subsection}
  {\normalfont\fontsize{12}{12}\itshape}{\thesubsection}{1em}{}  
  
\titleformat{\subsubsection}
  {\normalfont\fontsize{12}{12}\itshape}{\thesubsubsection}{1em}{}  

\usepackage{titling}
\newcommand{\subtitle}[1]{%
  \posttitle{%
    \par\end{center}
    \begin{center}\large#1\end{center}
    \vskip0.5em}%
}

\newcolumntype{L}[1]{>{\raggedright\let\newline\\\arraybackslash\hspace{0pt}}m{#1}}
\newcolumntype{C}[1]{>{\centering\let\newline\\\arraybackslash\hspace{0pt}}m{#1}}
\newcolumntype{R}[1]{>{\raggedleft\let\newline\\\arraybackslash\hspace{0pt}}m{#1}}

\usepackage{babel}

\makeatletter
\renewenvironment{abstract}{
  \begin{center}
    {\bfseries \large\abstractname\vspace{-1em}}
  \end{center}
  \begin{adjustwidth}{1.5cm}{1.5cm} 
  \noindent
}{%
  \end{adjustwidth}\par\vspace{3em}
}
\makeatother

\begin{document}

\title{\huge \onehalfspacing Assortative Mating, Inequality, and \\Rising Educational Mobility in Spain%
\thanks{\, We thank seminar participants at Banco de Espa\~na and Universidad Carlos III de Madrid. Ricard Grebol gratefully acknowledges financial support from the Swedish Research Council (Grant 2018-04581). Jan Stuhler acknowledges funding from the MICIU/AEI (PID2024-162138NB-I00 and CEX2021-001181-M). A previous version of this paper was presented under the title ``Educational inequality in Spain''. The opinions and conclusions expressed are solely those of the authors. }
}

\date{\today}

\author{%
\large
\begin{tabular}{@{}C{0.42\textwidth} C{0.42\textwidth}@{}}
\textbf{Ricard Grebol} &
\textbf{Margarita Machelett} \tabularnewline
{Uppsala University} &
{Banco de Espa\~na} \tabularnewline[1em]
\textbf{Jan Stuhler} &
\textbf{Ernesto Villanueva} \tabularnewline
{Universidad Carlos III de Madrid} &
{Banco de Espa\~na} \tabularnewline
\end{tabular}
}

\pagenumbering{gobble}
\maketitle

\begin{abstract}

\large
We study the evolution of intergenerational educational mobility and related distributional statistics in Spain. Over recent decades, mobility has risen by one-third, coinciding with pronounced declines in inequality and assortative mating among the same cohorts. To explore these patterns, we examine regional correlates of mobility, using split-sample techniques. A key finding from both national and regional analyses is the close association between mobility and assortative mating: spousal sorting accounts for nearly half of the regional variation in intergenerational correlations and also appears to be a key mediator of the negative relationship between inequality and mobility documented in recent studies.

\vspace{1em}
\noindent\textbf{Keywords:} intergenerational mobility, assortative mating, inequality, education
\par\vspace{0.5em}
\noindent\textbf{JEL Classification:} I24, J12, J62, N34, R11

\end{abstract}

\clearpage
\pagenumbering{arabic}
\setcounter{page}{1}
\doublespacing

\section{Introduction}

A key distributional characteristic is the extent of intergenerational mobility in a society, but evidence on how it varies over time or space -- and how it covaries with other characteristics -- remains limited, as many countries lack suitable data. Intergenerational studies require family data over two generations, but comparisons over time also require observing different cohorts, and regional comparisons need detailed geographic identifiers and large samples. 

In this paper, we study spatial and temporal trends in intergenerational educational mobility and related distributional characteristics in Spain. Combining several data sets, we first document long-term trends in inequality, mobility, and assortative mating for cohorts born from the mid-1920s through recent cohorts. To better understand these national patterns, we study ``regional correlates'' of mobility in the second part of the paper. We highlight two main findings. First, educational mobility has increased markedly in Spain. The intergenerational correlation in education has fallen by one third over the past four decades, and these gains in mobility coincided with reductions in inequality and assortative mating among the same cohorts. Second, much of the variation in intergenerational correlations -- both nationally and regionally, in the cross-section and over time -- can be accounted for by variation in assortative mating. Our contribution is thus twofold: we provide new evidence on long-run distributional trends in Spain, but also contribute to the broader debate of what factors drive variation in intergenerational mobility.

The main theme in both national and regional analyses is the strong association between intergenerational mobility and assortative mating. Conceptually, this link is intuitive: if children's outcomes are influenced by both parents, mobility will be lower in societies with less socio-economic mixing among spouses (e.g., \citealt{blinder1973model, kremer1997much, Ermischetal2006}). However, the strength of this relationship in our data is striking: nearly \textit{half} of the variation in intergenerational correlations across regions can be explained by differences in spousal sorting. In addition, assortative mating appears to be a key mediator of the negative association between cross-sectional inequality and intergenerational mobility, documented across countries (\citealt{Blanden2011Survey}, \citealt{CorakJEP2013}) and across regions within a country (\citealt{CHKS_Where_QJE,durlauf2022great, BattistonetalGatsby}).

To ensure broad coverage over both time and regions, we combine six data sets. Earlier birth cohorts are covered by the \emph{Encuesta Sociodemogr{\'a}fica} (ESD), conducted in 1991. The ESD provides detailed retrospective information on respondents' life courses and family backgrounds, and with 157,100 respondents, it offers an unusually large sample. For cohorts born between the early 1950s and the mid-1990s, we use all available waves (1977-2023) of the \emph{Spanish Labor Force Survey} (\emph{Encuesta de Poblaci{\'o}n Activa}, EPA). We also incorporate data from the 2001, 2011, and 2021 \emph{Censuses} to complement the time analysis and study regional variations. Finally, we use the \emph{Encuesta de Caracter{\'i}sticas Esenciales de la Poblaci{\'o}n y las Viviendas} (ECEPOV) to study mobility among the most recent cohorts. Importantly, both the ESD and ECEPOV report parental education regardless of whether children reside with their parents, allowing us to account for \emph{coresidence bias}, a central concern in the related literature (\citealt{Hilger2017}). 

In the first part of the paper, we study how (relative) educational mobility and related distributional characteristics evolved in Spain over the past 70 years. During this period, educational attainment expanded faster in Spain than in the US or in many other developed countries, as the country underwent major economic and political transformations. We document that these structural changes were accompanied by (i) a decrease in educational \emph{inequality}; (ii) a strong increase in \emph{intergenerational mobility}; (iii) a decrease in \emph{assortative mating}. We highlight the potential role of educational reforms in this process, especially the 1970 comprehensive school reform (\textit{General Education Law}), which coincided with the shifts in sorting and mobility documented in this study. The intergenerational correlation in schooling was nearly 0.5 in earlier cohorts, but declined by one-third for cohorts born after 1960, who attended school after the reform. 

The assortative patterns are particularly interesting. Consistent with earlier work (\citealt{fernandez2005love}), we confirm that assortative mating was much stronger in Spain than in other developed countries, with the spousal correlation in schooling reaching 0.7 for some cohorts. But we also find that the degree of sorting has changed considerably over time -- and that spousal sorting decreased in those cohorts affected by the 1970 comprehensive school reform, who also experienced increasing intergenerational mobility. These findings align with theoretical work that highlights a causal link between assortative and intergenerational mobility. However, while standard models imply a causal effect of assortative mating in the \emph{parent} generation on intergenerational mobility of their children (\citealt{blinder1973model, kremer1997much, Ermischetal2006,NybomZiegler2023}), we find that the shifts towards greater mobility and less assortative mating occur in the same cohorts. 

In the second part of our paper, we study ``regional correlates'' of mobility. Similar to \cite{CHKS_Where_QJE}, we divide the data into subprovincial units and major cities to analyze regional disparities in mobility and investigate the regional factors associated with low or high mobility levels. We find large regional differences, with the intergenerational correlation in schooling being 60\% larger among cities with low social mobility, such as C\'{a}diz or Toledo, as compared to cities with high mobility, such as Vigo or San Sebasti\'{a}n. Moreover, we find that such regional differences are highly persistent over time, not only in terms of the level of schooling, but also in its distributional properties as captured by the standard deviation, intergenerational correlation, and spousal correlation in schooling.

However, an explicit treatment of sampling error is crucial to arrive at this conclusion. While simpler statistics (such as mean schooling) are less sensitive, bivariate associations between ``complex'' regional statistics (such as intergenerational correlations) are heavily attenuated by sampling error. Moreover, in multivariate regressions, sampling error in one variable can contaminate the coefficients on others. These findings suggest that the recent literature on regional correlates is biased toward detecting simple correlates and may miss associations between mobility and more complex regional statistics. We address this problem using a split-sample IV estimator and show that it greatly affects our findings. For example, the estimated persistence of regional differences in assortative mating increases by a factor of four when switching from OLS to a split-sample IV estimator. 

We then study the relation between educational expansions and intergenerational mobility. Changes in the \textit{level} of schooling are not systematically linked to changes in mobility. However, educational expansions often coincide with changes in the \textit{dispersion} of schooling, and we find a strong negative correlation between inequality and intergenerational mobility. This evidence is in line with other recent studies documenting a negative association between inequality and mobility (dubbed the ``Great Gatsby Curve'', see \citealt{Blanden2011Survey}, \citealt{CorakJEP2013}, \citealt{CHKS_Where_QJE}, and \citealt{durlauf2022great}) across countries, and across regions within a country (\citealt{BattistonetalGatsby}). Adding to this literature, we show that this relationship holds not only across cross-sectional snapshots but also over time: conditional on region fixed effects, an increase in inequality is associated with a decrease in intergenerational mobility.

Finally, corroborating the findings of our national analysis, we find a robust statistical association between the strength of assortative mating in a region and its intergenerational mobility rate. This relationship is strong: across regions, nearly half of the variation in intergenerational correlations can be explained by differences in spousal sorting. After correcting for sampling error, a 0.1 increase in the spousal correlation in schooling among parents is associated with a 0.08 increase in the intergenerational correlation. Variation in assortative mating, therefore, appears to be a key factor both for understanding changes in mobility over time (national-level analysis) and differences and changes in mobility across regions (regional-level analysis). It also appears to be a key mediator in the negative association between inequality and mobility (the Great Gatsby Curve): using split-sample IV, spousal sorting can account for half of the association between inequality and immobility. 

Overall, our evidence is consistent with a vicious cycle between high inequality, strong assortative mating, and low mobility. More generally, it suggests that assortative patterns are a key to understanding differences in intergenerational mobility across populations. This finding is also relevant for policy, as assortative mating is potentially malleable by institutional settings. For example, \cite{holmlund2022preferences} shows that a comprehensive school reform in Sweden reduced assortative mating among affected cohorts, a finding that is consistent with the pattern observed for Spain. More generally, our arguments relate to recent work identifying socioeconomic segregation as a fundamental mechanism underlying the Gatsby Curve (\citealt{durlauf2022great}). If segregation fosters assortative matching (\citealt{Schwartz2013,mare2016educational}), policies aimed at reducing segregation should, through their effect on spousal sorting, also reduce inequality and increase mobility.

\textbf{Related literature.} Our results contribute to four distinct literatures. First, studies that track the long-run trends in intergenerational mobility, such as \cite{Hilger2017}, \cite{bailey2025changes}, and \cite{JacomeetalJPE} for the US, or \cite{KarlsonLanderso2025} for Denmark. These studies are data-intensive, requiring researchers to combine estimates from multiple data sets in a consistent manner. To provide such evidence for Spain, we need to address the potential issue of coresidence bias, which is also an important methodological concern in \cite{Hilger2017}. We discuss the optimal age to measure child education and the trade-off between selection and censoring bias that researchers face in this choice. In line with \cite{Hilger2017} and \cite{KarlsonLanderso2025}, we highlight the influence of educational reforms on long-run trends in educational inequality and mobility. Compared to the prior literature, we emphasize the important role of assortative mating in interpreting mobility trends. 

Second, we contribute to the literature on intergenerational mobility in Spain. Prior research, mainly in sociology, has focused on trends in occupational or social class mobility over time.\footnote{\, Earlier work found no systematic trend for men (\citealt{carabana1999estudios}), whereas more recent studies covering a longer time span find an increase in ``social fluidity'', the weakening association between parents' and children's social class, particularly among women (\citealt{fachelli2015we, gil2017intergenerational}).} Evidence on \textit{educational} mobility remains more limited. A major cross-country study on mobility trends by \cite{hertz2008} did not include Spain. However, \cite{dePablo2016} document rising educational mobility for cohorts born after 1965, a finding corroborated by a new \textit{Global Database on Intergenerational Mobility} (\citealt{van2024intergenerational}).\footnote{\, More generally, most sociological research points to a weakening link between parental background and child education in Spain (\citealt{ballarino2009persistent, esping2012asymmetries, di2012parental, gil2017intergenerational, barone2018educational, GilHernandezetal2020}), although some studies argue that ``inequality of educational opportunities'' remained fairly constant (\citealt{carabana2004educacion, fernandez2022como}).} Our analysis extends this literature in three ways. First, by combining multiple datasets, we trace mobility trends over a uniquely long horizon, covering cohorts born from the mid-1920s to the present. Second, large sample sizes allow us to describe regional differences and to pinpoint the magnitude and timing of shifts, especially the strong gains in educational mobility after the 1970 comprehensive school reform. Third, by tracking the co-movement between mobility and other distributional trends, on both national and regional levels, we can link rising mobility to a concurrent decline in assortative mating.\footnote{\, Economic research on Spanish mobility trends is limited. A notable exception is \cite{Guell10122014}, whose innovative surname-based estimator points to declining educational mobility in \textit{Catalonia}. While seemingly at odds with the finding of rising mobility in other studies, we show that part of the increase in mobility -- and most of the decline in assortative mating -- occurred in later cohorts not yet covered by the 2001 Census used in \cite{Guell10122014}.}

Third, we contribute to the literature on regional differences in intergenerational mobility. Following \cite{CHKS_Where_QJE}, numerous studies have explored whether such differences correlate with other local characteristics (see \citealt{HHBlandenDoepkeStuhler22}). We add to this literature in three ways. First, using a regional design, we document a strong link between intergenerational mobility and assortative mating. Differences in the latter can explain much of the variation in mobility, and also account for half of the association between inequality and mobility across regions. Second, leveraging the long coverage of our data, we show that the negative relationship between inequality and mobility (the ``Gatsby Curve'') holds not only in the cross-section but also when examining changes over time. Third, we make a methodological point, demonstrating that sampling error distorts the search for regional correlates by biasing different correlates very differently, and creates both attenuation and contamination biases. We address this problem using split-sample IV techniques. 

Fourth, we contribute to the literature on assortative mating and its relation to intergenerational mobility. Standard models imply that spousal sorting reduces mobility and increases inequality in the next generation (\citealt{blinder1973model, kremer1997much, FernandezRogerson2001}). 
However, causal evidence on this question remains scarce, due to the lack of exogenous variation in assortative mating.\footnote{\, As an interesting exception, see \cite{goni2022assortative}. Other work decomposes the contribution of spousal sorting to intergenerational persistence in Germany, the UK, and Sweden (\citealt{Ermischetal2006, holmlund2022marital}); documents parallel trends in assortative mating and inequality in the United States (\citealt{mare2016educational}); and quantifies the role of assortative mating in household income inequality across countries (\citealt{eika2019educational}).} We document a strong co-movement of spousal sorting, educational inequality, and intergenerational mobility in Spain. The near-simultaneous shifts in assortative mating and inequality suggest that the causal link may not run solely from spousal sorting to inequality in the \textit{next} generation, as standard models imply, but also from rising inequality to increased sorting within the \textit{same} generation (\citealt{fernandez2005love,mare2016educational}). While co-movements on the national level might be a coincidence, we show that differences in spousal sorting also explain much of the regional variation -- and regional changes -- in intergenerational mobility. Overall, assortative mating emerges as a key factor in understanding variation in intergenerational mobility, and the recent rise in educational mobility in Spain appears partly driven by a marked decline in spousal sorting.

\section{A simple model of assortative mating, inequality, and mobility}\label{sec:theory}

To help interpret our empirical results, we first provide a simple model that links inequality, assortative mating, and intergenerational mobility. Under standard assumptions, assortative mating increases inequality and reduces mobility (e.g., \citealt{kremer1997much}). Perhaps less obvious is the possibility of a reciprocal effect of inequality on assortative mating -- and therefore mobility. While such a reciprocal effect follows less mechanically, it is also intuitive. Economic considerations may become more salient in spousal search when inequality is higher (\citealt{fernandez2005love}), or inequality may increase segregation (\citealt{DurlaufSeshadri2018}) and reduce the likelihood that individuals with different educational backgrounds meet (\citealt{mare2016educational}).\footnote{\, As noted by \cite{mare2016educational}, ``\textit{... inequality has potentially profound effects on social organization, lowering the chances that persons from different socioeconomic strata come in contact with one another, raising the stakes for choosing to associate with some kinds of people relative to others, and altering the kinds of social arrangements that result}''.}

To illustrate these ideas, consider an overlapping-generations economy in which individuals form households, produce children, and transmit human capital to the next generation. Each individual $i$ is endowed with a scalar trait $e_i$, which corresponds to educational attainment in our application. Education in generation $t$ is distributed according to $e_i \sim F_t(e)$, with mean $\mu_t$ and variance $\sigma_t^2$. 

Individuals form couples through assortative matching on $e$. Let the degree of assortative mating in generation $t$ be summarized by a parameter $\rho_t \in [0,1]$, where $\rho_t = 0$ corresponds to random mating and $\rho_t = 1$ to perfect assortative mating, such that the expected education of spouse $j$ of individual $i$ satisfies
\begin{equation}
\mathrm{E}[e_j | e_i] = \mu_t + \rho_t (e_i - \mu_t).
\end{equation}

Finally, children's education depends on the average of their parents' education and an idiosyncratic shock:
\begin{equation}
e_i' = \lambda_t \frac{e_i + e_j}{2} + \varepsilon_i,
\end{equation}
where $\lambda_t \in (0,1)$ captures the strength of intergenerational transmission in generation $t$ and $\varepsilon_i$ is i.i.d.\ with mean zero and variance $\sigma_\varepsilon^2$.

\textbf{The effect of assortative mating on inequality and mobility.}
Under this structure, the variance of children's education in the next generation is
\begin{equation}
\sigma_{t+1}^2
= \lambda_t^2 \operatorname{Var}\left(\frac{e_i + e_j}{2}\right) + \sigma_\varepsilon^2
= \frac{\lambda_t^2(1+\rho_t)}{2}\sigma_t^2 + \sigma_\varepsilon^2,
\end{equation}
and the intergenerational correlation in education between one parent and their child equals
\begin{equation}
\frac{\operatorname{Cov}(e_i,e_i')}{\operatorname{Var}(e_i)} 
= \frac{\lambda_t (1+\rho_t)}{2}.
\end{equation}
An increase in assortative mating $\rho$ therefore increases the intergenerational correlation (i.e., reduces mobility) and raises inequality in the \textit{next} generation. These implications follow directly from the assumption that a child's education depends on the endowment or education of both parents.

\textbf{A reciprocal effect of inequality on assortative mating.}
Moreover, there might also be a reciprocal effect from inequality on assortative mating within the \textit{same} generation. To understand why, suppose that partners derive utility from the sum of their education as well as from an idiosyncratic match-specific component:
\begin{equation}
U_{ij} = e_i + e_j + \ell_{ij},
\end{equation}
where $\ell_{ij}$ captures non-educational aspects of match quality such as preferences or compatibility (``love'').

When educational inequality $\sigma_t$ is low, differences in years of schooling across potential partners are small. As a result, idiosyncratic match quality dominates mate choice, and matching by education is relatively weak. By contrast, when educational inequality is high, differences in schooling between potential partners are large, making partner education a more salient attribute in the marriage market. Individuals then become less willing to marry partners whose schooling differs substantially from their own -- assortative mating $\rho_t$ increases. \cite{fernandez2005love} develop an equilibrium model of marital sorting that formalizes this mechanism and generates a positive correlation between sorting and wage inequality. Alternatively, rather than operating through match utilities, inequality may increase segregation and reduce the likelihood that individuals with different educational backgrounds meet in the first place, increasing sorting (\citealt{mare2016educational}). 

\textbf{Dynamics.} Together, these mechanisms generate a feedback loop between inequality and assortative mating,
\[
\begin{cases}
\sigma_{t+1}^2 = f(\rho_t, \sigma_t^2), \\
\rho_{t} = g(\sigma_t^2),
\end{cases}
\qquad \text{with } f_\rho > 0,\ g_\sigma > 0.
\]
which in turn affects mobility. The model thus admits a vicious cycle in which high inequality, strong assortative mating, and low intergenerational mobility reinforce one another. However, it also implies that policies reducing educational inequality may, indirectly, weaken assortative mating and increase mobility.

\section{Institutional context and data}
\label{sec_data_method} 

\subsection{Spain's educational system in the 20th Century}
\label{sec_institutions} 

The Spanish educational system underwent major transformations in the 20th century, shaped by three key legislative acts: the Public Instruction Law in 1857 (\emph{Ley de Instrucci\'on P\'ublica}), the General Education Law in 1970 (\emph{Ley General de Educaci\'on, LGE}), and the General Organic Law of the Educational System in 1990 (\emph{Ley Org\'anica General del Sistema Educativo, LOGSE)}.\footnote{\, For further information on the legal framework of the Spanish educational system, see \href{https://www.boe.es/buscar/doc.php?id=BOE-A-1857-9551}{\emph{``Ley de Instrucci\'on P\'ublica''}}, \href{https://www.boe.es/buscar/doc.php?id=BOE-A-1970-852}{\emph{``Ley General de Educaci\'on'' (LGE)}}, and \href{https://www.boe.es/buscar/doc.php?id=BOE-A-1990-24172}{\emph{``Ley Org\'anica General del Sistema Educativo'', LOGSE}} (in Spanish).} Preceding our study period, the Public Instruction Law was introduced to address high illiteracy rates by creating a uniform education system that made primary education tuition-free and compulsory for children aged 6 to 9. Later reforms introduced a two-track compulsory school system and raised the compulsory schooling age to 12. A major overhaul came with the 1970 LGE, which extended compulsory education to age 14 and replaced the two-track system with a uniform eight-year comprehensive school (\textit{Educaci\'on General B\'asica, EGB}), postponing the choice between vocational and academic tracks from age 10 to 14. This reform was part of a broader set of comprehensive school reforms throughout Europe in the 1960s aimed at broadening educational access and promoting economic growth.\footnote{\, For example, see \citealt{meghir2005educational} for Sweden, \citealt{pekkarinen2009school} for Finland, \citealt{aakvik2010measuring} for Norway, or \citealt{pischke2006comprehensive} for England and Wales.} Finally, the 1990 LOGSE extended compulsory schooling to age 16 and established a unified structure for compulsory secondary education (\textit{Educaci\'on Secundaria Obligatoria, ESO}). After completing ESO, students could pursue an academic track (\textit{Bachillerato}) leading to university or a vocational track (\textit{Formaci\'on Profesional, FP}) providing professional qualifications. Additional institutional details are provided in Appendix \ref{apinstitutional}.

\subsection{Data sources} 
\label{sec_data} 

To provide a comprehensive analysis of educational inequality, mobility, and assortative mating across both time and regions, we combine six data sources. These data sources have distinct strengths and weaknesses, allowing us to address potential biases and test the robustness of our results.

First, we use the \textit{Encuesta Sociodemogr{\'a}fica} (ESD, Sociodemographic Survey), conducted in 1991 by the Spanish Statistical Office (\citealt{ine1991encuesta}). We use this retrospective survey to cover cohorts born between the mid-1920s and early 1960s, and to address key methodological issues that arise in mobility studies. The ESD offers three main advantages. (1) It is representative of the Spanish population and unusually large for such a detailed survey, with 157,100 individuals (excluding 1,166 foreign-born). (2) Its retrospective design provides rich life-course data on residential moves, employment histories, and key life events, such as the age of leaving the parental home or the start and completion dates of each educational qualification. (3) It includes extensive information on family members -- parents, siblings, spouses, and children -- independently of the respondent's household composition at the time of the interview. Crucially, it reports parental education even when respondents no longer lived with their parents, avoiding the coresidence biases that affect many mobility studies (\citealt{Hilger2017, emran2018measure, munoz2021measure}). This makes the ESD a valuable benchmark for assessing such biases in other datasets (see Section \ref{sec_method_issues}). 

Second, we use the \textit{Encuesta de Poblaci{\'o}n Activa} (EPA, Labor Force Survey), a quarterly labor force survey conducted by the Spanish Statistical Office (\citealt{ine2021epa}). Combining waves from 1977 to 2023, we can use these data to study cohorts born from the early 1950s to the mid-1990s, and to cross-check estimates that overlap with the ESD. Unlike the ESD, the EPA is a household survey and records parental education only for children still living with their parents. A key methodological issue we address below is whether these co-resident samples are representative of the wider population.

Third, we draw on microdata from the Spanish Population and Housing Censuses for 2001 (5\% sample), 2011 (10\% sample), and 2021 (10\% sample), respectively. We use these data to verify recent trends in educational inequality, as estimated in the EPA. Moreover, owing to their large size and detailed geographic coverage, these datasets are ideal for studying regional correlates of inequality and mobility among recent birth cohorts. Like the EPA, the Census records parental education only for coresident children.

Finally, we use the \emph{Encuesta de Caracter{\'i}sticas Esenciales de la Poblaci{\'o}n y las Viviendas} (ECEPOV, Population and Household Essential Characteristics Survey), a survey that is complementary to the 2021 Census (\citealt{ine2021ecepov}). Its main advantage is that it reports parental education information regardless of the child's coresidence status, allowing us to assess whether our estimates for recent cohorts are sensitive to coresidence bias. Its main limitations are a smaller sample size and limited information on respondents' partners' education.

\subsection{Variable definitions and comparability of sources} 
\label{sec_variables} 

Our primary outcome of interest is educational attainment, which offers several advantages. It is an important outcome in its own right and a central mediator of socioeconomic status in standard theories of intergenerational transmission (\citealt{Goldthorpe2014}). Moreover, because log earnings are approximately linear in years of schooling, education-based distributional statistics tend to be a good proxy for those in log permanent earnings (\citealt{kremer1997much}).\footnote{\, In line with this hypothesis, measures of educational and income mobility are closely related; for example, the correlation between estimates of income mobility and educational mobility is around 0.7 across developed countries (\citealt{Stuhler2018JRC}).} Education is typically completed at a young age, reducing exposure to life-cycle biases. And unlike tax data, it is not affected by reporting thresholds or misreporting, and can be tracked over longer periods. Still, our results regarding educational mobility may differ from broader trends in socioeconomic mobility.

In all datasets, the term \textit{spouse} refers to both marital and cohabiting unions, regardless of whether the relationship is legally formalized. We measure \textit{education} as the number of years of schooling completed. The level of detail varies across datasets: the ESD reports specific degrees, while the EPA, ECEPOV, and Census provide broader categories that group multiple qualifications. To ensure comparability across surveys and cohorts, we adapt the classification scheme of \cite{martinez1996construccion} into seven broad education categories (with assigned years of schooling): Illiterate (1), Literate (3), Primary Schooling (5), Secondary School (8), Academic high school and professional studies (11), Short college degree (15), and Long college degree (18). Details are provided in Appendix \ref{apvardefs1}. 

For our regional analysis, we disaggregate our samples using a subprovincial definition that separates all municipalities with at least 100,000 inhabitants in 1991 and aggregates the remaining observations from each province. For example, in C\'{a}diz province, we distinguish Algeciras, C\'{a}diz, and Jerez de la Frontera, as well as the rest of the province. This approach allows us to track educational mobility and its regional correlates consistently for 107 provinces and cities, across our main sources, and covering cohorts born from the 1930s to the mid-1990s. See Appendix \ref{apvardefs2} for a list of included municipalities. In Section \ref{extensions}, we compare our measures of educational mobility to estimates of income mobility from the ``Atlas de Oportunidades'' (\citealt{LlanerasAtlas2020}),  identifying 96 provinces and cities that appear in both sources.

\subsection{Methodological issues} 
\label{sec_method_issues} 

Our estimates are subject to two key methodological issues: (i) potential ``coresidence bias'' in the EPA or the censuses, and (ii) attenuation bias from sampling error when partitioning samples to study regional correlates. We outline these issues here and provide detailed evidence in the following sections. 

\textit{Coresidence bias.} Many countries lack suitable data sets that track family linkages and socioeconomic information across generations, complicating the analysis of long-term mobility trends. Instead, researchers often rely on censuses or household surveys that capture parent-child links only when children still live at home.\footnote{\, An alternative strategy is to use \textit{names} to track intergenerational mobility on the name rather than the individual level; see \cite{Clark2014book,Guell10122014,Pasermanetal2015} and \cite{BaroneMocetti2016} for prominent examples, and \cite{SantarvitaStuhlerEJ} for a survey of the recent literature.} Examples include \cite{Cardetalmobility2022} on intergenerational mobility in the early 20th-century US and \cite{derenoncourt2022can} on Black educational mobility between 1940 and 2015. \cite{alesina2021intergenerational} and \cite{Munoz2021} study mobility trends in Africa and Latin America using harmonized Census data, while \cite{van2024intergenerational} compile a global mobility database by combining retrospective survey data on parental education with household surveys of coresident children.

Coresidence restrictions may introduce bias if the mobility patterns of coresident children differ from those who have left the parental household. In particular, researchers face a trade-off between two sources of bias. Measuring education too early may induce \emph{censoring bias}, as some children have not yet completed schooling, while measuring it later increases \emph{selection bias}, since older coresident children constitute a smaller and potentially more selected group. The direction and magnitude of these biases depend on the age at measurement, context, and the mobility measure used (\citealt{emran2018measure}; \citealt{munoz2021measure}).\footnote{\, The literature on long-run income mobility faces similar methodological challenges, including linking errors, representativeness concerns, and comparability across time; see \cite{abramitzky2025intergenerational} for a review.}

However, coresidence bias is less of an issue in the Spanish context. First, children tend to co-reside with their parents for longer: 55\% of those born in the 1960s and 75\% of those born in the 1990s were still coresident at age 27, compared with only 23\% of U.S. children born in 1985 \citep{USeduc}. Second, education in Spain is typically completed earlier: by age 27, 95\% of the 1960 cohort and 86\% of the 1980 cohort had finished schooling.\footnote{\, Author's calculations from the EPA.} Moreover, the ESD and ECEPOV record parental education regardless of household composition, allowing us to assess coresidence bias directly. Using the ESD, we show that measuring education around age 27 minimizes the combined impact of censoring and selection bias (Appendix \ref{Appendix-Coresidence}). Consistent with this claim, ECEPOV estimates are similar whether or not we restrict the sample to coresident households, and results are robust to applying the bias-correction method of \cite{Hilger2017}.

\textit{Sampling error.} A second methodological issue arises when studying mobility at the regional level. Although our data sources are large, sampling error becomes a concern when disaggregating data into many regions. While not biasing the mobility statistics themselves, sampling error attenuates the associations between mobility and \textit{other} regional statistics (``regional correlates''). In Section \ref{sec:regional}, we show that this bias can be sizable, even in large data sources such as the Census. More critically, its magnitude varies depending on the correlate of interest: the bias is much larger for ``complex'' measures, such as spousal correlations, than for simpler ones, such as mean schooling. Moreover, in multivariate regressions, sampling error can introduce ``contamination bias'' that inflates some coefficients (\citealt{van2024teacher}). To remove bias from sampling error and provide reliable evidence on regional correlates, we employ a two-sample IV strategy.

\section{National level and trends}

In this section, we analyze the evolution of educational inequality in Spain over the past 70 years. Our analysis covers cohorts born in the mid-1920s, who attended primary school during the Second Spanish Republic (1931-1939), through those born in the mid-1990s, who completed their education in recent years. We study educational mobility, measured by the intergenerational correlation (IGC) and regression coefficient (IGR) in years of schooling, but also educational inequality and spousal sorting. We show that trends in inequality, sorting, and mobility are closely linked, consistent with the idea that the former are key determinants of the latter -- a hypothesis explored further in Section \ref{sec:regional}. 

\subsection{Trends in educational attainment and inequality}
\label{sec_trend_levels}

\begin{figure}[!htb]
\begin{subfigure}{0.82\textwidth}
  \centering
  \includegraphics[width=.995\linewidth]{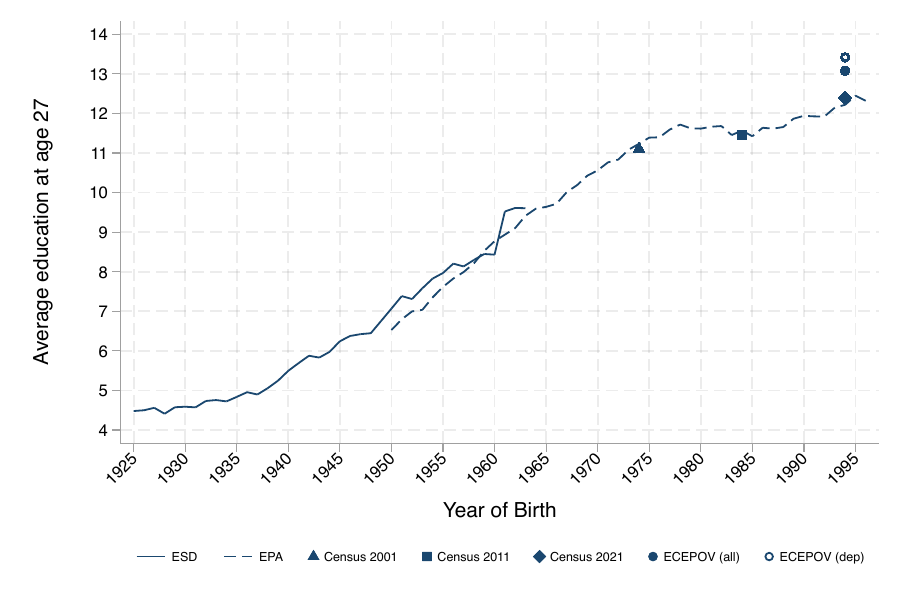}
  \subcaption{Average years of schooling}
  \label{fig:national_trends_education_a}
\end{subfigure}% 

\begin{subfigure}{0.82\textwidth}
  \centering
  \includegraphics[width=.995\linewidth]{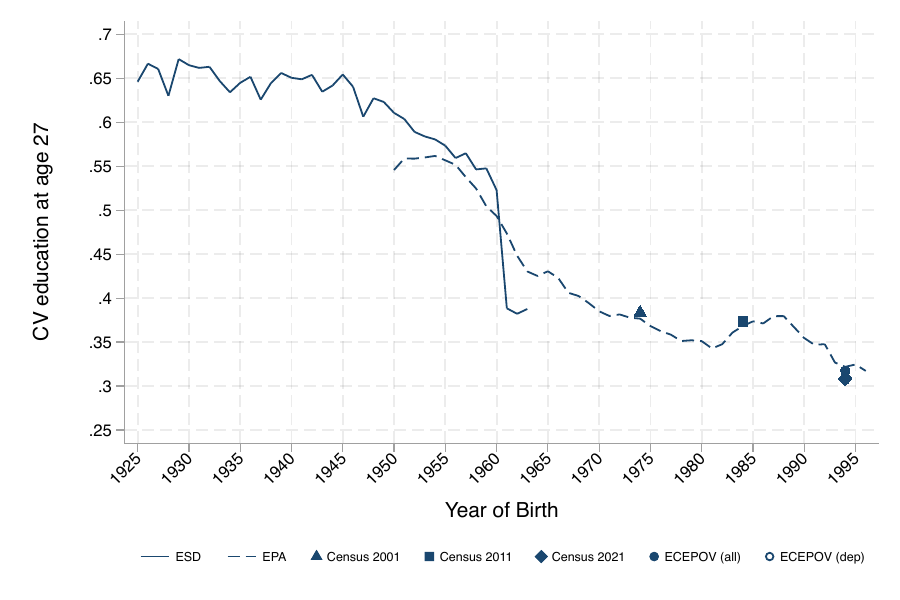}
  \subcaption{Coefficient of variation in years of schooling}
  \label{fig:national_trends_education_b}
\end{subfigure}
\caption{Trends in educational attainment in Spain}
\label{fig:national_trends_education}
\floatfoot{Notes: Average years of schooling and coefficient of variation (CV) at age 27 for cohorts born 1925-1996. Years of schooling are harmonized across six data sources (ESD, EPA, ECEPOV, and the 2001, 2011, and 2021 Censuses), see Appendix \ref{apvardefs1}. For ECEPOV, separate measures are shown for dependent children living with their parents (\textit{dep}) and for the full sample (\textit{all}).}
\end{figure}

Spain experienced a steady rise in educational attainment throughout the 20th century, mirroring trends in other advanced economies. Figure \ref{fig:national_trends_education_a} shows the average years of schooling at age 27 for cohorts born between 1925 and 1996. The 1925 cohort averaged only 4.5 years of education, less than a completed primary education. Although the reorganization of the education system under Franco's regime following the Spanish Civil War (1936-1939) delayed improvements that would otherwise have occurred earlier (\citealt{GrebolNavarro2025})\footnote{\, Following the Spanish Civil War, the regime dismissed about 10\% of primary teachers as part of an ideological purge, reversing education efforts undertaken by the Second Spanish Republic (1931-1939). Dismissal rates were even higher at upper levels of the education system (e.g., 23\% for full professors). Although the consolidation of the regime delayed progress for nearly a decade, legal changes in 1945 extending compulsory schooling to age 12 eventually raised attainment levels.}, educational attainment increased gradually for cohorts born from the 1940s onward. By the 1975 cohort, average schooling exceeded 11 years, corresponding to upper-secondary education. Attainment stalled for cohorts born in the late 1970s and early 1980s before resuming its upward trajectory, with early 1990s cohorts attaining an average of 13 years of schooling.\footnote{\, Throughout this period, ``dependent'' children residing in the parental household at age 27 had higher average schooling than ``independent'' children living outside (Figure \ref{fig:national_educ_bycoresidence}).} 

To assess changes in education inequality over time, we use the coefficient of variation (CV, \citealt{Atkinson1970}) as our main measure of dispersion. Figure \ref{fig:national_trends_education_b} shows that the CV remained relatively stable for cohorts born before 1960, educated under the two-track school system.\footnote{\, During this period, the standard deviation of schooling increased proportionally with average attainment (Figure \ref{fig:national_sdeduc_a}), such that the CV remained stable.} Inequality then fell sharply for cohorts born after 1960, who were subject to the comprehensive school reform of 1970. Rising attainment coincided with a more compressed distribution of outcomes, leading to a marked decline in educational inequality.

These findings are robust across our (harmonized) data sources. The ESD and EPA report similar levels and trends for cohorts born in the 1950s, with relatively stable CV levels (Figure \ref{fig:national_trends_education_b}) and rising standard deviations (Figure \ref{fig:national_sdeduc_a}) until the 1960 cohort, followed by a decline in educational inequality. The ESD shows a sharper drop, likely due to its more detailed reporting of educational categories around the 1970 educational reform (see Section \ref{sec_institutions}). % JS: We probably should have imputed fewer years of schooling for the first cohorts affected by the reform; see the Figure in Espadafor and Garcia-Sierra for details.
A similar difference is apparent in other mobility measures discussed below. For later cohorts, estimates align closely across sources. 

While other countries also saw rising attainment, Spain's progress was especially pronounced. Using data from \cite{barro2013new} to compare the evolution of schooling in Spain and the United States, US attainment was much higher for cohorts born between 1925 and 1950, but the gap narrowed considerably for later cohorts (see Figure \ref{USfigure}). %Reassuringly, the data in \cite{barro2013new} aligns closely with our findings on levels and trends of schooling across data sources, as shown in Figure \ref{fig:national_trends_education_a}.
A key driver was rising female attainment (Figure \ref{fig:national_educ_gender_a}). Women, who averaged half a year less schooling than men in the early 20th century, overtook men in the 1960s cohorts, with the gap exceeding one year for recent cohorts.\footnote{\, Educational inequality, as measured by the CV, follows similar trends for both genders (Figures \ref{fig:national_educ_gender_b}), though the post-1970 reform decline is slightly more pronounced for women. Trends in the standard deviation of schooling by gender are reported in Figure \ref{fig:national_sdeduc_b}.}

\subsection{Trends in intergenerational mobility}
\label{sec_trend_mobility}

Our main measure of relative mobility is the intergenerational correlation (IGC) between children's and fathers' years of schooling, $y_i$ and $y_i^{p}$.\footnote{\, We focus on fathers' rather than the mothers' or average parental education because mothers' educational attainment was low among earlier cohorts. Consistent with \cite{dePablo2016}, we find that mother-child correlations are smaller than father-child correlations in Spain but follow similar trends.} We also report the intergenerational regression coefficient (IGR), estimated as the slope coefficient in a regression of child schooling on paternal schooling. The two measures are thus related by:
\begin{equation}
IGR = IGC\times\frac{SD(y_i)}{SD(y_i^{p})},
\end{equation}
where $SD(y_i)$ and $SD(y_i^{p})$ are the standard deviations of child and father schooling, respectively. Although both measures are complementary, we prefer the IGC for studying long-run trends since it tends to be less volatile and depends less mechanically on changes in dispersion during periods of educational expansion (\citealt{hertz2008}).\footnote{\, The literature uses different measures for educational mobility, but the IGC and IGR are the most popular measures in the economic literature. They provide only a linear approximation, but the conditional expectation of children with respect to fathers' education is approximately linear in Spain (see Figure \ref{fig:child_father_educ_densities}).} We minimize coresidence bias by measuring children's education at age 27, as discussed in Section \ref{sec_coresidencebias} and Appendix \ref{Appendix-Coresidence}. We also show that estimates from sources restricted to coresident children (e.g., Census) closely match those obtained from sources that are not (e.g., ECEPOV).

\begin{figure}[htbp]
\begin{subfigure}{0.82\textwidth}
    \centering
    \includegraphics[width=.995\linewidth]{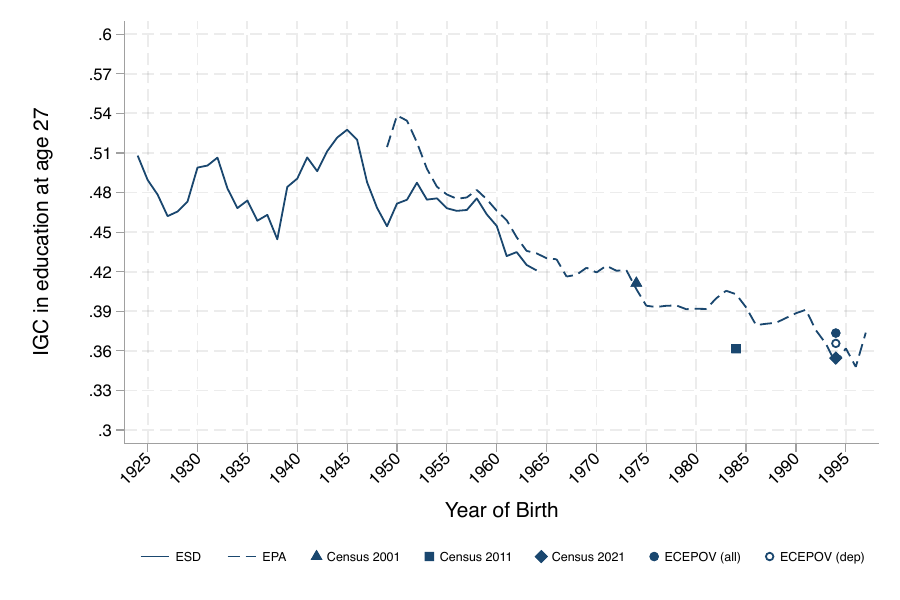}
    \subcaption{IGC in education}
    \label{fig:national_mobility_a}    
\end{subfigure}

\begin{subfigure}{0.82\textwidth}
    \centering
    \includegraphics[width=.995\linewidth]{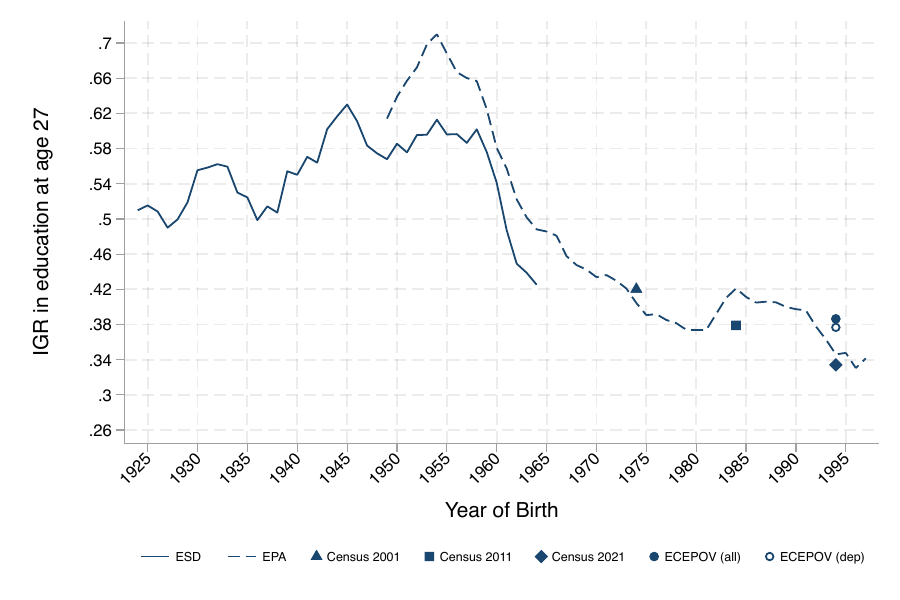}
    \subcaption{IGR in education}
    \label{fig:national_mobility_b}    
\end{subfigure}
\caption{Trends in intergenerational educational mobility in Spain}
\label{fig:national_mobility}
\floatfoot{Notes: Three-year moving average of the intergenerational correlation (IGC) and regression coefficient (IGR) in years of schooling, measured at child age 27. Years of schooling are harmonized across six data sources (ESD, EPA, ECEPOV, and the 2001, 2011, and 2021 Censuses). For ECEPOV, separate estimates are shown for dependent children living with their parents (\textit{dep}) and for the full sample (\textit{all}).}
\end{figure}

Figure \ref{fig:national_mobility_a} shows the time trend in the IGC from cohorts born in 1925 until cohorts that recently completed education. Educational mobility remained relatively stable during the first half of the analysis period: the IGC hovered between 0.45 and 0.50 for cohorts born between 1925 and 1960, with minor fluctuations that also reflect sampling error. The IGR (Figure \ref{fig:national_mobility_b}) rose slightly over the same period, from 0.50 for the 1925 cohort to 0.60 for late-1950s cohorts, consistent with the rising dispersion of schooling (Figure \ref{fig:national_sdeduc_a}). Compared to other Western European countries, Spain thus had low educational mobility; for instance, \cite{hertz2008} report an average IGC of 0.39 for 1950s cohorts in Western European countries, while \cite{van2024intergenerational} report an average IGR of 0.37 in high-income countries. 

However, from around the 1961 cohort onward -- the first affected by the 1970 comprehensive school reform -- mobility increased markedly. The IGC declined from 0.47 for the 1960 cohort to 0.35 for the 1995 cohort; the IGR dropped even more sharply, from 0.57 to 0.33. Notably, this large gain in educational mobility (reduced persistence) unfolded over nearly four decades, spanning the final years of Franco's regime, Spain's democratic transition, and its admission to the European Union in 1986, and into the present.

Our finding of increasing educational mobility is consistent across data sources and alternative mobility measures. ESD and EPA track similar declines, though EPA levels are slightly higher. EPA estimates align closely with the point estimates from the Censuses and ECEPOV, reinforcing the conclusion that coresidence bias is limited. Mobility trends are also robust to the age at measurement (trends at ages 24 and 30 are similar, see Figures \ref{fig:national_IGC_age} and \ref{fig:national_IGR_age}), to the coding of parental education, and to the use of alternative mobility measures.\footnote{\, In our baseline specification, we focus on the father's education, as the mother's education was very low in earlier cohorts. If we use the highest parental education instead, the IGC remains the same for earlier cohorts (when the maximum is typically the fathers' education) but exhibits a flatter post-1960 decline (Figure \ref{fig:robustness_IGC_parmax}). The time trend remains very similar when using Spearman's rank rather than Pearson's correlation coefficient (Figure \ref{fig:robustness_IGC_parmax_rank}): while the rank correlation is slightly lower, it closely tracks the IGC over time, suggesting that the observed trend is not driven solely by changes in marginal distributions.}

Our finding of a decline in educational persistence for cohorts affected by the 1970 LGE reform may be surprising, as the reform's effect on inequality and mobility has been contested. Recently, \cite{espadafor2025intergenerational} found no significant differences in the extent to which children from high- and low-SES families benefited from the reform. One potential explanation for the divergence from our results is the larger sample sizes in our data sources, which allow us to trace mobility trends with greater precision.\footnote{\, Based on data from the \textit{Survey of Health, Ageing and Retirement in Europe}, the analysis sample in \cite{espadafor2025intergenerational} includes 2,072 observations for cohorts 1953-1963, compared with 29,849 observations in the ESD and 35,375 in the EPA for the same cohorts.} Another is the longer time horizon of our analysis, which can reveal more gradual changes in persistence.

The decline in educational persistence in Spain mirrors, but exceeds, global trends. \cite{van2024intergenerational} report that persistence fell only modestly in high-income countries -- on average, the IGR fell from 0.37 for cohorts born in the 1950s to 0.33 for those born in the 1980s. % Figure 15 in their paper 
And while comprehensive school reforms have reduced persistence in other countries (e.g., \citealt{Holmlund2006,meghir2005educational}), the magnitude of change in Spain is striking: between the 1950s and 1980s cohorts, the IGR fell by 0.24 points, about six times the 0.04-point decline observed in the high-income countries considered by \cite{van2024intergenerational}, aligning Spain's mobility levels closer to those observed in other European countries.\footnote{\, \cite{KarlsonLanderso2025} document a similarly strong increase in educational mobility in Denmark, where major school reforms benefited children from disadvantaged backgrounds. They further show that college expansions reduced mobility, a pattern not -- at least so far -- evident in Spain.}

We thus find that both educational inequality (Figure \ref{fig:national_trends_education_b}) and persistence (Figure \ref{fig:national_mobility_a}) fell for cohorts affected by the 1970 comprehensive school reform, and continued to improve in subsequent decades. A similar association between inequality and mobility trends (``Great Gatsby Curve over time'') has been found in other contexts, including the US (\citealt{BattistonetalGatsby}) and historical China (\citealt{Shiue2025}). In Section \ref{subsec:region_expansion}, we show that it holds across Spanish regions as well, both in levels and changes, consistent with the theoretical link between inequality and mobility (Section \ref{sec:theory}). 

\begin{figure}[t!]
\centering
\includegraphics[width=.82\linewidth]{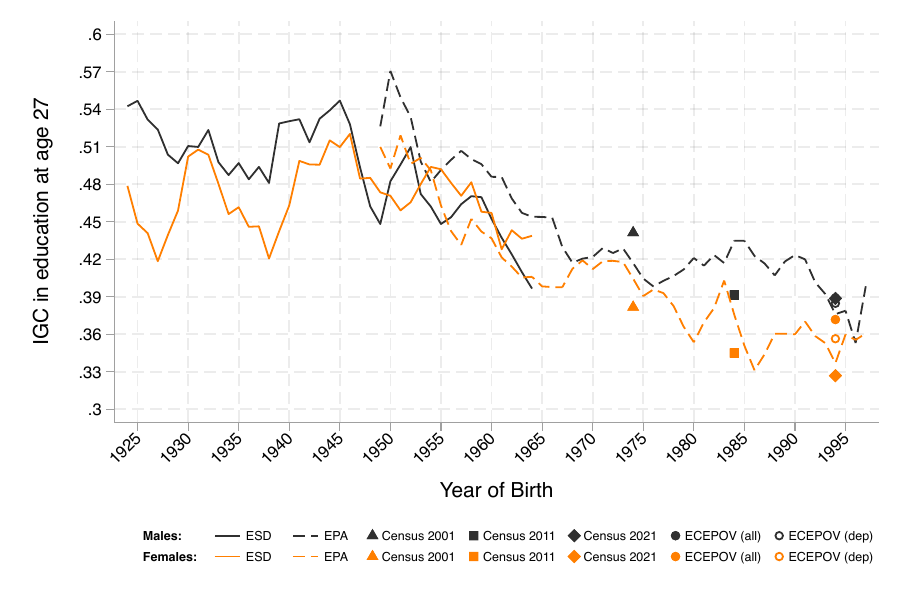}
\caption{Trends in the intergenerational correlation in education in Spain by gender}
\label{fig:national_IGC_gender}
\floatfoot{Notes: Three-year moving average of the intergenerational correlation (IGC) for male (black) and female children (orange), based on child schooling at age 27 and father's schooling for the parent generation. Years of schooling are harmonized across six data sources (ESD, EPA, ECEPOV, and the 2001, 2011,
and 2021 Censuses). For ECEPOV, separate estimates are shown for dependent children living with their parents (\textit{dep}) and for the full sample (\textit{all}).}
\end{figure}

Figure \ref{fig:national_IGC_gender} shows that while the pre-reform levels were similar, the subsequent decline in educational persistence was more pronounced for women than men. Among recent cohorts, women's mobility exceeds men's by about 0.05. The IGR exhibits a similar post-reform pattern.\footnote{\, The IGR rose strongly for women for the 1925-1960 cohorts (Figure \ref{fig:national_IGR_gender}), reflecting increasing variance in women's educational attainment (Figure \ref{fig:national_sdeduc_b}).}  This gender divergence is consistent with prior sociological evidence documenting larger mobility gains for Spanish women born after 1970 and a weakening of the link between social background and educational attainment \citep{gil2017intergenerational}. 

\subsection{Trends in assortative mating}
\label{sec_trend_assortative}

A key factor influencing intergenerational mobility is assortative mating. When spouses sort on education, children are exposed to the combined advantages or disadvantages of both parents rather than mixed parental backgrounds, decreasing mobility, and increasing inequality (see Section \ref{sec:theory}). However, there exists little systematic evidence on how assortative mating has changed over time. To provide such evidence for Spain, we measure assortative mating as the spousal correlation in years of schooling, as is standard in the literature \citep{greenwood2014marry,eika2019educational,gihleb2020educational}.\footnote{\, Evidence from extended kins suggests that the spousal correlation in schooling  understates the strength of assortative mating in the population (\citealt{ColladoOrtunoStuhlerKinship}). However, our interest here is not in levels but in changes over time.} 
To ensure that most spousal links have formed, we measure the spousal correlation when at least one partner is aged 35, but results remain similar if measured at ages 30 or 40 (see Figure \ref{fig:national_AM_age}).

\begin{figure}[t!]
\centering
\includegraphics[width=.82\linewidth]{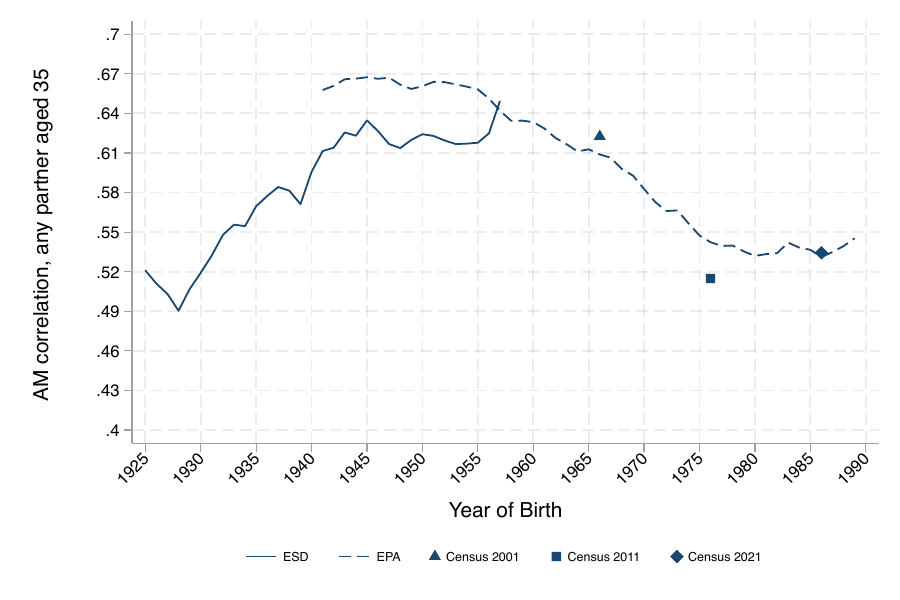}
\caption{Trends in assortative mating in education in Spain}
\label{fig:AM}
\floatfoot{Notes: Three-year moving average of the spousal correlation in years of schooling, for pairs in which at least one partner is aged 35. Years of schooling are harmonized across five data sources (ESD, EPA, and the 2001, 2011, and 2021 Censuses).}
\end{figure}

Figure \ref{fig:AM} shows the evolution of assortative mating in Spain across cohorts. Spousal sorting was remarkably strong for cohorts born in the 1950s and early 1960s, with correlations in years of schooling of around 0.62 in the ESD and 0.66 in the EPA. The finding of comparatively strong sorting aligns with previous studies. In particular, \cite{fernandez2005love} report a spousal correlation of 0.69 for Spain, compared to correlations between 0.4 and 0.6 for other European countries.\footnote{\, \cite{fernandez2005love} use the 1990 Expenditure and Income Survey from INE, restricting the sample to men aged 36-45 (born 1945-54). Using the 2001 Spanish Census, \cite{esteve2006changes} also find high levels of assortative mating across cohorts born between 1920 and 1969, and \cite{tolstova2021marital} confirm that Spain had higher spousal correlations than other developed countries.}

However, assortative patterns have changed significantly over time. Spousal correlations were lower for earlier cohorts, rose significantly for those born in the 1930s and 1940s\footnote{\, Our finding that assortative mating rises when female education expanded (see Figure \ref{fig:national_educ_gender_a}) is consistent with models in which education expansion enlarges the marriage market for higher-educated individuals and strengthens sorting on education levels (\citealt{NybomZiegler2023}).}, and then stabilized at high levels. After decades of strong sorting in education, the spousal correlation began to decline for cohorts born in the early 1960s, and especially those born in the late 1960s and early 1970s. For the most recent cohorts, the spousal correlation is about 0.55, closer to levels observed in other European countries. 

Interestingly, this decline in assortative mating coincides with reductions in educational inequality and intergenerational correlations, as shown in Figures \ref{fig:national_trends_education} and \ref{fig:national_mobility}. While this parallel evolution of key distributional statistics is in line with theoretical priors (Section \ref{sec:theory}), co-movements on the national level might be coincidental rather than causal. In the next section, we study \textit{regional} differences and changes to confirm a robust association between assortative mating and other distributional characteristics.

\section{Regional correlates}
\label{sec:regional}

We next examine how intergenerational mobility varies across Spanish \textit{regions}. While such regional patterns are interesting in their own right, our main goal is to show how mobility relates to other distributional characteristics, such as the strength of assortative mating. Following an approach pioneered by \cite{CHKS_Where_QJE}, we correlate our mobility measures with other local characteristics (``\textit{regional correlates}''). Using the broad time span of our data allows us to study not only cross-sectional variation but also changes over time.

Specifically, we estimate equations of the form
\begin{equation}
\label{eqn:regional}
y_{rt}=\alpha_t+\beta x_{rt}+\varepsilon_{rt}
\end{equation}
where $y_{rt}$ and $x_{rt}$ are regional statistics for region of birth $r$ and cohort $t$, with $r$ defined at a subprovincial level, distinguishing 107 regions and cities (see Section \ref{sec_variables}). To probe whether a causal interpretation is plausible, we also estimate first-difference versions of equation (\ref{eqn:regional}) that abstract from persistent regional characteristics (i.e., region fixed effects).

For this analysis, we use the Encuesta Sociodemogr\'{a}fica (ESD) and the 2001-2021 Censuses.\footnote{\, We do not use EPA for this analysis due to its less detailed geographic information.} The ESD reports parents' education regardless of coresidence, and allows us to compile decadal regional statistics for cohorts born in the 1930s-1960s (up to 1965, as we restrict the sample to individuals aged 26 or older at the survey date). In addition, we compile regional statistics from the 2001, 2011, and 2021 Censuses, which report parental education only for coresident children. As already mentioned (and discussed further in Section \ref{sec_coresidencebias}), coresidence bias can be minimized by measuring education around age 27; to increase sample size, we include children aged 26-28. Thus, in the ESD, the index $t$ refers to decades, while in the Census it refers to three-year birth groups.\footnote{\, Specifically, we distinguish cohorts born in the 1930s, 1940s, 1950s, 1960-1965 (ESD), 1973-75 (Census 2001), 1983-85 (Census 2011), and 1993-95 (Census 2021). The intervals between each ``period'' $t$ are therefore roughly one decade.} 

We first study whether educational expansions -- gains in the average level of schooling -- are associated with changes in intergenerational mobility. They are, but the sign of this relationship depends critically on how educational expansions affect inequality: mobility tends to increase when expansions compress the schooling distribution (as in comprehensive school reforms, Section \ref{sec_trend_mobility}) but falls when they widen inequality. This pattern is in line with the observation that more unequal regions tend to have lower mobility, a relationship that also holds in Spain (the ``Great Gatsby Curve'', Section \ref{subsec:region_greatgatsby}). We add to this literature by showing that the link between inequality and mobility holds not only in the cross-section, but also when studying changes over time. 

We then show that assortative correlations are highly predictive of intergenerational mobility (Section \ref{subsec:region_assortative}): regions with stronger parental sorting exhibit lower mobility. Again, this relationship holds not only in the cross-section, but also in changes, when abstracting from time-invariant regional differences. As a qualitative finding, this association is unsurprising given the conceptual link between parental sorting and mobility (Section \ref{sec_trend_assortative}), though we are not aware of prior region-level evidence on this question. What is striking, however, is its strength: in the 2021 Census, nearly \textit{half} of the variation in intergenerational correlations can be explained by differences in assortative mating. For example, the average intergenerational correlation of regions in the top decile of spousal sorting is around 0.38, compared to just 0.26 in the bottom decile -- or nearly 50\% higher. 

Moreover, regional differences in assortative mating also turn out to be a key mediator, explaining almost half of the well-documented association between inequality and immobility (the ``Great Gatsby Curve'') across regions. Taken together, our evidence points to a potential vicious cycle: higher inequality results in stronger spousal sorting, which lowers mobility and, in turn, feeds back into higher inequality. 

Before turning to these substantive questions, we show that sampling error is a key methodological issue for studying regional correlates (Section \ref{subsec:region_patterns}). Previous work highlights that subnational mobility estimates can be quite noisy (\citealt{durlauf2022great}, \citealt{MogstadWilhelm23}). Although sampling error in the dependent variable $y_{rt}$ in eq. (\ref{eqn:regional}) does not bias our coefficient estimates, error in the regressor $x_{rt}$ introduces attenuation bias. Crucially, sampling error affects different regional correlates differently, as more ``complex'' statistics (e.g., assortative mating) are measured less precisely than simpler ones (e.g., average schooling). In multivariate regressions, sampling error can also introduce an upward ``contamination'' bias. To capture the true magnitude of regional correlates, we use a split-sample instrumental variable technique \citep{fuller2009measurement}, 
showing that it produces dramatically different estimates.

\subsection{Regional patterns and sampling error}
\label{subsec:region_patterns}

\begin{figure}[h!]
\begin{subfigure}{0.95\textwidth}
    \centering
    \includegraphics[width=.80\linewidth]{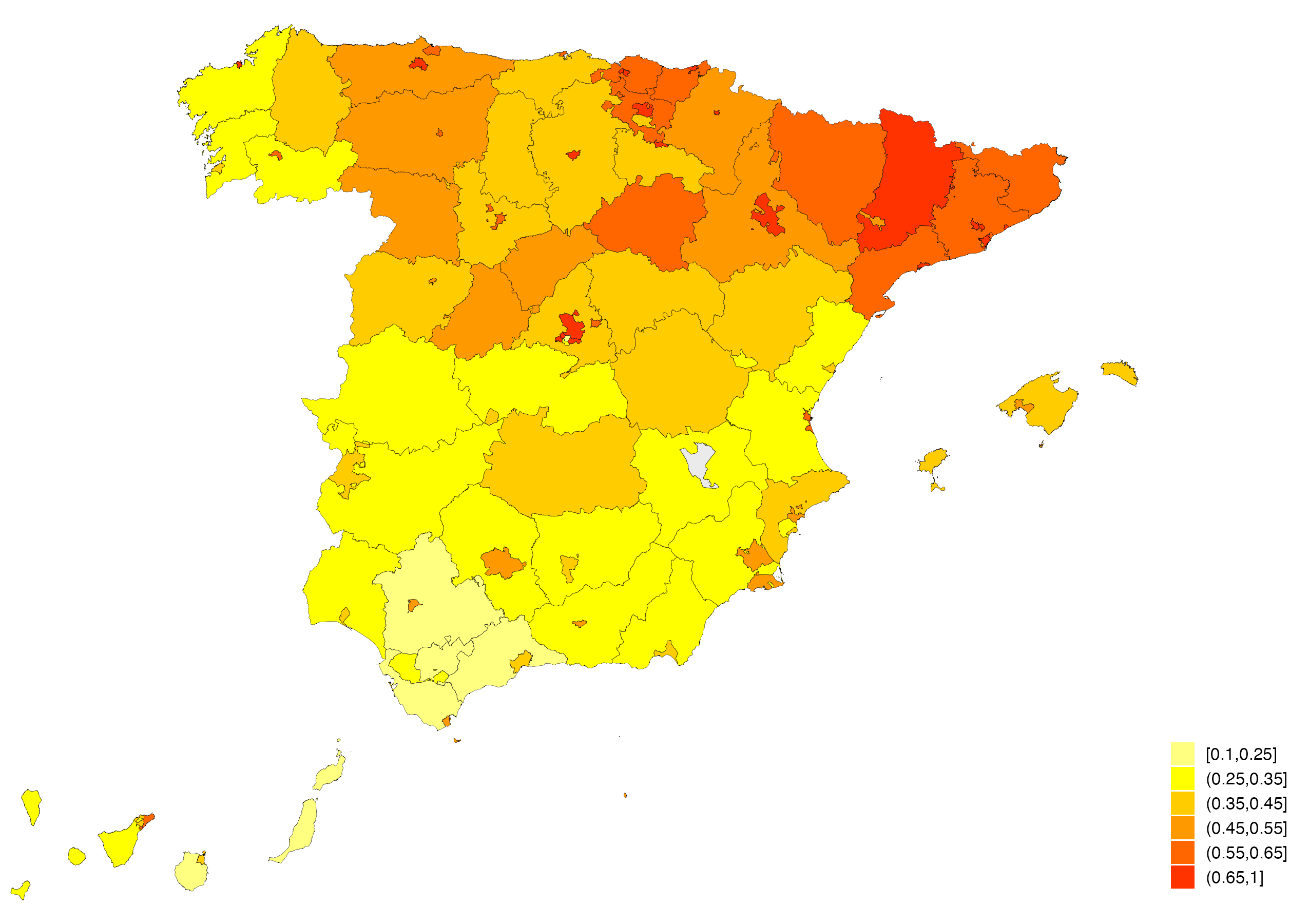}
    \subcaption{Secondary school attendance (1950-59 cohorts)}
    \label{fig:map_av}    
    \medskip
    \medskip
    \medskip
\end{subfigure}
\begin{subfigure}{0.95\textwidth}
    \centering
    \includegraphics[width=.80\linewidth]{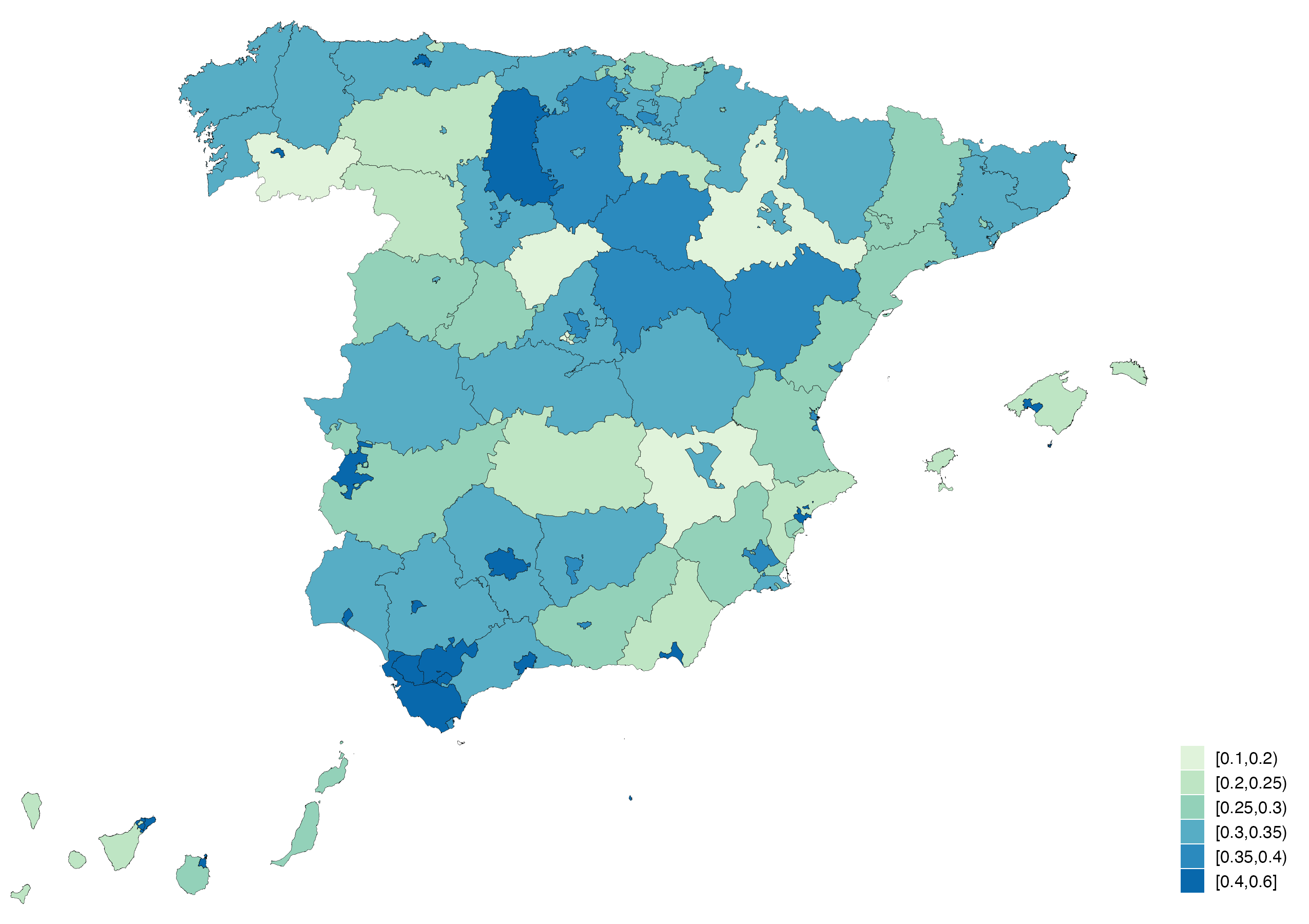}
    \subcaption{Intergenerational correlation in schooling (2021 Census)}
    \label{fig:map_igc}    
\end{subfigure}
\caption{ Regional differences in schooling levels and mobility}
\label{fig:maps}
\floatfoot{Notes: Panel (a) shows the share of students born in the 1950s who attended secondary school across 107 regions and cities (ESD). Panel (b) shows the intergenerational correlation in years of schooling for children born 1993-95 (2021 Census).}
\end{figure}

\begin{figure}[t]
\begin{subfigure}{0.49\textwidth}
    \centering
    \includegraphics[width=.99\linewidth]{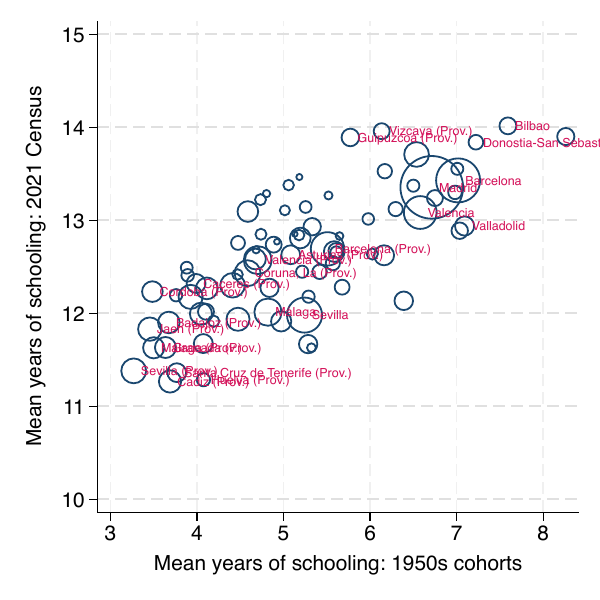}
    \subcaption{Persistence of regional gaps in education}
    \label{fig:persist_mean}    
\end{subfigure}
\begin{subfigure}{0.49\textwidth}
    \centering
    \includegraphics[width=.99\linewidth]{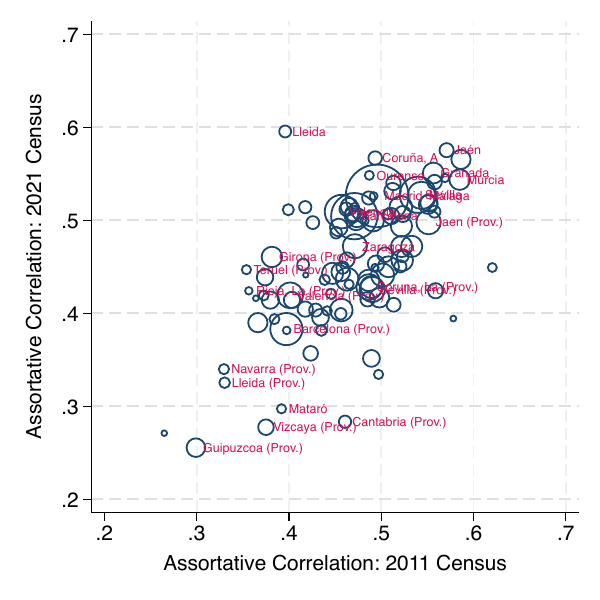}
    \subcaption{Persistence of assortative correlations}
    \label{fig:persist_AM}    
\end{subfigure}
\caption{Persistence of regional patterns}
\label{fig:persistence_census}
\floatfoot{Notes: Panel (a) compares mean years of schooling for cohorts born in the 1950s (ESD) and those born in 1993-95 (aged 26-28 in the 2021 Census). Panel (b) compares spousal correlations in years of schooling for cohorts born 1973-77 (aged 34-38 in the 2011 Census) and 1983-87 (2021 Census) across 107 regions. The size of each circle is proportional to the number of underlying observations in the 2021 Census (only regions with $n>50$ plotted).}
\end{figure}

Spain has large regional imbalances in educational attainment. Despite rapid nationwide gains in schooling (see Figure \ref{fig:national_trends_education_a}), those relative gaps have remained remarkably stable throughout the 20th century. For illustration, Figure \ref{fig:map_av} shows that for cohorts born in the 1950s, secondary school attendance was below 20\% in Andaluc\'{i}a but reached nearly 80\% in Madrid and provinces in the North-East. Comparing average schooling for cohorts born in the mid-1990s (2021 Census) with those born in the 1950s (ESD), Figure \ref{fig:persist_mean} shows that these disparities endured. Cities like M\'{a}laga or Sevilla had consistently low levels of educational attainment, while Barcelona or Bilbao have among the highest levels of attainment.

To study these patterns more systematically, Table \ref{tab:persistence} reports the coefficient from a regression of the specified regional statistic $y_{rt}$ -- the mean, standard deviation, intergenerational correlation, or spousal correlation in years of schooling -- on their lagged value from the previous decade (i.e., $x_{rt}=y_{rt-1}$). As described earlier, we distinguish 107 regions (Figure \ref{fig:maps}), using the ESD for earlier cohorts and the 2001, 2011, and 2021 Censuses for later ones. Column 1 shows that differences in average education are highly persistent, with a slope coefficient of $\hat{\beta}=0.768$ and an $R2$ of nearly one. In contrast, differences in educational inequality, here measured as the standard deviation of years of schooling, appear far less persistent ($\hat{\beta}=0.452$; column 3). However, this partly reflects that in small samples, the standard deviation is estimated less accurately than the mean, leading to more substantial attenuation bias. 

% Table: Persistence
\begin{table}[t]
\begin{centering}
\caption{Regional persistence in educational distributions}
\label{tab:persistence}
\par\end{centering}
\begin{centering}
\begin{tabular}{lccccccccccc}
\toprule 
 & \multicolumn{2}{c}{$y_{rt}=Mean(edu)$} &  & \multicolumn{2}{c}{$y_{rt}=SD(edu)$} &  & \multicolumn{2}{c}{$y_{rt}=IGC(edu)$} &  & \multicolumn{2}{c}{$y_{rt}=AM(edu)$}\tabularnewline
\cmidrule{2-3} \cmidrule{3-3} \cmidrule{5-6} \cmidrule{6-6} \cmidrule{8-9} \cmidrule{9-9} \cmidrule{11-12} \cmidrule{12-12} 
 & (1) & (2) &  & (3) & (4) &  & (5) & (6) &  & (7) & (8)\tabularnewline
\midrule
$y_{rt-1}$ & 0.768*** & 0.804*** &  & 0.452*** & 0.577*** &  & 0.285*** & 0.528*** &  & 0.204*** & 0.999*** \\
 & (0.018) & (0.025) &  & (0.029) & (0.056) &  & (0.035) & (0.109) &  & (0.041) & (0.374) \\
Split-IV & - & \checkmark &  & - & \checkmark &  & - & \checkmark &  & - & \checkmark \\
N & 637 & 633 &  & 636 & 626 &  & 631 & 612 &  & 626 & 611 \\
$R^2$ & 0.950 & . &  & 0.672 & . &  & 0.301 & . &  & 0.267 & . \\
\bottomrule
\end{tabular}
\vspace{-2mm}
\par\end{centering}
\floatfoot{Notes: Coefficient estimates from regressions of the indicated regional statistic (mean, SD=standard deviation, IGC=intergenerational correlation, and AM=spousal correlation in years of schooling) on decadal lags and time fixed effects. ESD and 2001, 2011, and 2021 Censuses, 107 regions.}
\end{table}

Put differently, our statistics $x_{rt}$ are generated regressors that capture both true regional variation and noise from sampling variability. To isolate the former, we implement a simple split-sample procedure (e.g., \citealt{Goldschmidtetal2017}). Specifically, we divide the sample into two randomly drawn groups (stratified by gender, year of birth, survey, and region) and estimate $x_{rt}$ separately for each, labeled $x_{rt,1}$ and $x_{rt,2}$. We then regress $x_{rt,1}$ on $x_{rt,2}$ (``first stage''). If there were no sampling or measurement error, the slope coefficient $\hat{\beta}_\text{first}$ from this regression should be close to one; in contrast, if $x_{rt}$ reflects \emph{only} noise, the coefficient should be near zero. 

Our results confirm substantial variation in precision across the different indicators. For mean education levels, regressing $x_{rt,1}$ on $x_{rt,2}$ yields a coefficient close to one ($\hat{\beta}_\text{first}=0.96$). For the standard deviation of schooling, this first-stage coefficient is much smaller ($\hat{\beta}_\text{first}=0.70$), reflecting that standard deviations are less precisely estimated in small samples than means. The influence of sampling error is even more pronounced for ``complex'' regional statistics such as the intergenerational correlation ($\hat{\beta}_\text{first}=0.40$) or the spousal correlation in years of schooling ($\hat{\beta}_\text{first}=0.25$). 

Studies searching for ``regional correlates'' are therefore prone to attenuation bias from sampling error, and the extent of this bias will vary across indicators and data sources. For example, first-stage coefficients are higher in the larger Census data than the ESD, and especially so for the more complex regional statistics such as the assortative correlation ($\hat{\beta}_\text{first}=0.28$ in Census vs. $\hat{\beta}_\text{first}=0.09$ in ESD). Unless based on very large samples, studies may therefore miss correlational patterns in less precisely estimated indicators.\footnote{\, The influential study by \cite{CHKS_Where_QJE} is less affected, as their core sample contains nearly 10 million children across 741 commuting zones (though fewer than 250 children are observed in 32 commuting zones).}

To correct for sampling error, we use a split-sample IV estimator, regressing the statistic of interest on $x_{rt,1}$, using $x_{rt,2}$ as an instrument. As expected, the resulting slope coefficients $\beta_{SSIV}$ are systematically larger than the naive OLS estimates $\beta_{OLS}$, but the magnitude of the increase varies strongly across indicators. Persistence rises only slightly for mean education (column 2 of Table \ref{tab:persistence}), but more substantially for educational inequality (column 4). Still, even after accounting for measurement error, we find that regional differences in educational levels are more persistent ($\hat{\beta}_\text{SSIV}=0.80$) than differences in dispersion ($\hat{\beta}_\text{SSIV}=0.58$). 

Regional differences in intergenerational (columns 5-6) and assortative correlations (measured in the parent generation, columns 7-8) are also quite persistent, once sampling error is taken into account. The lag coefficient for the intergenerational correlation increases by more than 80\%, from $\hat{\beta}_{OLS}=0.29$ to $\hat{\beta}_{SSIV}=0.53$. The increase is even more dramatic, from $\hat{\beta}_{OLS}=0.20$ to nearly one, for the assortative correlation. Figure \ref{fig:persist_AM} illustrates that regional differences in assortative mating remain highly persistent in recent Census data: estimated spousal correlations for each municipality in the 2011 Census align closely with those in the 2001 Census.

Below, we show that these enduring gaps in the strength of assortative mating help explain why some regions exhibit higher educational mobility than others.

In sum, we observe strong regional differences not just in the average level of education, but also in inequality, mobility, and assortative measures. Those regional differences are highly stable over time. We next explore how these regional characteristics are interrelated, highlighting the importance of accounting for sampling error to capture those ``regional correlates''. 

\subsection{Educational expansions and the ``Great Gatsby Curve''}
\label{subsec:region_expansion}
\label{subsec:region_greatgatsby}

Spain experienced rapid educational expansions during the 20th century, especially for cohorts born after 1940 (see Figure \ref{fig:national_trends_education_a}). An interesting question is whether such expansions have strengthened social mobility, possibly explaining the large drop in intergenerational correlations in Spain in recent decades (Figure \ref{fig:national_mobility_a}). To explore this, we first examine whether regional differences in schooling levels correlate with differences in intergenerational mobility. We then use the long time span of our data to study whether regional \textit{expansions} of schooling can explain changes in mobility over time. 

\begin{figure}
\begin{centering}
\caption{The ``Great Gatsby Curve'' in education}
\includegraphics[scale=0.8]{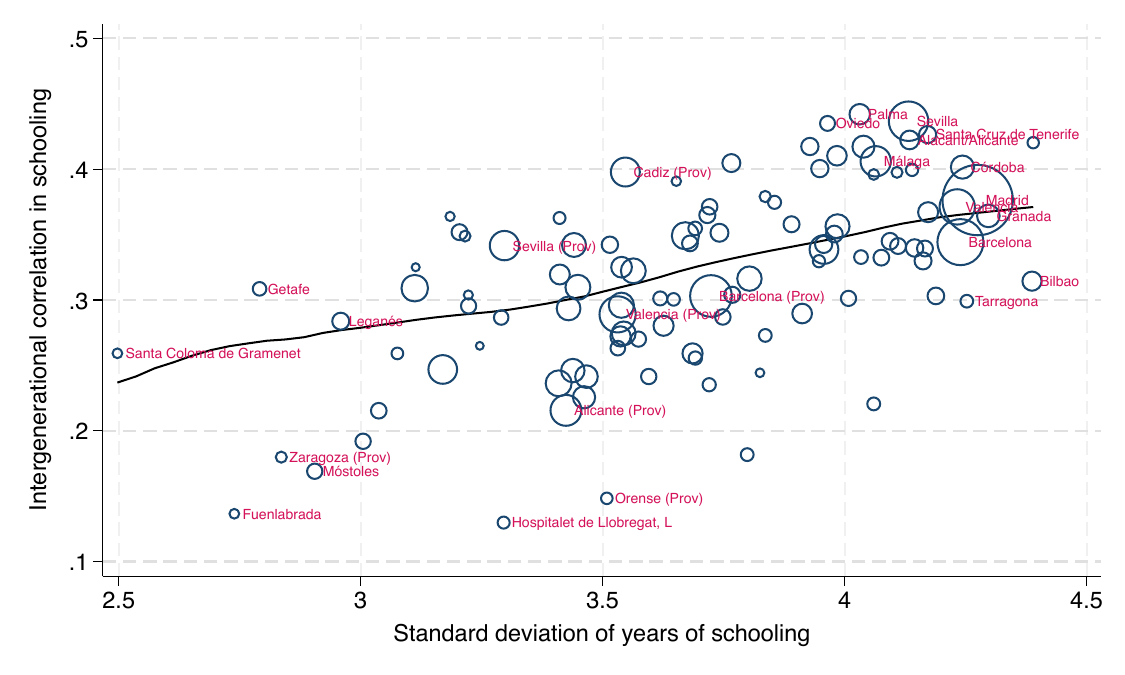}
\label{fig:GG_municipalities_census}
\par\end{centering}
\floatfoot{Notes: Parent-child correlation and standard deviation of fathers' years of schooling by region, 2021 Census. The size of each circle is proportional to sample size (only regions with $n>50$ plotted). The black line is a local polynomial fit.} 
\end{figure}

Table \ref{tab:eduexpansion} summarizes our findings. The average \textit{level} of education in the parent generation has a small positive association with the intergenerational correlation (column 1), but this association reverses sign once we control for the parental standard deviation of schooling (column 2). By contrast, this standard deviation shows a strong and robust positive association with intergenerational correlations. This association becomes even more pronounced when using a split-sample estimator to address sampling error (column 3). While Table \ref{tab:eduexpansion} pools all cohorts, Figure \ref{fig:GG_municipalities_census} illustrates this pattern for more recent cohorts using the 2021 Census. The intergenerational correlation in years of schooling is nearly 50\% larger in the municipalities with the greatest educational inequality (such as C\'ordoba) compared to those with the lowest dispersion in education (such as Legan\'{e}s). We find similarly strong associations in the ESD and earlier censuses.

This positive association between educational inequality and intergenerational persistence aligns with previous research linking inequality and mobility across countries, both for income (\citealt{CorakJEP2013}) and education (\citealt{HHBlandenDoepkeStuhler22}). Known as the ``Great Gatsby Curve'', this pattern has also been documented across regions within country (e.g. \citealt{CHKS_Where_QJE,GuellPellizzariPicaMora2018}) and over time (\citealt{wardAERintergenerational,JacomeetalJPE}). Theoretically, a causal link between inequality and immobility may operate in both directions. On the one hand, greater immobility will mechanically increase inequality in the child generation (\citealt{Berman2021vb,durlauf2022great}). On the other hand, greater socio-economic inequality in the parent generation may lead to more unequal investments in child education (e.g., \citealt{HHBlandenDoepkeStuhler22}), reinforcing persistence. 

\begin{table}[t!]
\begin{centering}
\caption{Educational expansion and intergenerational mobility}
\label{tab:eduexpansion}
\begin{tabular}{lccccccc}
\toprule 
 & \multicolumn{7}{c}{Dependent variable: $y_{rt}=IGC(edu)$} \tabularnewline
 & \multicolumn{3}{c}{Levels} & & \multicolumn{3}{c}{Changes} \tabularnewline
 \cmidrule{2-4} \cmidrule{6-8}
 & (1) & (2) & (3) & & (4) & (5) & (6) \tabularnewline
\midrule
$Mean(edu_\textit{father})$ & 0.020*** & -0.017*** & -0.030*** & ~ & 0.036*** & -0.006 & -0.014  \\
 & (0.002) & (0.004) & (0.006) & ~ & (0.009) & (0.010) & (0.033)   \\ 
$SD(edu_\textit{father})$ & ~ & 0.081*** & 0.122*** & ~ & ~ & 0.067*** & 0.194***  \\ 
 & ~ & (0.006) & (0.011) & ~ & ~ & (0.010) & (0.038)  \\ 
Split-IV & - & - & \checkmark &  & - & - & \checkmark \\ 
Observations & 739 & 739 & 726 & ~ & 631 & 631 & 615  \\ 
$R^2$ & 0.270 & 0.438 & . & ~ & 0.085 & 0.163 & .  \\
\bottomrule
\end{tabular}
\vspace{-2mm}
\end{centering}
\floatfoot{Notes: Coefficient estimates from regressions of the intergenerational correlation (IGC) in years of schooling on the mean and standard deviation of fathers' schooling and time fixed effects. Columns 1-3 are estimated in levels using decadal data, columns 4-6 are estimated in first (decadal) differences. Data sources: ESD and the 2001, 2011, and 2021 Censuses, covering 107 regions. Robust standard errors in parentheses, {*}{*}{*} p\textless 0.01, {*}{*} p\textless 0.05, {*} p\textless 0.1.}
\end{table}

Our evidence contributes to this literature in two ways. First, we confirm that an educational Great Gatsby Curve also holds across regions in Spain, corroborating the view that this relationship is systemic rather than incidental. Second, and more importantly, we can use the long time span of our data to study whether this relationship also holds when considering \textit{changes} over time, abstracting from time-constant differences between regions. We find that the relationship between inequality and immobility is similarly strong in changes (columns 5-6 of Table \ref{tab:eduexpansion}) as in levels: regions experiencing rising educational inequality tend to see an increase in the intergenerational correlation (i.e., a decline in educational mobility), and vice versa. 

Comparable evidence on such ``temporal Great Gatsby Curves'' is still scarce, as noted by \cite{durlauf2022great} in their comprehensive review of the literature. As important exceptions,  \cite{Branden2019} provides detailed evidence on the Great Gatsby Curve in both levels and changes in Sweden, while  \cite{BattistonetalGatsby} relate changes in income inequality to changes in intergenerational income mobility across US counties. We find a pronounced temporal Great Gatsby Curve in education for Spain. In addition, we highlight that its strength -- especially of its temporal version that difference out time-constant regional factors -- becomes much more pronounced when correcting for sampling error (see columns 3 and 6 in Table \ref{tab:eduexpansion}).

Once inequality is held constant, educational expansions  -- as measured by changes in the average level of education -- show no systematic relationship with changes in intergenerational mobility (columns 5 and 6). Our findings thus suggest that the effect of educational expansions on social mobility hinges primarily on whether they widen or narrow the dispersion in educational outcomes, which in turn depends on where in the educational distribution the expansion occurs.\footnote{\, For example, \cite{manzano2024intensity} show that in Spain, the expansion of secondary education initially increased inequality in tertiary attainment, but later reduced it as the expansion progressed.} Expansions that reduce inequality in the parent generation tend to be followed by mobility gains.

This link between inequality and immobility may also explain the trends in intergenerational mobility in Spain during the second half of the 20th century (Figure \ref{fig:national_mobility_a}). While average schooling increased steadily (Figure \ref{fig:national_trends_education_a}), dispersion first remained steady or increased, then fell sharply following the 1970 comprehensive school reform (Section \ref{sec_institutions}), while intergenerational mobility increased. Overall, the pattern in Spain seems to be in line with evidence that comprehensive school reforms increased mobility in other countries (e.g., \citealt{meghir2005educational} for Sweden or \citealt{pekkarinen2009school} for Finland). 

\subsection{Assortative mating, inequality and mobility}
\label{subsec:region_assortative}

\begin{figure}[h!t]
\begin{subfigure}{1\textwidth}
    \centering
    \includegraphics[scale=0.8]{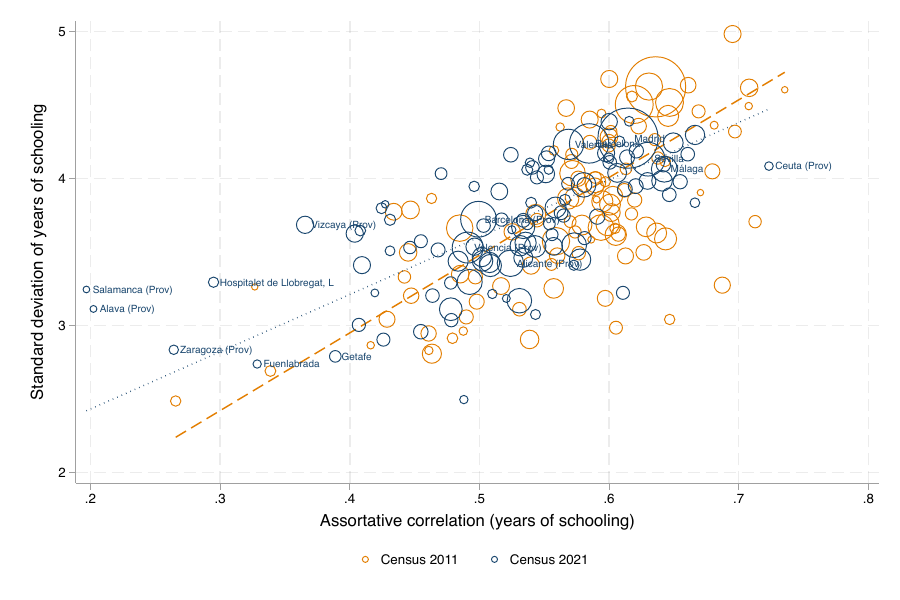}
    \subcaption{Assortative mating and inequality}
    \label{fig:AM_SD} 
\end{subfigure}  \\
\begin{subfigure}{1\textwidth}
    \centering
    \includegraphics[scale=0.8]{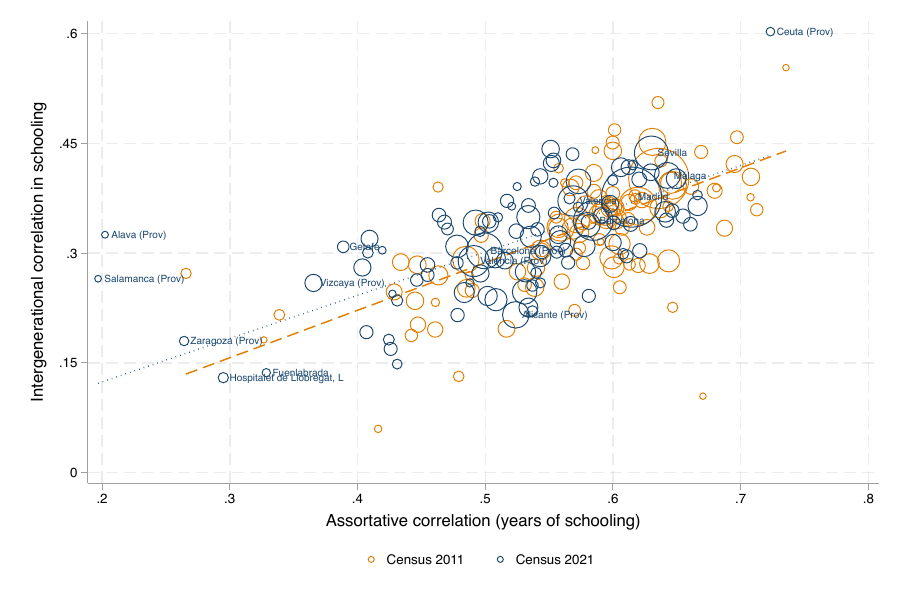}
    \subcaption{Assortative mating and intergenerational mobility}
    \label{fig:AM_IGC}     
\end{subfigure}     
\caption{Assortative mating, inequality and intergenerational mobility}
\label{fig:SD_and_IGC_vs_AM_census}
\floatfoot{Notes: Standard deviation of fathers' years of schooling, intergenerational correlation, and assortative correlation in parents' schooling by region, 2011 and 2021 Census. Circle size is proportional to sample size (only regions with $n>50$ plotted).}
\end{figure}

Assortative mating in education has changed greatly over time in Spain. Could the decline in spousal sorting among cohorts born after 1960 (Figure \ref{fig:AM}) explain declining educational inequality and rising mobility among the same cohorts?\footnote{\, Notably, a co-movement in the national trends of assortative mating and inequality has also been observed in the United States (\citealt{mare2016educational}).} A co-movement between these distributional characteristics at the national level could be coincidental rather than causal. To provide more targeted evidence, we study whether regional differences in the strength of assortative mating correlate with regional differences in inequality and mobility, and whether regional \textit{changes} in sorting are linked to changes in inequality and mobility. 

Figure \ref{fig:SD_and_IGC_vs_AM_census} shows that assortative correlations are indeed a strong predictor of differences in educational inequality and persistence across regions. Figure \ref{fig:AM_SD} shows a strong positive relationship between assortative mating in the parent generation and the standard deviation of fathers' years of schooling. We observe a strong and highly significant positive correlation ($p<0.001$) in both the 2011 and 2021 censuses. Regions with high educational inequality, such as Sevilla or M\'{a}laga, also tend to exhibit strong spousal sorting. Thus, assortative matching is closely related to cross-sectional inequality. 

Figure \ref{fig:AM_IGC} illustrates that assortative mating is also closely linked to intergenerational (im-)mobility. The strength of this relationship is striking: across regions, nearly \textit{half} of the variation in intergenerational correlations can be explained by differences in assortative mating, even if sampling error is not yet accounted for. In both Censuses, a 0.1 increase in the assortative correlation in parents' years of schooling is associated with a 0.06 rise in the intergenerational correlation. 

As discussed in Section \ref{sec:theory}, a causal link between inequality and assortative mating could operate in both directions. Greater inequality may lead to stronger spousal sorting (inequality $\rightarrow$ sorting). Stronger sorting among parents, in turn, may increase inequality among their children (sorting $\rightarrow$ inequality). This latter effect of sorting on inequality follows naturally from standard models of intergenerational transmission: assortative mating leads to a more unequal distribution of parental advantages across households, thereby increasing inequality in the next generation (\citealt{blinder1973model, kremer1997much, FernandezRogerson2001,Ermischetal2006}).\footnote{\, Assortative mating may affect children's outcomes via different pathways. For example, educational assortative mating increases inequality in household income (\citealt{eika2019educational}), thereby affecting educational investments, but it may also affect intergenerational skill transfers (\citealt{NybomZiegler2023}).} A reciprocal effect of inequality on assortative mating follows less immediately, but \cite{DurlaufSeshadri2018} note that ``\textit{there are good reasons to believe that greater inequality [...] increases segregation}'', and segregation makes it more likely for people to marry within their ``own type'' (\citealt{mare2016educational}). The strong empirical associations that we document here are suggestive of such mechanisms, although we do not have a targeted causal design to isolate the impact of any single factor.

\begin{table}
\begin{centering}
\caption{Assortative mating and inequality}
\label{tab:AM_inequality}
\par\end{centering}
\centering{}%
\begin{tabular}{lc@{\hspace{0.3cm}}c@{\hspace{0.3cm}}p{0.01cm}c@{\hspace{0.3cm}}c@{\hspace{0.4cm}}p{0.1cm}c@{\hspace{0.3cm}}c@{\hspace{0.3cm}}p{0.01cm}c@{\hspace{0.3cm}}c}
\toprule 
 & \multicolumn{5}{c}{Dependent variable: $SD(edu_\textit{child})$} & & \multicolumn{5}{c}{Dependent variable: $AM(edu_\textit{parent})$}\tabularnewline
& \multicolumn{2}{c}{Levels} & & \multicolumn{2}{c}{Changes} & & \multicolumn{2}{c}{Levels} & & \multicolumn{2}{c}{Changes} \tabularnewline
\cmidrule{2-3} \cmidrule{5-6} \cmidrule{8-9} \cmidrule{11-12}
 & (1) & (2) & & (3) & (4) & & (5) & (6) & & (7) & (8) \tabularnewline
\midrule
$AM(edu_\textit{parent})$ & 0.869*** & 2.310*** & & 0.210 & -0.324 & ~ & ~ & & ~ & ~ &   \\ 
~ & (0.171) & (0.782) & & (0.130) & (1.653) & ~ & ~ & & ~ & ~ &   \\ 
$SD(edu_\textit{father})$ & ~ & ~ & & ~ & ~ & ~ & 0.035*** & 0.037*** & & 0.033*** & 0.064**  \\ 
~ & ~ & ~ & & ~ & ~ & ~ & (0.005) & (0.007) & & (0.008) & (0.031)  \\ 
Split-IV & - & \checkmark & & - & \checkmark &  & - & \checkmark & & - & \checkmark \\
Observations & 736 & 723 & & 626 & 613 & ~ & 736 & 725 & & 626 & 614  \\ 
$R^2$ & 0.592 & . & & 0.724 & . & ~ & 0.294 & . & & 0.185 & . \\ 
\midrule
\end{tabular}
\vspace{-2mm}
\floatfoot{Notes: Regressions of the standard deviation of years of schooling in the child generation on the spousal correlation in schooling in the parent generation (columns 1-4) or regression of the spousal correlation in schooling in the parent generation on the standard deviation of fathers' schooling (columns 5-8) and time fixed effects. Columns 1-2 and 5-6 are estimated in levels using decadal data, columns 3-4 and 7-8 are estimated in first (decadal) differences. ESD and 2001, 2011, and 2021 Censuses, 107 regions. Robust standard errors in parentheses, {*}{*}{*} p\textless 0.01, {*}{*} p\textless 0.05, {*} p\textless 0.1.} 
\end{table}

Table \ref{tab:AM_inequality} explores this relation between assortative mating and inequality in more depth, pooling earlier cohorts from the ESD with more recent cohorts from the Census. We first study whether stronger assortative mating among parents is associated with higher inequality in the next generation (sorting $\rightarrow$ inequality). Column 1 confirms that this association is highly statistically significant, and it becomes nearly three times stronger when we use split-sample IV to address sampling error (column 2). The association loses significance when we consider \textit{changes} in inequality and assortative mating (column 3). A likely interpretation is that the signal-to-noise ratio in our regional measures of assortative mating -- already modest in levels (see Table \ref{tab:persistence}) -- is too low in changes. Indeed, the association between the two split-sample estimates of changes in assortative mating (``first stage'', see Section \ref{subsec:region_patterns}) is small and statistically insignificant. Consequently, split-sample IV estimates in changes are too imprecise to be informative (column 4). 

In the right panel of Table \ref{tab:AM_inequality}, we study the reciprocal mechanism -- whether educational inequality predicts assortative mating (inequality $\rightarrow$ sorting). We find a strong relationship both in levels (columns 5-6) and in changes, that is, when abstracting from region fixed effects (columns 7-8).\footnote{\, Because the standard deviation can be estimated more precisely than the spousal correlation (see Table \ref{tab:persistence}), the specification in changes has sufficient power when using the former but not when using the latter as a regressor.} As expected, this association becomes more pronounced when we use split-sample IV to address sampling error. Our evidence thus supports the hypothesis that greater inequality induces stronger assortative mating. We also find evidence for the reverse effect of assortative mating on inequality, although this analysis lacks power when studying regional changes rather than levels. 

\begin{table}
\begin{centering}
\caption{Intergenerational mobility, assortative mating and inequality}
\label{tab:IGC_AM_inequality}
\par\end{centering}
\centering{}%
\begin{tabular}{lcccc@{\hspace{0.3cm}}p{0.1cm}cccc}
\toprule 
 & \multicolumn{9}{c}{Dependent variable: $y_{rt}=IGC(edu)$} \tabularnewline
& \multicolumn{4}{c}{Levels} & & \multicolumn{4}{c}{Changes} \tabularnewline
\cmidrule{2-5} \cmidrule{7-10} 
 & (1) & (2) & (3) & (4) & & (5) & (6) & (7) & (8) \tabularnewline
\midrule
$SD(edu_\textit{father})$ & 0.060*** & ~ & 0.050*** & 0.036*** & ~ & 0.064*** & ~ & 0.057*** & 0.002  \\ 
~ & (0.004) & ~ & (0.004) & (0.010) & ~ & (0.008) & ~ & (0.008) & (0.079)  \\ 
~ & [0.519] & ~ & [0.470] & [0.345] & ~ & [0.335] & ~ & [0.299] &  [0.011] \\ 
$AM(edu_\textit{parent})$ & ~ & 0.413*** & 0.278*** & 0.794*** & ~ & ~ & 0.250*** & 0.200*** & 1.758*  \\ 
~ & ~ & (0.036) & (0.032) & (0.207) & ~ & ~ & (0.047) & (0.046) & (1.017)  \\ 
~ & ~ & [0.407] & [0.276] & [0.793] & ~ & ~ & [0.255] & [0.203] &  [1.913] \\ 
Split-IV & - & - & - & \checkmark & ~ & - & - & - & \checkmark  \\ 
Observations & 739 & 736 & 736 & 723 & ~ & 631 & 626 & 626 & 613  \\ 
$R^2$ & 0.418 & 0.336 & 0.473 & . & ~ & 0.163 & 0.118 & 0.200 & . \\ 
\midrule
\end{tabular}
\vspace{-2mm}
\floatfoot{Notes: Regression of the intergenerational correlation in years of schooling on the standard deviation of fathers' schooling and the spousal correlation in schooling in the parent generation. Columns 1-4 are estimated in levels using decadal data, columns 5-8 are estimated in first (decadal) differences. ESD and Census 2001, 2011, and 2021, 107 regions. Robust standard errors in parentheses, {*}{*}{*} p\textless 0.01, {*}{*} p\textless 0.05, {*} p\textless 0.1. Standardized beta coefficients are reported in brackets.} 
\end{table}

Finally, Table \ref{tab:IGC_AM_inequality} shows how intergenerational correlations relate to both inequality and assortative mating in the parent generation. Column 1 confirms that higher inequality is associated with lower mobility (i.e., higher intergenerational correlations), as already shown in Table \ref{tab:eduexpansion}. Column 2 shows that a higher spousal correlation in the parent generation is also associated with lower mobility; this reaffirms the evidence from Figure \ref{fig:AM_IGC} for the entire sample. Column 3 shows that, when included jointly, both inequality and assortative mating remain strong predictors of regional variation in the intergenerational correlation. These associations are robust to differencing out regional fixed effects by estimating in changes (columns 5-7).

A comparison of standardized beta coefficients (in square brackets) in columns 3 and 7 would suggest that inequality is a stronger predictor of mobility than assortative mating. However, when using the split-sample IV approach to address sampling error, the coefficient on the assortative correlation increases substantially -- by a factor of three -- while the coefficient on the standard deviation in schooling \textit{declines} (compare columns 3 and 4, or columns 7 and 8). This pattern illustrates that sampling error can have complex consequences in multivariate regressions: while it attenuates coefficients in bivariate settings, it can cause an upward ``contamination bias'' on other coefficients (\citealt{van2024teacher}). Here, because the spousal correlations are especially noisy (see Table \ref{tab:persistence}), their coefficient is severely downward-biased, which in turn induces an upward contamination bias in the coefficient on the (more precisely measured) standard deviation of schooling. 

After correcting for sampling error, assortative mating is more than twice as important a predictor of intergenerational correlations as inequality. A one-standard-deviation increase in the spousal correlation is associated with a 0.79-standard-deviation rise in the intergenerational correlation, compared with 0.35 for a one-standard-deviation increase in our measure of inequality (column 4). Moreover, once sampling error is addressed, differences in assortative mating explain nearly \textit{half} of the association between inequality and immobility (the ``Great Gatsby Curve'', see Section \ref{subsec:region_greatgatsby}) across regions. Specifically, applying split-sample IV to column 1 of Table \ref{tab:IGC_AM_inequality} yields a coefficient of 0.066, which shrinks to 0.036 when conditioning on assortative mating in column 4. 

Assortative mating thus appears to be a key mediator of the link between inequality and intergenerational mobility that has been documented in the recent literature. And taken together, our findings point to a potential vicious cycle: higher inequality results in stronger spousal sorting, which lowers mobility and, in turn, feeds back into higher inequality. While we observe these associations at both levels and over time, our research design is only suggestive; more targeted designs are needed to confirm their causal interpretation. 

\section{Extensions and robustness}
\label{extensions}

In this section, we provide additional evidence and robustness tests. We provide additional evidence on coresidence bias, examine gender differences in the distributional characteristics of interest, and compare our measure of educational mobility with external measures of income mobility.

\subsection{Robustness to coresidence bias}
\label{sec_coresidencebias}

A key challenge in intergenerational studies is coresidence bias in data sources such as the Census, where parental education is observed only for children still living with their parents. As discussed in Section \ref{sec_method_issues}, researchers face a trade-off between measuring education at a younger age -- when most children are still coresident, but educational outcomes may be censored -- or a later age, when education is completed but coresident children represent a smaller and potentially more selective group. In our main analysis, we chose age 27 as the optimal age to measure educational attainment. To motivate this choice, we use that the \textit{Encuesta Sociodemogr\'{a}fica} (ESD) reports on when children leave their parental home, how their educational attainment changes over time, and parental education regardless of coresidence status. This allows us to compare ``true'' mobility estimates with those based on coresident samples at different ages, and to find the optimal age at measurement. We provide a summary here and relegate details to Appendix \ref{Appendix-Coresidence}.

We distinguish two biases: (1) \emph{selection bias}, which arises at older ages due to differences between coresident and independent children, and (2) \emph{censoring bias}, which occurs at younger ages when educational attainment may not yet be complete. The overall bias can thus be minimized in the mid-20s -- when most children have completed their education, but many still live with their parents. Appendix \ref{Appendix-Coresidence-Optage} shows that, in the Spanish context, measuring education around age 23-27 minimizes overall bias. More importantly, the remaining bias appears stable over cohorts.\footnote{\, Figures \ref{fig:national_IGC_age} and \ref{fig:national_IGR_age} confirm that mobility trends in the IGC and IGR remain similar when measuring education at ages 24 or 30, despite differences in estimated mobility levels. This robustness is partly due to Spain's high coresidence rates, which limit the potential for selection bias (see Figure \ref{fig:national_share_dependence_24_27_30}).} To buffer against censoring bias in recent cohorts with higher educational attainment, we select age 27 as our preferred measurement age.\footnote{\, Choosing a sufficiently high measurement age is important when studying mobility \emph{trends} over long periods, as otherwise censoring bias would tend to grow as educational attainment rises.}

As an additional test, we implement the correction method proposed by \cite{Hilger2017}, which %estimates the conditional expectation function (CEF) of child education conditional on parental education based 
relies on two assumptions: (i) ``parallel trends'', assuming that educational differences between dependent and independent children can be captured by a constant (i.e., the CEF for independent children has a different intercept but the same slope as the CEF for dependent children), and (ii) ``smooth cohorts'', assuming that the share of independent children in each parental educational group can be approximated using group shares from younger cohorts. We assess the performance of this correction method in Spain by comparing bias-corrected estimates of the IGR to those derived from the ESD, which includes parental education regardless of coresidence status. Appendix \ref{Appendix-Coresidence-Correcting} provides details. While the method effectively reduces selection bias at younger ages, it does not address censoring at younger or, in our data, selection at older ages.\footnote{\, See Figure \ref{fig:IGR_hilger_dependent_true}. The underlying ``smooth cohort'' assumption breaks down as parental education rises rapidly in recent cohorts (Figure \ref{fig:IGR_hilger_smooth}). Moreover, the method does not correct for censoring, the main source of bias at younger ages.} Our baseline estimates are thus based on education at age 27, minimizing bias from both selection and censoring.\footnote{\, Figure \ref{fig:IGR_hilger_dependent_true_cohort_bias27vs30} validates this approach by comparing the bias in IGR estimates from the coresident sample at ages 27 and 30 against the benchmark IGR estimates (including both dependent and independent children) at age 30 (when education is completed for nearly everyone).}

\subsection{Gender differences}

In cohorts from the 1950s onwards, women's educational attainment surpassed men's (see Fig \ref{fig:national_educ_gender_a}). As of the 2021 Census, women aged 26-28 had, on average, 1.3 more years of schooling than their male counterparts. Indeed, we find a gender gap favoring women across all 107 regions in our sample. However, its size varies considerably, from below ten months in Palma and the province of Huelva to around two years in the Lugo and Albacete provinces. 

\begin{figure}[ht]
\begin{subfigure}{0.495\textwidth}
    \centering
    \includegraphics[width=.99\linewidth]{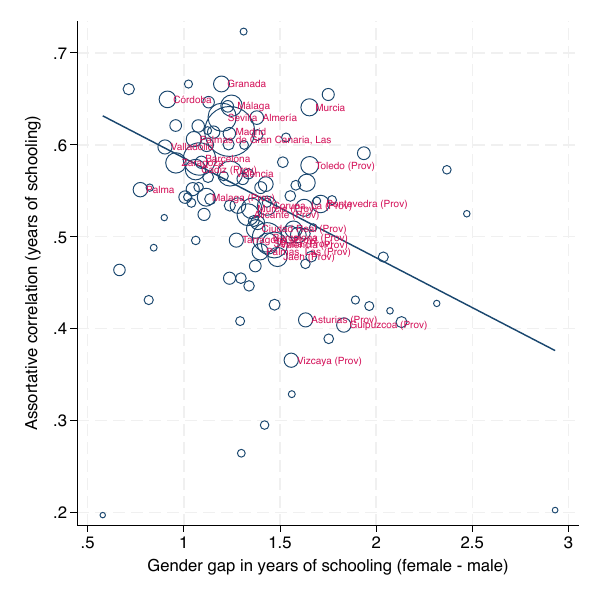}
    \subcaption{Assortative mating and gender gaps in education}
    \label{fig:AM_gender_gaps} 
\end{subfigure}    
\begin{subfigure}{0.495\textwidth}
    \includegraphics[width=.99\linewidth]{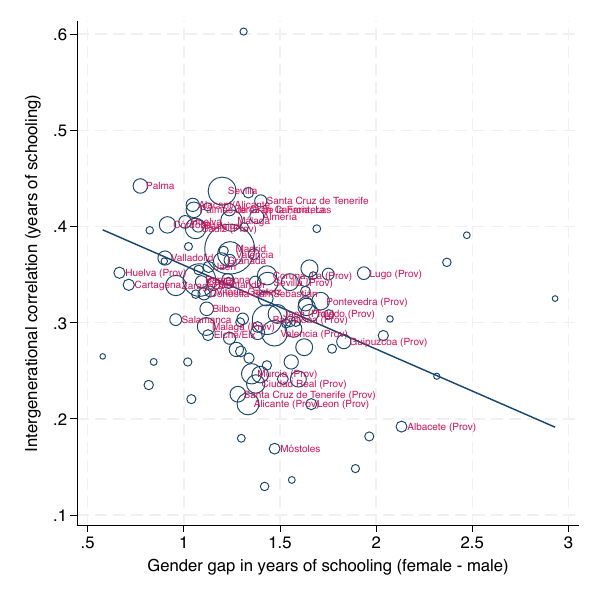}
    \subcaption{Intergenerational mobility and gender gaps in education}
    \label{fig:IGC_gender_gaps}     
\end{subfigure}     
\caption{Gender gaps in education}
\label{fig:AM_and_IGC_vs_gender_gaps}
\floatfoot{Notes: Assortative correlation in parents' schooling and father-child correlation for children aged 26-28, 2021 Census. Circle size is proportional to sample size (only regions with $n>50$ plotted).}
\end{figure}

Interestingly, these gender gaps in educational attainment are predictive of assortative and intergenerational correlations. As shown in Figure \ref{fig:AM_gender_gaps}, regions where the gender gap is more pronounced tend to have lower assortative correlations. Figure \ref{fig:IGC_gender_gaps} shows that such regions also have lower intergenerational correlations (i.e., higher mobility). These patterns are robust to considering the ratio of women's to men's schooling rather than the difference, or to controlling for other regional characteristics, such as the mean or dispersion of schooling.

One potential interpretation of these patterns is that the same structural factors that benefit women relative to men's educational attainment also reduce assortative and intergenerational inequality. In this view, the patterns in Figure \ref{fig:IGC_gender_gaps} are just a reflection of other structural differences between regions. Alternatively, women's educational attainment might have a direct effect on assortative mating and mobility, perhaps because higher relative education alters their behavior in the partner selection process. While our research design does not allow us to distinguish between these specific hypotheses, our observations support the broader conclusion that regional differences in intergenerational mobility can be well explained by other regional characteristics.

\subsection{Educational vs. income mobility}

To study intergenerational mobility over a broad time period and across multiple data sources, our analysis focused on mobility in \textit{education}. As education is considered a key link in the transmission of socioeconomic status from one generation to the next \citep{Goldthorpe2014}, one might expect that regions with high educational mobility also exhibit high mobility in \textit{income}. To provide direct evidence on this hypothesis, we link our educational measures to external estimates of income mobility. 

Due to a lack of suitable data sources, there is little evidence on income mobility in Spain. The best-known national-level estimates by \cite{CerviniPla2015} are based on an indirect, two-sample IV approach that predicts parents' income from their education and occupation. However, a recent project by \cite{LlanerasAtlas2020} offers detailed regional estimates of income mobility using linked tax records (the ``\textit{Atlas de Oportunidades}'') for cohorts born between 1984 and 1990. In related work, \cite{soria2025intergenerational} use linked tax records to document strong intergenerational persistence at the top of the income distribution and substantial variation in income mobility across regions.\footnote{\, While based on administrative tax records, the microdata is hosted by a private foundation that has not shared it with other researchers. However, \cite{LlanerasAtlas2020} provide detailed regional estimates of income mobility that we use for our analysis.} As these estimates are based on relatively short income spans, and subject to selection bias, they may not accurately reflect the level of income mobility in Spain \citep{Polavieja2020}. However, we are here primarily interested in mobility differences across regions, for which the \textit{Atlas} remains the best available data source.

By combining municipality- and province-level statistics provided by \cite{LlanerasAtlas2020}, we can construct a measure of intergenerational mobility in income ranks at the same level of aggregation as available in our data sources.\footnote{\, We merge these estimates of income mobility with our estimates of educational mobility (for 96 regions, as described in Section \ref{sec_variables}). Using the 2011 Census, we can measure educational mobility for the 1983-85 birth cohorts, which overlap with the 1984-90 birth cohorts covered by the \textit{Atlas de Oportunidades}. We also merge our measures of the standard deviation and the spousal correlation in schooling.} Table \ref{tab:Atlas} compares this measure of income mobility with our education-based measures. Column 1 shows that the slope coefficient in a regression of the intergenerational rank correlation in income on the intergenerational correlation in schooling (IGC) is positive and statistically significant. This coefficient increases further when we address sampling error in the IGC using a split-sample IV estimator (column 2). In the next columns, we show that across regions, the income rank correlation is also associated with the standard deviation (SD) and the spousal correlation in schooling (assortative mating, AM). 

This evidence confirms a positive association between income and educational mobility across Spanish regions. While this association appears modest, its magnitude is difficult to interpret due to measurement error in short income data and potential selection in the Spanish tax data (\citealt{NybomStuhlerJHR2017,Polavieja2020}).\footnote{\, Similarly, \cite{FletcherJan2019} and \cite{NybomStuhlerSMR} find positive but modest associations between income and educational mobility across regions in the US and Sweden.} Nevertheless, the pattern corroborates our qualitative findings: regions with higher inequality and stronger spousal sorting exhibit lower intergenerational mobility, whether measured by education or income.

\begin{table}
\begin{centering}
\caption{Educational vs. income mobility}
\label{tab:Atlas}
\par\end{centering}
\centering{}%
\begin{tabular}{lcccccc}
\toprule 
 & \multicolumn{6}{c}{Dependent variable: Rank correlation in income}\tabularnewline
\cmidrule{2-7}
 & (1) & (2) & (3) & (4) & (5) & (6) \tabularnewline
\midrule
$IGC(edu)$ & 0.131*** & 0.180*** &  &  &  &  \\
 & (0.028) & (0.050) &  &  &  &  \\
$SD(edu_\textit{father})$ &  &  & 0.012** & 0.014** &  &  \\
 &  &  & (0.005) & (0.005) &  &  \\
$AM(edu_\textit{parent})$ &  &  &  &  & 0.090*** & 0.122* \\
 &  &  &  &  & (0.030) & (0.048) \\
IV &  & Split-IV &  & Split-IV &  & Split-IV \\
Observations & 96 & 96 & 96 & 96 & 96 & 96 \\
$R^2$ & 0.146 & . & 0.087 & . & 0.065 & . \\
\midrule
\end{tabular}
\vspace{-2mm}
\floatfoot{Notes: Regression of the intergenerational rank correlation in income as measured in the \textit{Atlas de Oportunidades} \citep{LlanerasAtlas2020} and the intergenerational correlation, standard deviation, and assortative correlation in years of schooling, for 96 regions and cities. Robust standard errors in parentheses, {*}{*}{*} p\textless 0.01, {*}{*} p\textless 0.05, {*} p\textless 0.1.} 
\end{table}

\section{Conclusions}

By combining multiple sources, we traced the evolution of educational mobility and related distributional statistics in Spain over an exceptionally long period. Our central finding is strong co-movement among inequality, spousal sorting, and mobility. This observation aligns with standard intergenerational models (\citealt{blinder1973model, kremer1997much}), which predict that increased sorting in one generation reduces mobility and raises inequality in the \textit{next} generation. However, we also found that educational inequality and assortative mating co-move within the \textit{same} generation, consistent with the idea that inequality widens social distance and lowers the likelihood of marrying outside one's own educational stratum (\citealt{mare2016educational}).

While a causal link between inequality, assortative mating, and mobility is theoretically plausible, their co-movement on the national level might be a mere coincidence. To further corroborate this link, we used the large sample size and long coverage of our samples to study associations across regions and over time. Differences in spousal sorting explain nearly half of the regional variation in intergenerational correlations. They also explain half of the association between inequality and mobility across regions. Assortative mating thus emerges as a crucial link in understanding the ``Gatsby Curve'', the negative association between inequality and mobility documented in previous research, and as a key factor underlying intergenerational persistence.  

Taken together, our evidence is consistent with a vicious cycle between high inequality, strong assortative mating, and low mobility. On a more optimistic note, this relationship implies that policies that reduce educational inequality could, as a side effect, also weaken assortative mating and increase mobility. Indeed, the decline in inequality, sorting, and immobility in Spain begins with cohorts born in the early 1960s, who were the first exposed to a major comprehensive school reform, and educational reforms have been shown to have major distributional effects in other countries (\citealt{meghir2005educational,holmlund2022preferences}).

\pagebreak 
\onehalfspacing
\bibliographystyle{ecta}
\bibliography{Library_arXiv.bib}
\clearpage
\doublespacing
\appendix

\section*{Appendix}
\section{Variable Definitions and Harmonization}\label{apvardefs}
\setcounter{figure}{0}
\renewcommand{\thefigure}{A\arabic{figure}}
\subsection{Educational attainment}\label{apvardefs1}

To provide comparable estimates of educational mobility across five decades, we harmonize education categories across surveys and waves. This subsection describes how education is coded in each dataset and how we harmonized these categories to create a comparable measure of educational attainment.

The ESD records education in broad categories such as illiterate, literate, and pre-school, and further distinguishes primary, post-compulsory professional education, academic secondary, and postgraduate studies. Notably, it disaggregates most degrees conferred in Spain during the 20th century, providing a comprehensive overview of the country's educational system until 1991. For instance, it specifies the specific type of secondary academic education pursued by respondents, such as bachillerato, bachillerato elemental, bachillerato superior, Preu, BUP, and COU. 
To align the ESD with other sources and cohorts, we follow the classification scheme proposed by \cite{martinez1996construccion}, which maps detailed degree information into broader categories. Moreover, ESD codifies parental and spouse education into eight categories.\footnote{\, Post-compulsory academic secondary education (Bachiller Superior and BUP) and short/long post-compulsory professional education (Formaci\'{o}n Profesional, FP) are considered in the same broad category.} The ESD parental and spouse education information is given retrospectively by the surveyed individual and, therefore, unrelated to the children's co-residence status.

The EPA also records information on the highest academic degree attained, but uses different codification methods across waves. Waves from 1977 to 1986 provide eight categories\footnote{\, These waves do not distinguish short post-compulsory professional education (\emph{Formaci\'{o}n Profesional de grado medio}), long post-compulsory professional education (\emph{Formaci\'{o}n Profesional de grado superior}), and post-compulsory academic secondary education (\emph{Bachillerato}); these education levels are therefore reported in joint categories in the early EPA waves.}, while ten categories from 1987 to 1991, and 1992 onwards provide more detail. The main caveat is that parental education is only observed for co-resident children. 

The Censuses and ECEPOV also record information on the highest degree attained, with the same disaggregation by children, spouses, and parents. The 2001 census aggregates educational information into ten categories for each household member, while the 2011 census, 2021 census, and ECEPOV include twelve categories. Censuses only provide information on parental education for co-resident children.

To harmonize across sources, we collapse categories into a seven-level scheme. This reduction is required because the ESD and early EPA waves report fewer distinctions. Specifically, we merge codes for Bachillerato Superior, BUP, FP1 (media), and FP2 (superior).

The final harmonized categories and the corresponding years of school assigned to them are: Illiterate (1), Literate (3), Primary Schooling  (5), Secondary School (8), Academic high school and professional studies (11), Short college degree (15), Long college degree (18).

\subsection{Place of birth and residence}\label{apvardefs2}

For our regional analysis, the residence and place of birth variables are disaggregated using a subprovincial definition: all municipalities with at least 100,000 inhabitants in 1991 are treated separately, while the remaining areas within each province (NUTS 3) are aggregated into a single unit. This allows us to consistently track educational mobility and its regional correlates across sources and cohorts, yielding 107 observations: 55 large municipalities, 50 aggregated provinces, and 2 autonomous cities.

The 55 municipalities with over 100,000 in 1991 included in the regional analysis are: Alacant, Alcal\'{a} de Henares, Albacete, Alcorc\'{o}n, Algeciras, Almer\'{i}a, Badalona, Badajoz, Barakaldo, Barcelona, Bilbao, Burgos, C\'{a}diz, Cartagena, Castell\'{o} de la Plana, C\'{o}rdoba, Coru\~{n}a (A), Donostia - San Sebasti\'{a}n, Elx, Fuenlabrada, Getafe, Gij\'{o}n, Granada, Hospitalet de Llobregat (L'), Huelva, Ja\'{e}n, Jerez de la Frontera, Le\'{o}n, Legan\'{e}s, Lleida, Logro\~{n}o, Madrid, M\'{a}laga, Matar\'{o}, M\'{o}stoles, Murcia, Ourense, Oviedo, Palma, Palmas de Gran Canaria (Las), Pamplona - Iru\~{n}a, Sabadell, Salamanca, San Crist\'{o}bal de La Laguna, Santa Coloma de Gramenet, Santa Cruz de Tenerife, Santander, Sevilla, Tarragona, Terrassa, Valencia, Valladolid, Vigo, Vitoria-Gasteiz, Zaragoza.

The 50 (rest of) provinces are: Albacete, Alacant, Almer\'{i}a, Araba - \'{A}lava, Asturias, \'{A}vila, Badajoz, Balears (Illes), Barcelona, Bizkaia, Burgos, C\'{a}ceres, C\'{a}diz, Cantabria, Castell\'{o}, Ciudad Real, C\'{o}rdoba, Coru\~{n}a (A), Cuenca, Girona, Gipuzkoa, Granada, Guadalajara, Huelva, Huesca, Ja\'{e}n, Le\'{o}n, Lleida, Lugo, Madrid, M\'{a}laga, Murcia, Navarra, Ourense, Palencia, Palmas (Las), Pontevedra, Rioja (La), Salamanca, Santa Cruz de Tenerife, Segovia, Sevilla, Soria, Tarragona, Teruel, Toledo, Val\`{e}ncia, Valladolid, Zamora, Zaragoza.

The 2 autonomous cities are Ceuta and Melilla.

\clearpage
\section{Coresidence bias}\label{Appendix-Coresidence}
\setcounter{figure}{0}
\renewcommand{\thefigure}{B\arabic{figure}}

This Appendix discusses the ``coresidence bias'' in samples reporting parental education only for individuals living with their parents. We provide evidence on the optimal age for measuring child education in coresident samples\footnote{\, We use the terms dependent and coresident interchangeably to refer to individuals residing in their parents' household.}, test the robustness of our estimates to variations around that age, and validate our findings using a correction method suggested by \cite{Hilger2017}.

\subsection{The optimal age of measurement in coresident samples}\label{Appendix-Coresidence-Optage}

We distinguish two sources of coresidence bias: 
(1) \emph{selection bias}, arising at older ages from differences between dependent and independent children, and 
(2) \emph{censoring bias}, occurring at younger ages because early educational attainment may not reflect final achievement. 
In earlier cohorts, low overall educational levels mitigate censoring bias (Figure \ref{fig:national_trends_education_a}). Selection bias, however, affects all cohorts: as fewer children remain coresident with age, those who do are increasingly more selected.

\begin{figure}
\centering
\caption{Coresidence share measured at ages 24, 27, and 30.}
\includegraphics[width=.75\linewidth]{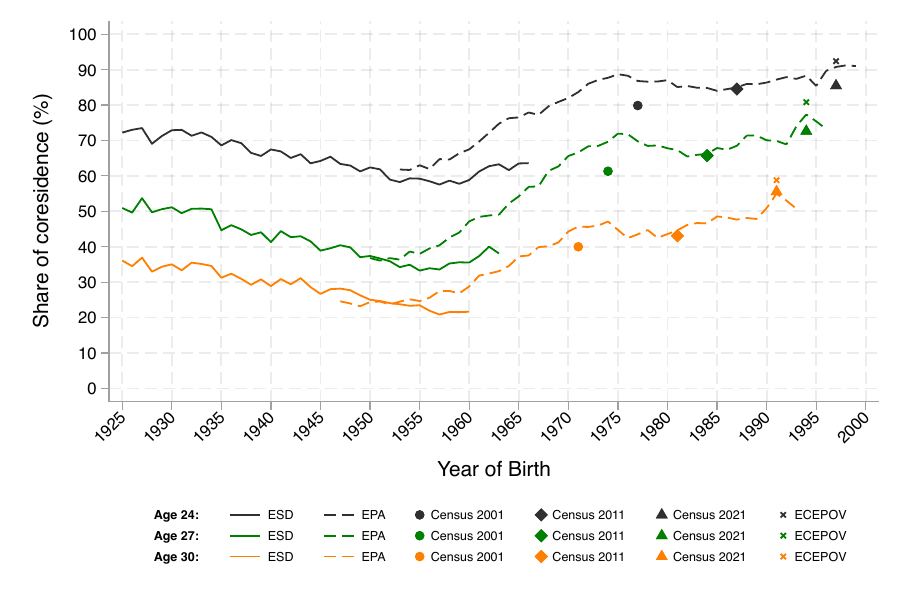}
\label{fig:national_share_dependence_24_27_30}
\floatfoot{Notes: The percentage of children residing in their parental household (\textit{coresident/dependent children}) is presented by cohort and age at measurement. For the ESD dataset, coresidence shares are estimated using retrospective self-reported ages of leaving the parental household. For the EPA, Censuses, and ECEPOV samples, the analysis is restricted to individuals observed within the relevant cohort and age at the time of measurement.}
\end{figure}

Coresidence rates in Spain are high throughout the analysis period, limiting the scope for selection bias. Figure \ref{fig:national_share_dependence_24_27_30} shows that 35 to 75\% of children still live with their parents at age 27, with males more likely to be coresident (Figure \ref{fig:national_sharedependence_gender}). Importantly, coresidence shares are particularly high for recent cohorts, for which we must rely on coresident samples when using EPA or Census data. Our main figures confirm the limited scope for selection bias at our preferred age of measurement (27) in recent cohorts by comparing estimates from ECEPOV for all individuals (both coresident and independent) and for coresident individuals.

\begin{figure}[H]
\begin{subfigure}{0.495\textwidth}
    \centering
    \includegraphics[width=.99\linewidth]{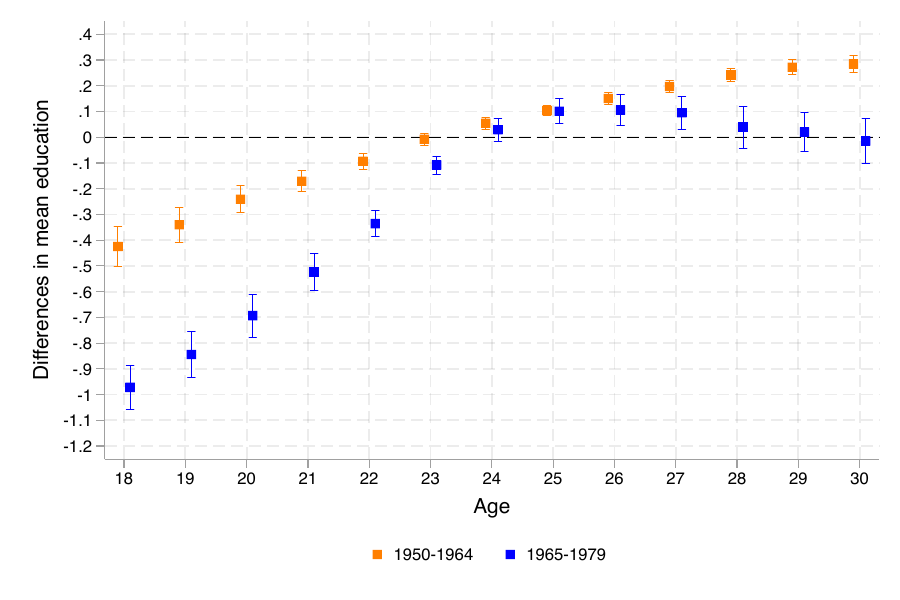}
    \caption{Differences}
\end{subfigure}
\begin{subfigure}{0.495\textwidth}
    \centering
    \includegraphics[width=.99\linewidth]{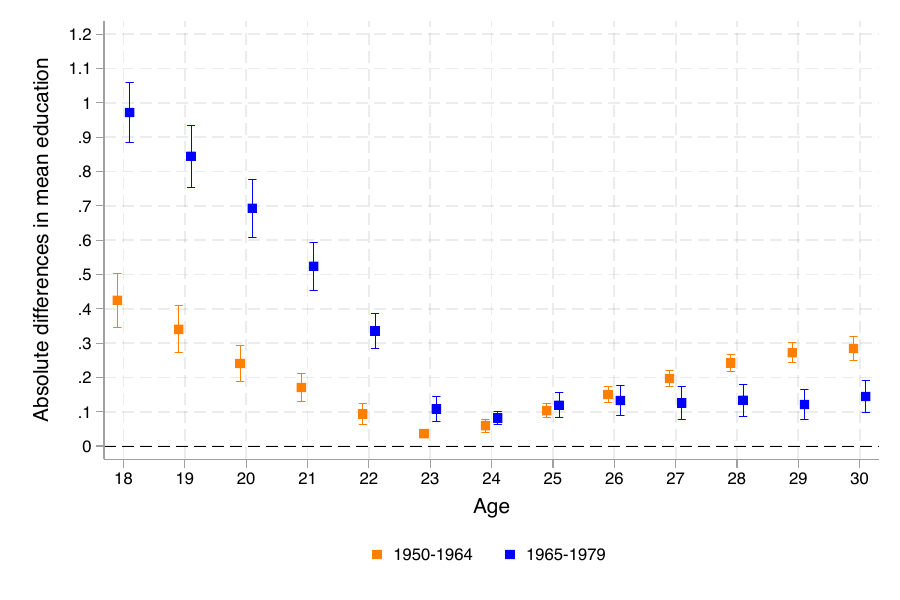}
    \caption{Absolute differences}
\end{subfigure}
\caption{Coresidence bias in education by age of measurement}
\label{fig:coresidencebias}
\floatfoot{Notes: Differences and absolute differences in mean education between using the education measured at a particular age (x-axis) for individuals living at their parents' household (\textit{coresident/dependent}) and the corresponding education at age 30 for all individuals (\textit{coresident} and \textit{independent}). We report the averages and standard errors of the annual (absolute) differences, pooled across two subperiods: 1950-1964 and 1965-1979.}
\end{figure}

In Figure \ref{fig:coresidencebias}, we show the difference (panel a) and absolute difference (panel b) between the average years of schooling for coresident children at a specified age (x-axis) and the average years of schooling for all (coresident and independent) children measured at age 30, when education is generally completed. To study how changes in the marginal distribution affect our estimates, we analyze two subgroups of ``fictitious'' survey years: 1950-1964 and 1965-1979\footnote{\, We use ``fictitious'' survey years instead of birth cohorts. This choice shows the implications of the interplay between survey or census data collected in different calendar years and the use of various educational measurement ages when estimating intergenerational mobility. For instance, the combination of the survey year 1965 and age 20 corresponds to the same 1945 cohort as the combination of the survey year 1975 and age 30.}, with the latter period characterized by significantly higher educational attainment (see Section \ref{sec_institutions}).
As expected, the increase in educational attainment in the 1965-1979 subgroup is associated with a larger censoring bias when measuring education at younger ages, as evidenced by the substantial gap in Figure \ref{fig:coresidencebias}. In contrast, the 1950-1964 subgroup shows a higher mean education among coresidents at older ages, resulting in selection bias, which is less pronounced in the later 1965-1979 period.

\begin{figure}[H]
\begin{subfigure}{0.495\textwidth}
    \centering
    \includegraphics[width=.99\linewidth]{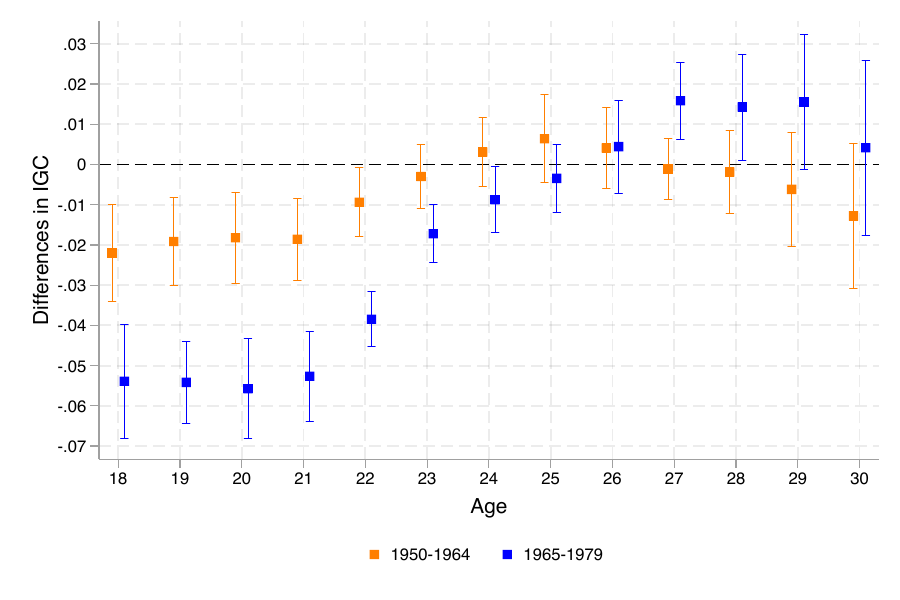}
    \caption{Differences}
\end{subfigure}
\begin{subfigure}{0.495\textwidth}
    \centering
    \includegraphics[width=.99\linewidth]{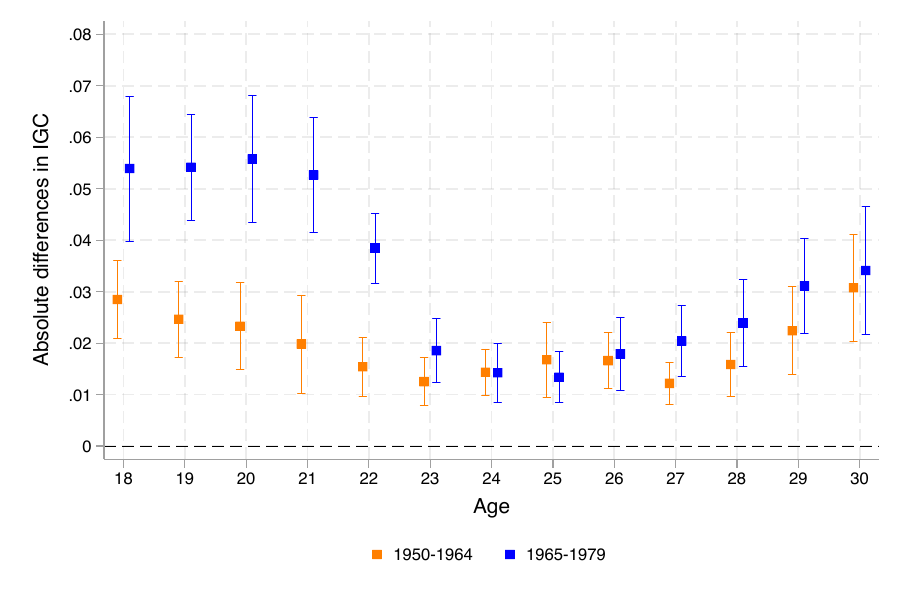}
    \caption{Absolute differences}
\end{subfigure}
\caption{Coresidence bias in the IGC by age of measurement} 
\label{fig:coresidence_bias} 
\floatfoot{Notes: Differences and absolute differences between the intergenerational correlation (IGC) using the education measured at a particular age for individuals living at their parents' home (\textit{coresident/dependent}) and the corresponding correlation at age 30 for all individuals (\textit{coresident} and \textit{independent}). We report the averages and standard errors of the annual (absolute) differences pooled together in two subperiods: 1950-1964 and 1965-1979.}  
\end{figure}

Figure \ref{fig:coresidence_bias} shows the resulting bias in the parent-child correlation in years of schooling when using only the coresident sample in the ESD. We report the difference (panel a) and absolute difference (panel b) between the correlation when measuring child schooling at the specified age (x-axis) for coresident children only and the ``benchmark'' correlation for all children (coresident and independent) measured at age 30, when education is generally completed. We again analyze two subgroups measured in the ``fictitious'' survey years 1950-1964 and 1965-1979 separately. 
% START (optional expression)
The differences and absolute differences in the intergenerational correlation (IGC) at a particular age of measurement for the two subgroups are estimated as follows:

\[
\left\{
\begin{aligned}
    &\text{Diff } IGC_{a}^{1950-1964} = 
    \frac{1}{T_{1950-1964}} \sum_{y=1950}^{1964} 
    \left(IGC^{ \text{dep}}_{y,a} - IGC^{ \text{dep+ind}}_{y,30}\right) \\[0.4em]
    &\text{Diff } IGC_{a}^{1965-1979} = 
    \frac{1}{T_{1965-1979}} \sum_{y=1965}^{1979} 
    \left(IGC^{ \text{dep}}_{y,a} - IGC^{ \text{dep+ind}}_{y,30}\right)
\end{aligned}
\right.
\]

\[
\left\{
\begin{aligned}
    &\text{Abs Diff } IGC_{a}^{1950-1964} = 
    \frac{1}{T_{1950-1964}} \sum_{y=1950}^{1964} 
    \left|IGC^{ \text{dep}}_{y,a} - IGC^{ \text{dep+ind}}_{y,30}\right| \\[0.4em]
    &\text{Abs Diff } IGC_{a}^{1965-1979} = 
    \frac{1}{T_{1965-1979}} \sum_{y=1965}^{1979} 
    \left|IGC^{ \text{dep}}_{y,a} - IGC^{ \text{dep+ind}}_{y,30}\right|
\end{aligned}
\right.
\]

where $a$ is the age of measurement, $y$ is the ``fictitious'' survey year, $dep$ refers to dependent children, $ind$ to independent children, and $T_{1950-1964}$ and $T_{1965-1979}$ denote the number of survey years available in each subperiod.

In both subgroups, the coresidence bias is minimized between ages 23 to 27, compared to earlier or later ages. As already explained, this pattern arises due to the interplay of two types of biases: selection bias, which is pronounced at older ages, and censoring bias, which is highest at younger ages. Therefore, the overall bias is minimized in the mid-20s when most children have completed their schooling, but coresidence rates remain high. Between ages 23 and 27, the bias is low (around 0.01-0.02 in absolute size) and, importantly for our purposes, stable across the different time periods. 

However, Figure \ref{fig:coresidence_bias} also highlights why censoring becomes a more severe issue when studying mobility \textit{trends}. Education becomes increasingly censored as educational attainment increases over cohorts, mechanically decreasing the intergenerational correlation. Consequently, the censoring bias at younger ages is expected to be more severe in recent than in earlier cohorts (i.e., the bias itself trends over birth cohorts). A comparison of the two subgroups of ``fictitious'' survey years in Figure \ref{fig:coresidence_bias} (panel a) confirms this pattern. To provide sufficient buffer against this censoring issue for more recent cohorts, we choose age 27 as our preferred age of measurement.

% Coresidence bias by age in the Encuesta Sociodemografica 

Nevertheless, to ensure that our \textit{mobility trends} are not dependent on the selection of age 27 for measurement, we replicate the IGC and IGR results measuring education at ages 24 and 30 in Figures \ref{fig:national_IGC_age} and \ref{fig:national_IGR_age}. Level differences are observed depending on the chosen age, as expected due to the biases discussed, but the overall trends remain consistent across the different age specifications.  % In Appendix Figure X, we show the contribution of Bias 1/2 ... check the code! 

To conclude, in settings with low educational attainment, researchers can mitigate the coresidence bias by measuring educational outcomes at an early age, when most children still reside with their parents (\citealt{alesina2021intergenerational, derenoncourt2022can}). As attainment rises, censoring bias becomes more relevant, requiring older ages for measurement. Researchers should balance selection and censoring biases based on context, particularly when analyzing mobility trends over extended periods.

\subsection{Correcting for coresidence bias}\label{Appendix-Coresidence-Correcting}

We have established that coresidence bias is minimal at the selected age of 27. In this subsection, we nevertheless evaluate whether a correction method proposed by \cite{Hilger2017} could be an effective alternative to address coresidence bias in Spain. The correction method targets what we refer to as \emph{selection bias} without directly addressing \emph{censoring bias}. However, by mitigating selection bias, which is more pronounced at older ages, this method might enable researchers to select a later age of measurement, thereby indirectly reducing censoring bias as well. 

The method aims to overcome the limitation that in census or survey data, parental education is unobserved for ``independent'' children who moved out from the parental home, and to approximate the CEF $E[y_{i t} \mid y_{i t-1}]$, where $y_{i t}$ represents the educational attainment of individual $i$ in generation $t$, and $y_{i t-1}$ denotes the educational attainment of the parents. 

To understand the method, we first decompose the average years of schooling $y_{a y_{t-1}}$ for children of a specific age $a$ and parental education group $y_{t-1}$ as 
\begin{align}
y_{a y_{t-1}} = d_{a y_{t-1}} y_{a y_{t-1}}^{D} + (1 - d_{a y_{t-1}}) y_{a y_{t-1}}^{I} \label{eq:3}
\end{align}
where  $d_{a y_{t-1}}=\frac{N_{a y_{t-1}}^{D}}{N_{a y_{t-1}}^{D}+N_{a y_{t-1}}^{I}}$ is the dependency rate, i.e. the share of children at age $a$ who co-reside with their parents, $y_{a y_{t-1}}^{D}$ is the average schooling for dependent children, and $y_{a y_{t-1}}^{I}$ is the average schooling for independent children. 

The correction method relies on two key assumptions: (i) the ``parallel trends'' assumption, which assumes that the difference in educational attainment between dependent and independent children is a constant unobserved parameter $\rho$ across parental educational groups, $y_{a y_{t-1}}^I - y_{a y_{t-1}}^D=\rho$ $\forall y_{t-1}$, and (ii) the ``smooth cohorts'' assumption, to approximate the number of independent children (and therefore the dependency rate) in each parental educational group based on the corresponding parental group shares among younger children, who still co-reside with their parents.\footnote{\, Specifically, \cite{Hilger2017} assumes that the parental group shares are identical for children aged 17 and those aged 26-29 in each observation year. Similarly, we measure parental group shares at age 16 and extrapolate these shares to our target ages within each observation year.}

For estimating the average schooling for independent children in each parental schooling group, the unobserved parameter $\rho$ is estimated as
\begin{align}
\hat{\rho} = y^{I} - \left(\sum_{g=1}^{G} \frac{\hat{N}_{a g}^{I}}{\hat{N}_{a}^{I}} y_{a g}^{D}\right)  \label{eq:4}
\end{align}
where $\frac{\hat{N}_{a g}^{I}}{\hat{N}_{a}^{I}}$ represents the estimated share of independent children in parental educational attainment group $g$, and $y^{I}$ is the overall average schooling for independent children. Using this estimate, we can estimate the average schooling for independent children in each parental group,  $\hat{y}_{a y_{t-1}}^{I}=y_{a y_{t-1}}^D+\hat{\rho}$, and in turn, the overall educational attainment by parental groups $\hat{y}_{a y_{t-1}}$ using eq. \eqref{eq:3}. This process yields non-parametric estimates of the CEF $E[y_{i t} \mid y_{i t-1}]$, grouped into bins of parental educational attainment.

If we have data that reports parental background regardless of coresidence status, one can test the ``parallel trends'' assumption by estimating the regression
\begin{align}
y_{i t}=\alpha+\beta y_{i t-1}+\rho d_{i}+\gamma y_{i t-1} d_{i}+\varepsilon_{i t}  \label{eq:5}
\end{align}

where $y_{it}$ and $y_{it-1}$ are the education of the child and parent in family $i$, respectively, and $d_i$ is an indicator that equals one for independent children and zero otherwise. The coefficient $\beta$ captures the impact of father's education attainment on children's education, $\rho$ captures a level shift in the CEF of independent children, and $\gamma$ captures differences in the slope between independent and dependent children with respect to father's education. The parallel trends assumption is plausible if the interaction term $\gamma$ is small relative to the main effect of parental background as captured by $\beta$. 

\begin{figure}[h]
\begin{subfigure}{0.496\textwidth}
    \centering{
    \includegraphics[width=\linewidth]{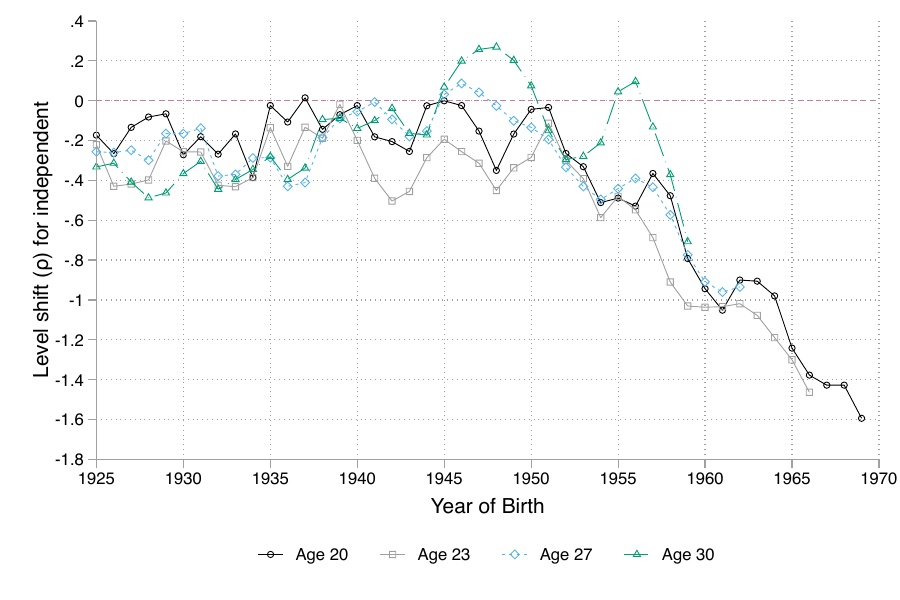} }
\end{subfigure}
\begin{subfigure}{0.496\textwidth}
    \centering{
    \includegraphics[width=\linewidth]{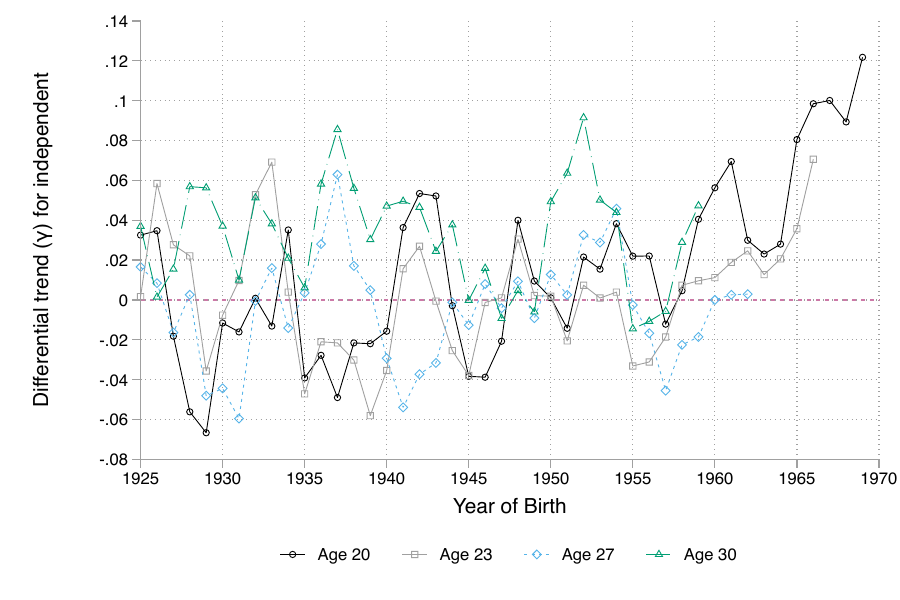}}
\end{subfigure}
 \caption{Estimated parameters $\rho$ and $\gamma$ by cohort and age of measurement}
 \label{fig:hilger_parameters}  
 \floatfoot{Notes: The left panel shows $\rho$ (intercept shift), while the right panel depicts $\gamma$ (differential slope) from equation \ref{eq:5}, estimated at ages 20, 23, 27, and 30 across cohorts. Each cohort estimate is smoothed using a 3-year moving average, incorporating data from adjacent cohorts. A negative $\rho$ indicates independent children attain less schooling than dependent children, controlling for parental education. Positive $\gamma$ reflects a steeper schooling gradient relative to parental background among independent children.}
\end{figure}

Using the ESD, we estimate eq.\eqref{eq:5} at ages 20, 23, 27, and 30 for cohorts between 1925 to 1970, and plot estimates of $\rho$ and $\gamma$ in Figure \ref{fig:hilger_parameters}. Estimates of $\rho$ are negative across most cohorts and ages, indicating a negative intercept in the CEF of independent children. Notably, we observe a substantial decrease in $\rho$ for cohorts starting from the 1950s, attributable to educational expansion. With an increase in the average level of education, the dependency rate increases (see Figure \ref{fig:national_share_dependence_24_27_30}), and independent children become more negatively selected in educational attainment than dependent children. The interaction term $\gamma$ varies depending on the age at which education is measured and the cohort. For example, while estimates of $\gamma$ fluctuate around zero at age 27, they tend to be positive at age 30, implying a violation of the assumption of ``parallel trends''. %In cases where $\gamma$ is positive, the CEF of independent children becomes steeper compared to dependent children as individuals grow older.
Moreover, while $\gamma$ is close to zero for ages 20 and 23 in earlier survey years, it becomes consistently positive for cohorts affected by rapid educational expansion in the 1970s and 80s. This combination of a negative, decreasing $\rho$ alongside a positive, increasing $\gamma$ reflects a widening educational attainment gap between independent and dependent children of lower-educated parents. Conversely, this gap remains relatively narrow for children with higher-educated parents. This pattern might reflect differences in the underlying causes of non-dependency between lower-educated and higher-educated households, such as relocating to another city to work or to attend university.

\begin{figure}[h]
\centering
 \includegraphics[scale=1]{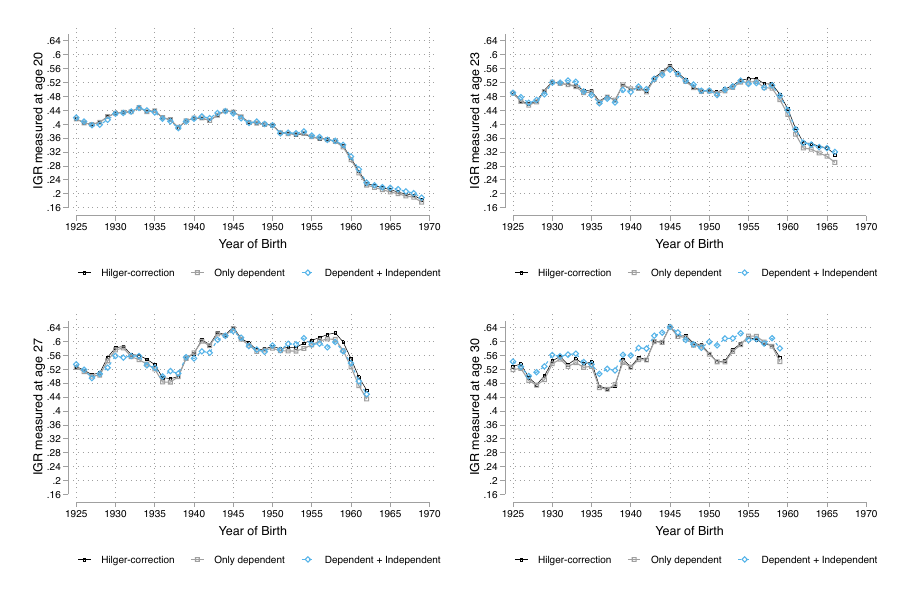}
 \caption{Hilger bias-corrected, only-dependent, and benchmark (dependent + independent) IGR estimates}
 \label{fig:IGR_hilger_dependent_true} 
  \floatfoot{Notes: Hilger bias-corrected IGR estimates (black circles), IGR estimates based solely on the dependent children subsample (grey squares), and benchmark IGR estimates using both dependent and independent children (blue triangles) are reported for ages of measurement of 20, 23, 27, and 30 across cohorts. Each cohort estimate is smoothed using a 3-year moving average, incorporating data from adjacent cohorts.}
\end{figure}

Figure \ref{fig:IGR_hilger_dependent_true} illustrates the performance of the correction method at each age, comparing ``raw'' estimates of the IGR from dependent (coresident) samples, bias-corrected estimates, and the benchmark estimates based on both dependent and independent children. Moreover, Figures \ref{fig:hilger_bias_obs} and \ref{fig:hilger_bias_correction} provide additional details on the extent of bias and bias correction across survey years. Figure \ref{fig:hilger_bias_obs} illustrates the ``dependent bias'', calculated as the difference between the coresident and the benchmark estimate (dependent and independent). Figure \ref{fig:hilger_bias_correction} illustrates the extent to which the correction method reduces bias. 

We find that the bias-corrected estimates tend to be closer to the benchmark than the raw estimates, but the method's effectiveness varies by child age and cohort. At ages 20 and 23, the method performs as intended, with the bias-corrected estimates aligning more closely with the benchmark than the raw estimates based solely on coresident children. However, at this age, the gap between raw and benchmark estimates is small to begin with, so the bias correction makes little difference.\footnote{\, This may be surprising, given that Figure \ref{fig:hilger_parameters} shows a strong level shift in education for independent children. However, combining equations (\ref{eq:3}) and (\ref{eq:5}) under the parallel-trends restriction ($\gamma \approx 0$) implies that the population slope satisfies
$\partial y_{a y_{t-1}} / \partial y_{t-1} = \beta - \rho\,\partial d_{a y_{t-1}} / \partial y_{t-1}$, 
so the bias of the dependent-only slope $\beta$ equals 
$\rho\,\partial d_{a y_{t-1}} / \partial y_{t-1}$.
When dependency rates are essentially flat across parental education groups ($\partial d_{a y_{t-1}} / \partial y_{t-1} \approx 0$), this bias is negligible even if $|\rho|$ is sizable, as is the case for ages 20 and 23.} As illustrated in the previous section, the main source of bias at such a young age is not a selection bias from focusing on co-resident samples but the censoring bias from measuring education before it is completed -- and this bias is not addressed by the correction method.

At our preferred age of measurement, age 27, the estimates based solely on coresident children are already similar to the benchmark estimates (dependent and independent). This follows the pattern shown in Figure \ref{fig:coresidence_bias}, and illustrates that the coresidence issue can be effectively addressed by choosing an ``optimal'' age at measurement, at least in our setting where coresidence shares remain relatively high. At this age, applying the bias-correction method results in only negligible changes, and not always in the right direction. Moreover, at age 30, the Hilger-corrected IGR estimates fail to close the gap between the estimates based on coresident children and the benchmark IGR estimates (dependent + independent). This is partly due to the limited magnitude of the $\rho$ parameter and the lack of clear ``parallel trends'' as shown in Figure \ref{fig:hilger_parameters}, but also due to the violation of the ``smooth cohorts'' assumption.

In Figure \ref{fig:IGR_hilger_smooth}, we evaluate the assumption of ``smooth cohorts'' by comparing fathers' educational group shares for children of different ages (16 vs. 30) in the same ``fictitious'' survey year. In recent cohorts, the method fails to correct for selection bias in Spain among older age groups (27 and 30) because it violates the smooth-cohorts assumption. Unlike the US, Spain experienced a sharper increase in educational attainment among post-1950 cohorts, as shown in Figure \ref{USfigure}. Such rapid changes in the marginal distribution of education may invalidate the ``smooth cohorts'' assumption when the distance between the target age (e.g., 27 or 30) and the ages used for parental education group share estimation (e.g., 16) is large. The method performs well when the age gaps are narrower (e.g., 20 vs. 16 or 23 vs. 16) as shown in Figure \ref{fig:IGR_hilger_dependent_true}. However, at these ages, the main concern is censoring bias, not selection bias, and the correction method is not designed to address it. 

\begin{figure}[h] % Average absolute bias for age 27 is 0.15 and for age 30 is 0.24
    \centering
    \includegraphics[width=0.7\linewidth]{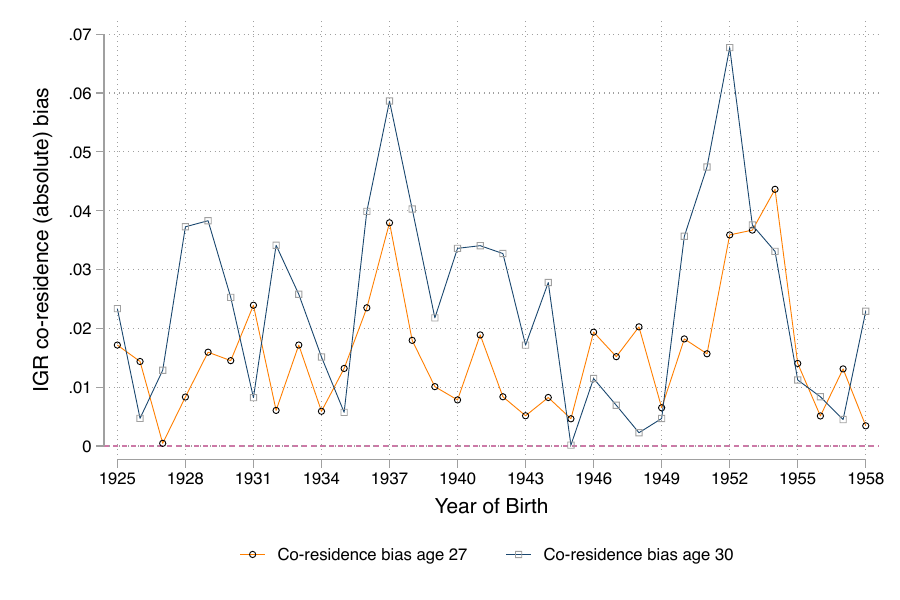}
 \caption{Absolute bias in IGR estimates for dependent-only samples by cohort and age of measurement}
 \label{fig:IGR_hilger_dependent_true_cohort_bias27vs30}  
  \floatfoot{Notes: Absolute bias difference between ``true'' IGR estimates (dependent + independent children at age 30) and IGR estimates based on dependent children only, measured at ages 27 and 30. For comparability, we limit the analysis to cohorts born from 1925 to 1958. The average absolute bias is 0.015 at age 27 and 0.024 at age 30, suggesting age 27 is the preferred measurement age to minimize coresidence bias.}
\end{figure}

Finally, in Figure \ref{fig:IGR_hilger_dependent_true_cohort_bias27vs30}, we compare the absolute estimated bias when using the coresident sample at age 27 and 30 to the ``true'' IGR estimated at age 30 for both dependent and independent children (i.e., using a sample that should not suffer from censoring or selection). Note that the coresident sample at age 27 combines both censoring and selection bias, whereas at age 30 it is entirely driven by selection. For cohorts born between 1925 and 1958, the average absolute bias for age 27 is 0.015, and for age 30, it is 0.024. The absolute bias at age 30 is 60\% higher than at age 27, confirming that age 27 is the age that minimizes the overall bias, as already discussed in Section \ref{Appendix-Coresidence-Optage}. 

Based on this evidence, we do not apply this correction method when estimating intergenerational mobility using the EPA and censuses' coresident samples. Although the correction method reduces selection bias when measuring education in the early 20s, the primary source of bias at those ages is censoring, which the correction method does not address. Instead, we focus on our preferred age of measurement, 27, minimizing the overall bias (from both selection and censoring). At this age, the correction method does not yield consistent improvements, due to violations of the underlying parallel trends and smooth cohort assumptions. In addition, we perform robustness checks using ECEPOV, which reports parental education for all children (coresident and independent) for the most recent birth cohorts in our sample.  

\clearpage
\section{Appendix Figures}
\setcounter{figure}{0}
\renewcommand{\thefigure}{C\arabic{figure}}

\begin{figure}[ht]
\centering
\includegraphics[width=.7\linewidth]{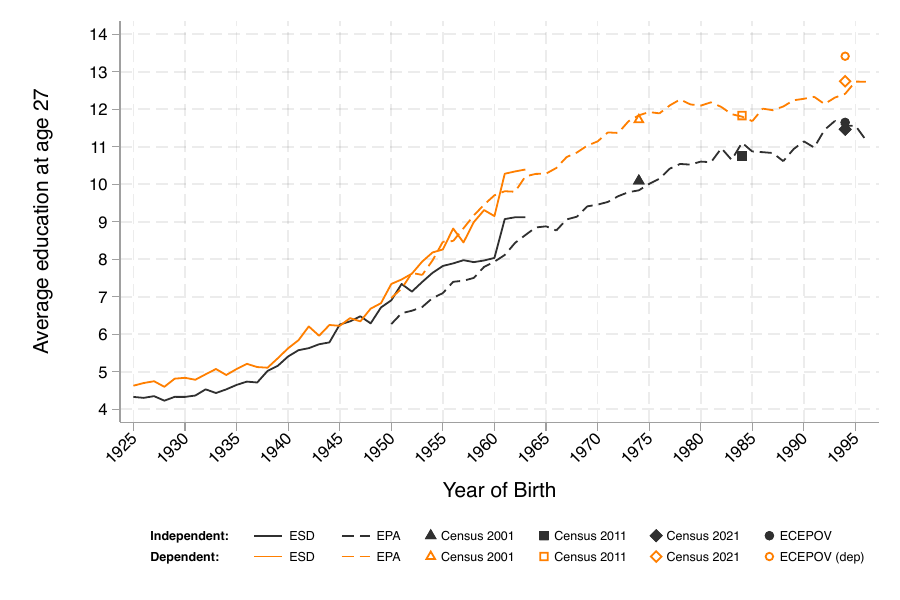}
\caption{Trends in educational attainment in Spain by coresidence status}
\label{fig:national_educ_bycoresidence}
 \floatfoot{Notes: The figure shows the average years of schooling at age 27, disaggregated by coresidence status, for cohorts born between 1925 and 1996. Black lines represent independent children (living outside the parental household), while orange lines represent dependent children (residing in the parental household at age 27). Data are harmonized across six sources: ESD, EPA, ECEPOV, Census 2001, Census 2011, and Census 2021. See Appendix \ref{apvardefs1} for data harmonization details.}
\end{figure}

\begin{figure}[ht]
\label{fig:national_sdeduc}
\begin{subfigure}{0.7\textwidth}
    \centering
    \includegraphics[width=.995\linewidth]{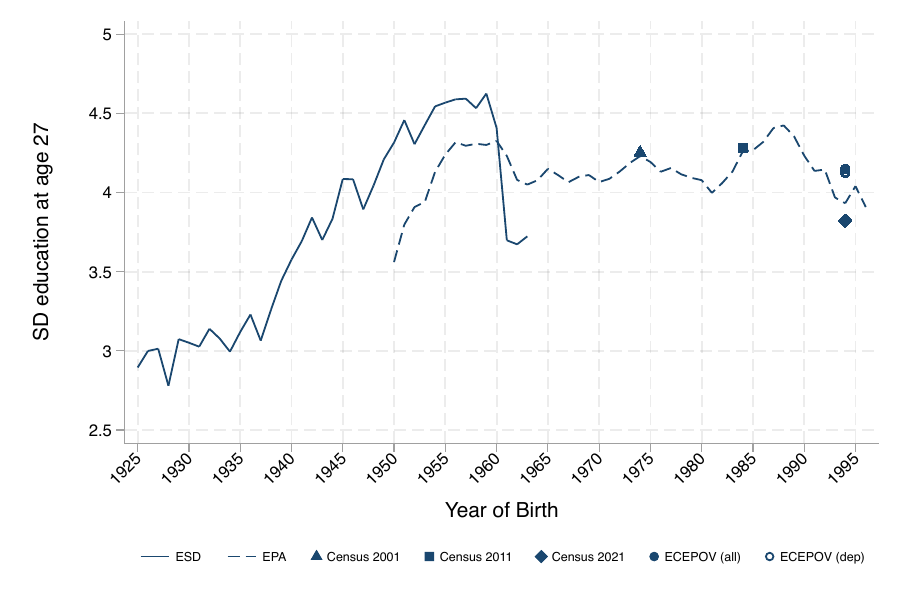}
    \subcaption{SD in education}
\label{fig:national_sdeduc_a}
\end{subfigure}
\begin{subfigure}{0.7\textwidth}
    \centering
    \includegraphics[width=0.995\linewidth]{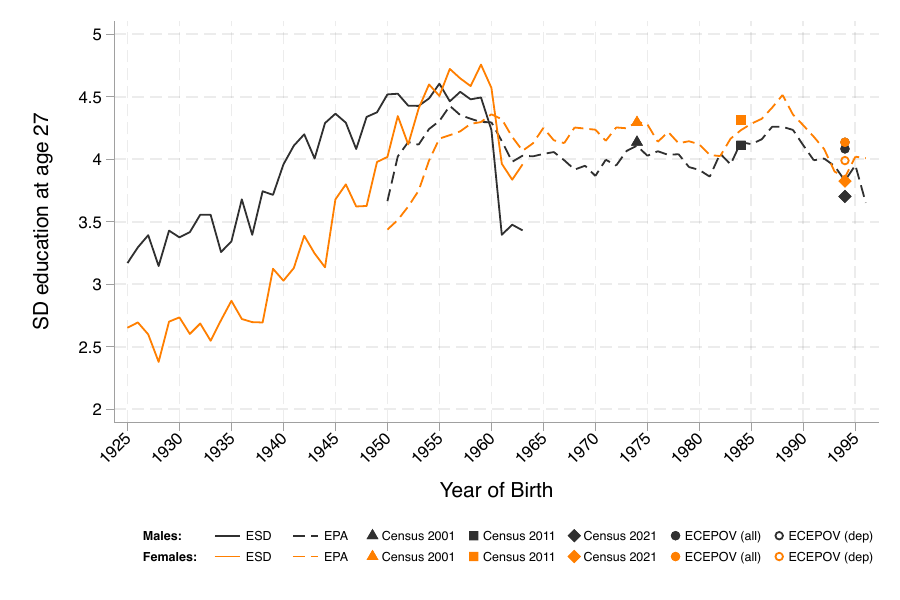}
    \subcaption{SD in education by gender}
\label{fig:national_sdeduc_b}
\end{subfigure}
\caption{Trends in educational dispersion in Spain}
\label{fig:sd27}
 \floatfoot{Notes: Panel (a) depicts the standard deviation (SD) in years of schooling at age 27 for cohorts born between 1925 and 1996. Panel (b) shows the standard deviation (SD) in years of schooling at age 27 for the same cohorts, disaggregated by gender. Black lines represent estimates for male children, while orange lines represent estimates for female children. Years of schooling are harmonized across six data sources: ESD, EPA, ECEPOV, Census 2001, Census 2011, and Census 2021. For ECEPOV, separate estimates are provided for dependent children residing with their parents (\textit{dep}) and the full sample (\textit{all}). See Appendix \ref{apvardefs1} for data harmonization details.}
\end{figure}

\begin{figure}[ht]
\centering
\includegraphics[width=.7\linewidth]{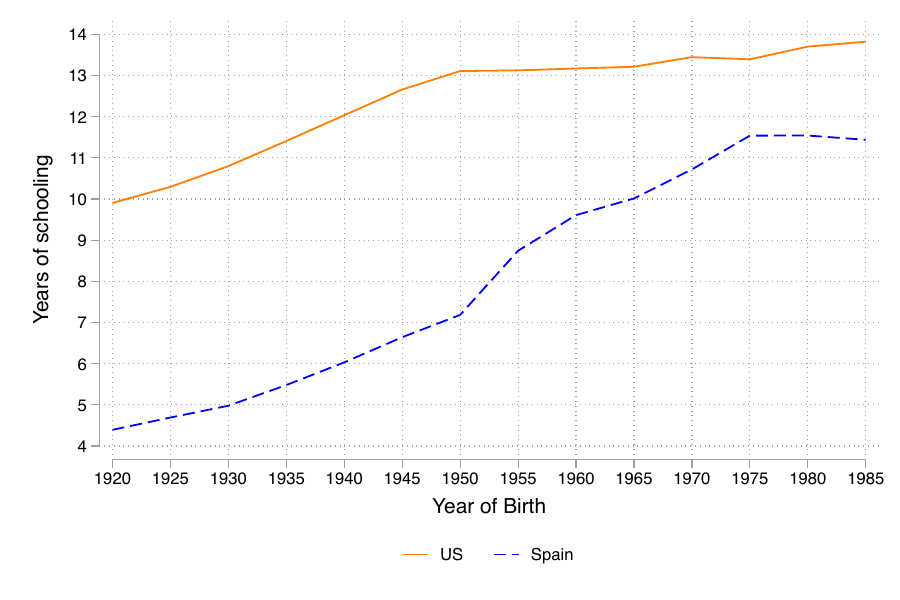}
\caption{Trends in educational attainment in Spain and the United States }
\label{USfigure}
 \floatfoot{Notes: Average years of schooling for cohorts born between 1920 and 1985 in Spain and the United States, based on the \cite{barro2013new} educational attainment dataset. Estimates are based on individuals aged 25 to 34, measured between 1950 and 2015. Because of the 25-to-34 age group aggregation, cohort estimates (x-axis) measure a 10-year moving smoothed cohort window, incorporating data from four earlier and five later cohorts. For instance, the 1920 cohort includes individuals born in 1920 (measured at age 30), 1916 (measured at age 34), and 1925 (measured at age 25).}
\end{figure}

\begin{figure}[!htb]
\begin{subfigure}{0.7\textwidth}
  \centering
  \includegraphics[width=\linewidth]{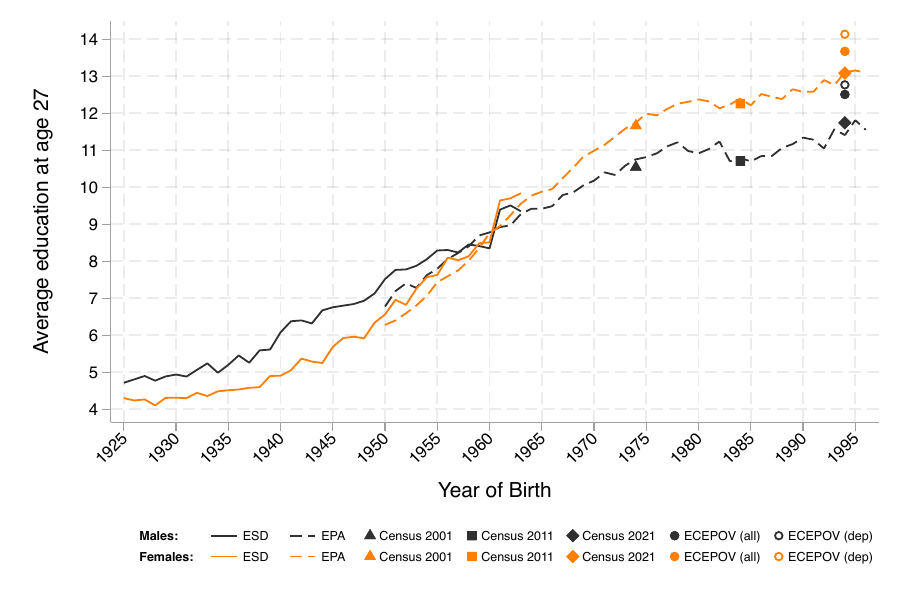}
  \subcaption{Average years of schooling by gender}
    \label{fig:national_educ_gender_a}
\end{subfigure}%

\begin{subfigure}{0.7\textwidth}
  \centering
  \includegraphics[width=\linewidth]{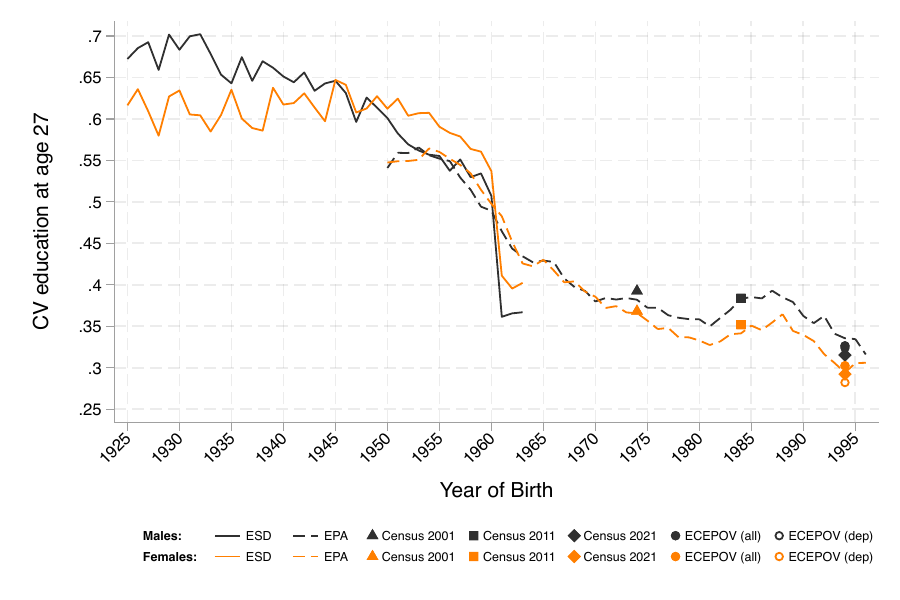}
  \subcaption{Coefficient of variation in years of schooling by gender}
    \label{fig:national_educ_gender_b}
\end{subfigure}
\caption{Trends in educational attainment in Spain by gender}
\label{fig:national_educ_gender}
 \floatfoot{Notes: Panel (a) illustrates the average years of schooling at age 27 for cohorts born between 1925 and 1996, disaggregated by gender. Panel (b) shows the coefficient of variation (CV) in years of schooling at age 27 for the same cohorts, disaggregated by gender. Black lines represent estimates for male children, while orange lines represent estimates for female children. Years of schooling are harmonized across six data sources: ESD, EPA, ECEPOV, Census 2001, Census 2011, and Census 2021. For ECEPOV, separate estimates are provided for dependent children residing with their parents (\textit{dep}) and the full sample (\textit{all}). See Appendix \ref{apvardefs1} for data harmonization details.}
\end{figure}

\begin{figure}[!htb]
    \centering
    \begin{subfigure}{0.39\textwidth}
        \centering
        \includegraphics[width=\linewidth]{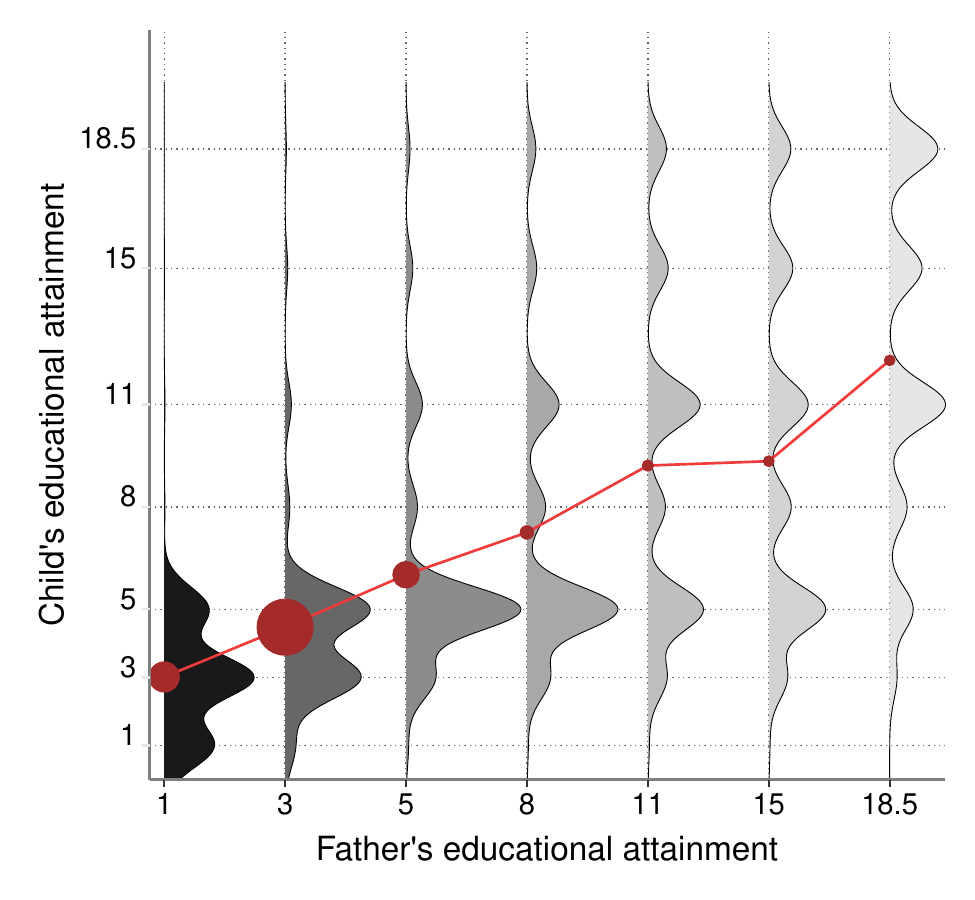}
        \subcaption{ESD (Cohorts 1925--1940)}
        \label{fig:child_father_educ_densities_esd_early}
    \end{subfigure}
    \hspace{1.5cm}
    \begin{subfigure}{0.39\textwidth}
        \centering
        \includegraphics[width=\linewidth]{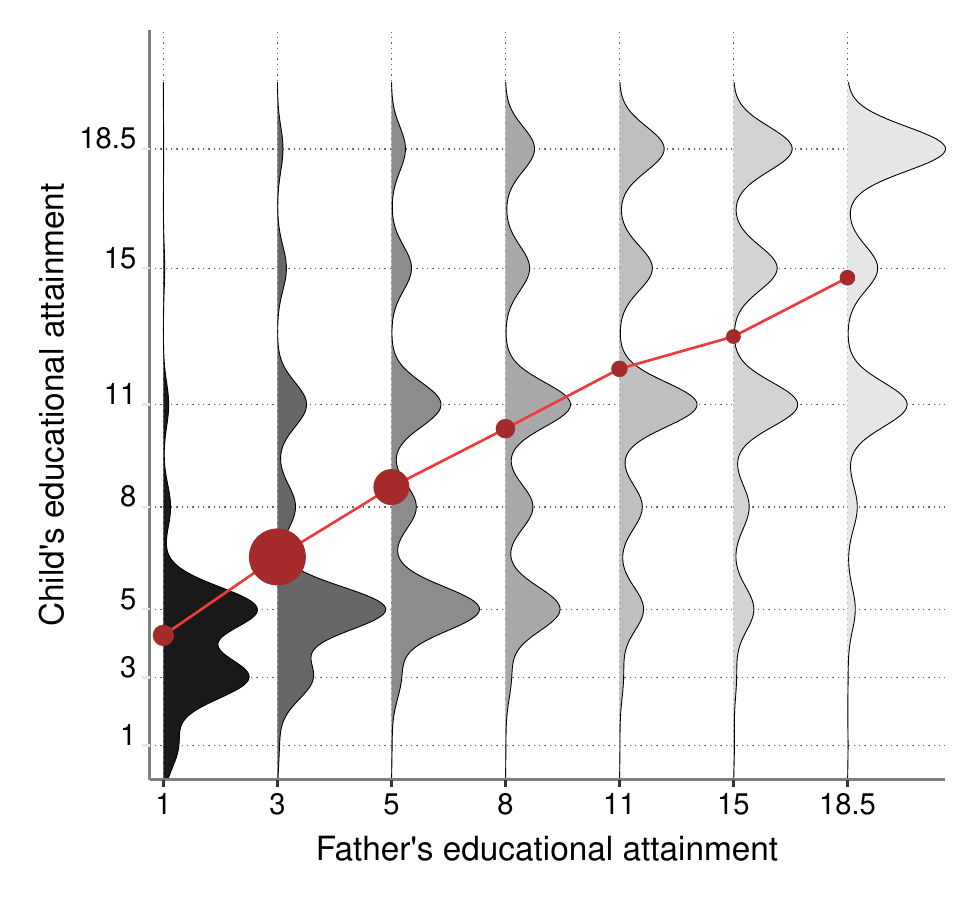}
        \subcaption{ESD (Cohorts 1945--1960)}
        \label{fig:child_father_educ_densities_esd_late}
    \end{subfigure}

    \begin{subfigure}{0.39\textwidth}
        \centering
        \includegraphics[width=\linewidth]{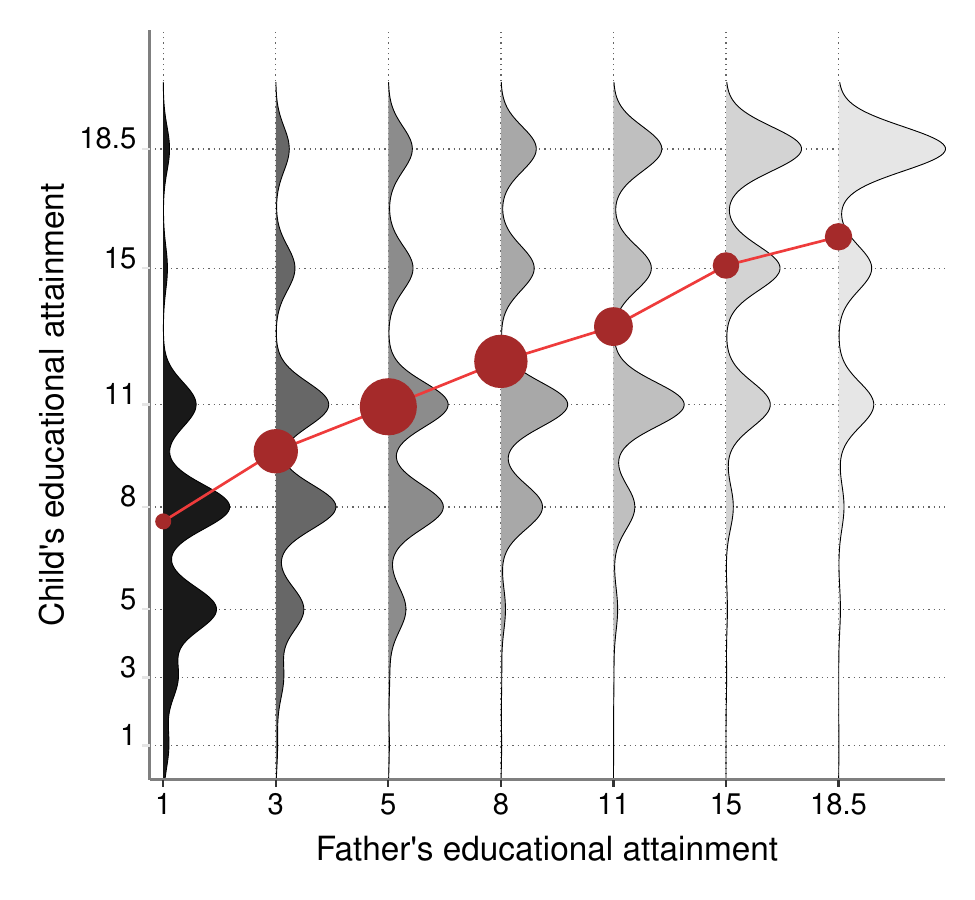}
        \subcaption{Census 2001 (Cohort 1974)}
        \label{fig:child_father_educ_densities_cen01}
    \end{subfigure}
    \hspace{1.5cm}
    \begin{subfigure}{0.39\textwidth}
        \centering
        \includegraphics[width=\linewidth]{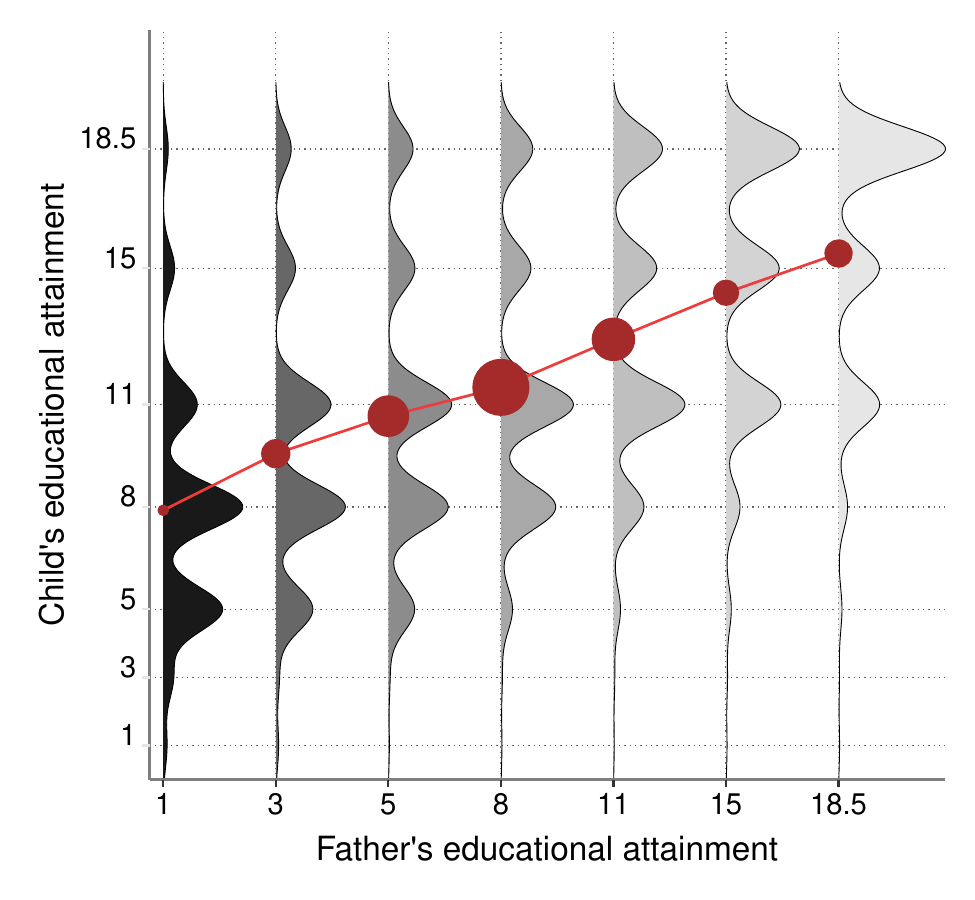}
        \subcaption{Census 2011 (Cohort 1984)}
        \label{fig:child_father_educ_densities_cen11}
    \end{subfigure}

    \begin{subfigure}{0.39\textwidth}
        \centering
        \includegraphics[width=\linewidth]{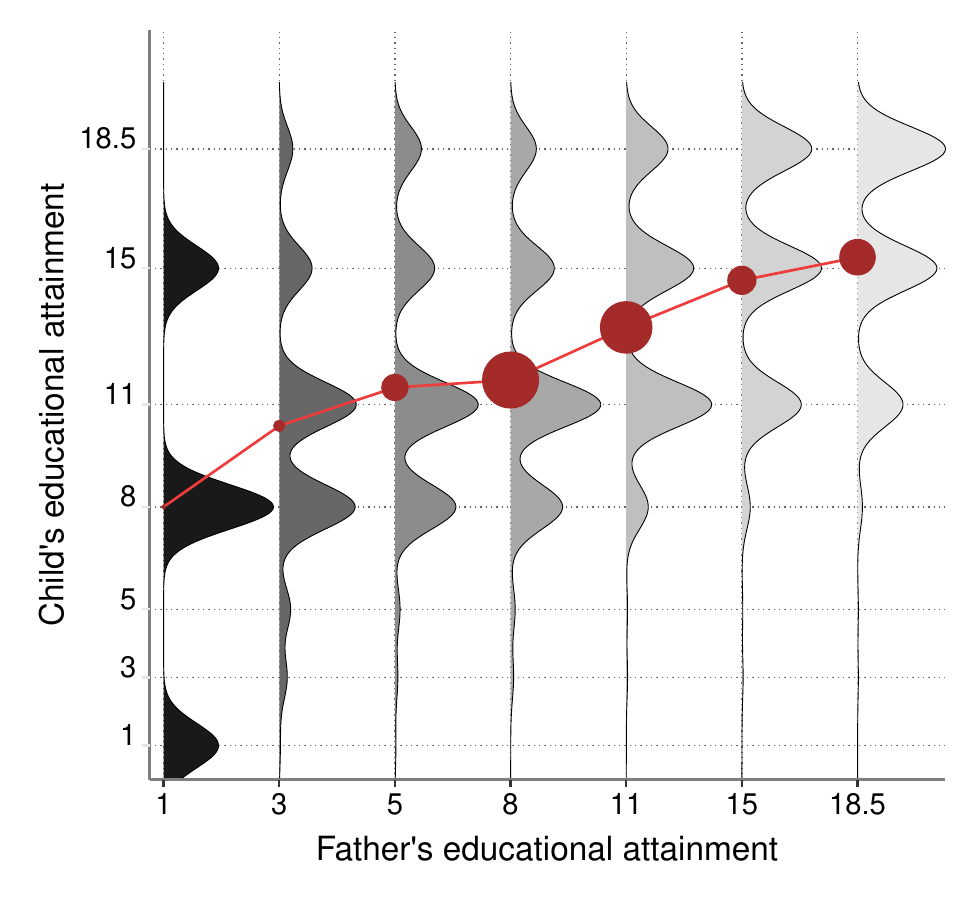}
        \subcaption{Census 2021 (Cohort 1994)}
        \label{fig:child_father_educ_densities_cen21}
    \end{subfigure}
    \hspace{1.5cm}
    \begin{subfigure}{0.39\textwidth}
        \centering
        \includegraphics[width=\linewidth]{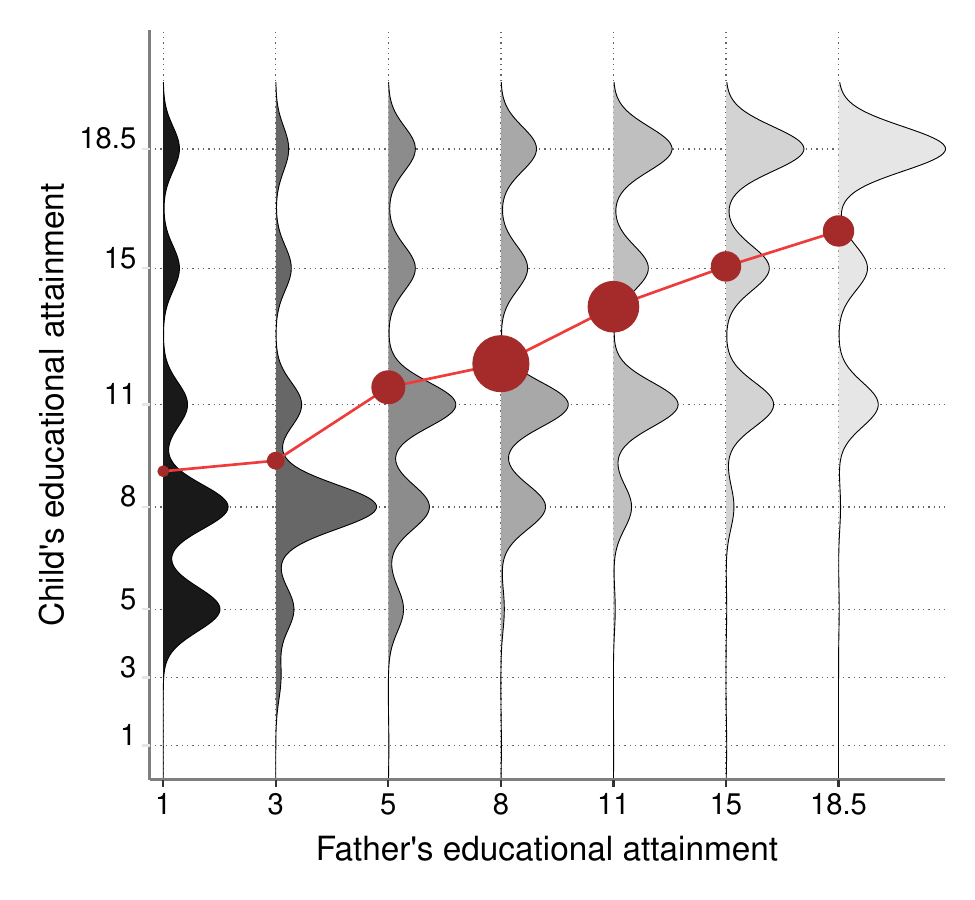}
        \subcaption{ECEPOV (Cohort 1994)}
        \label{fig:child_father_educ_densities_ecepov}
    \end{subfigure}

    \caption{Conditional distribution of children's years of completed education by father's education}
    \label{fig:child_father_educ_densities}
    
    \floatfoot{Notes: The figure plots the conditional distribution of children's education by father's years of completed education, measured when the child is age 27. Grey distributions represent smoothed conditional densities of children's educational attainment at each level of father's education. The red line and markers indicate the conditional mean, with marker size proportional to the share of fathers at that level of education. Panels (a) and (b) use data from the Sociodemographic Survey (ESD), and panel (f) uses ECEPOV; these samples include both co-resident and independent children. Panels (c), (d), and (e) use Census data (2001, 2011, and 2021) restricted to co-resident children living with their parents.}
\end{figure}

\begin{figure}[ht]
\begin{subfigure}{0.7\textwidth}
    \centering
    \includegraphics[width=.995\linewidth]{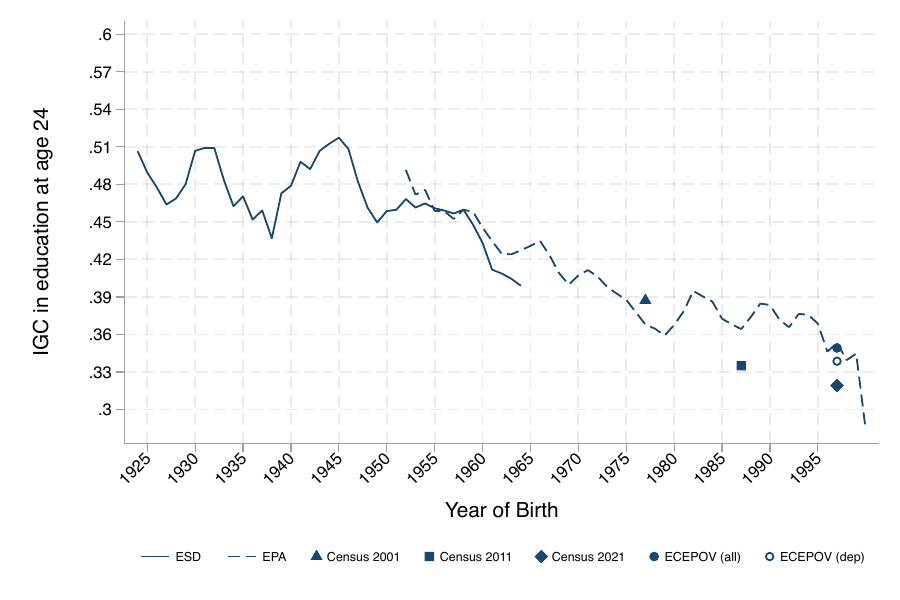}
    \subcaption{IGC in education at age 24}
\end{subfigure}
\begin{subfigure}{0.7\textwidth}
    \centering
    \includegraphics[width=.995\linewidth]{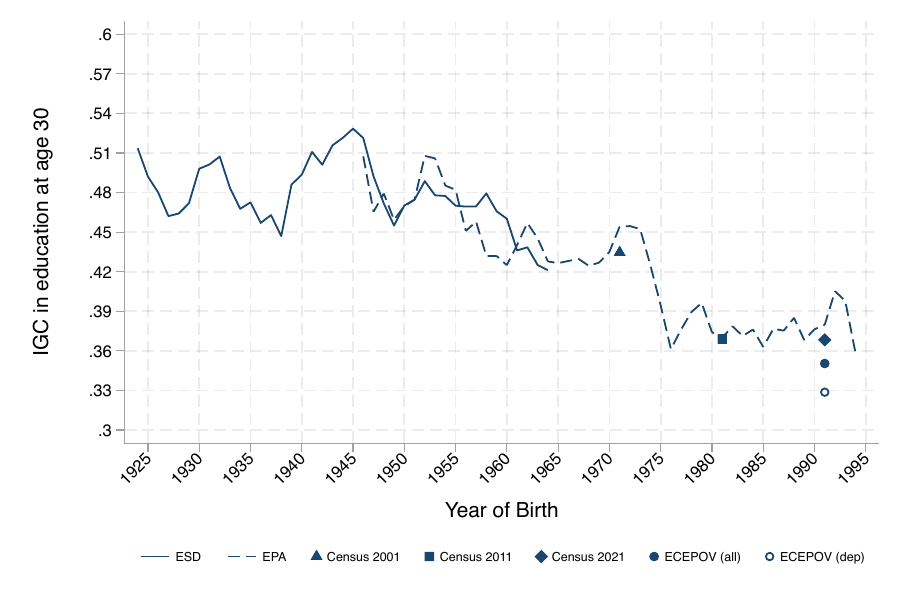}
    \subcaption{IGC in education at age 30}
\end{subfigure}
\caption{Trends in the intergenerational correlation in education at different ages}
\label{fig:national_IGC_age}
\floatfoot{Notes: 3-year moving average of the intergenerational correlation coefficient (IGC) in years of schooling, measured at the children's age of 24 (Panel a) and 30 (Panel b). Years of schooling are harmonized across six data sources: ESD, EPA, ECEPOV, Census 2001, Census 2011, and Census 2021. For ECEPOV, separate estimates are provided for dependent children residing with their parents (\textit{dep}) and the full sample (\textit{all}). See Appendix \ref{apvardefs1} for data harmonization details.}
\end{figure}

\begin{figure}[ht]
\begin{subfigure}{0.7\textwidth}
    \centering
    \includegraphics[width=.995\linewidth]{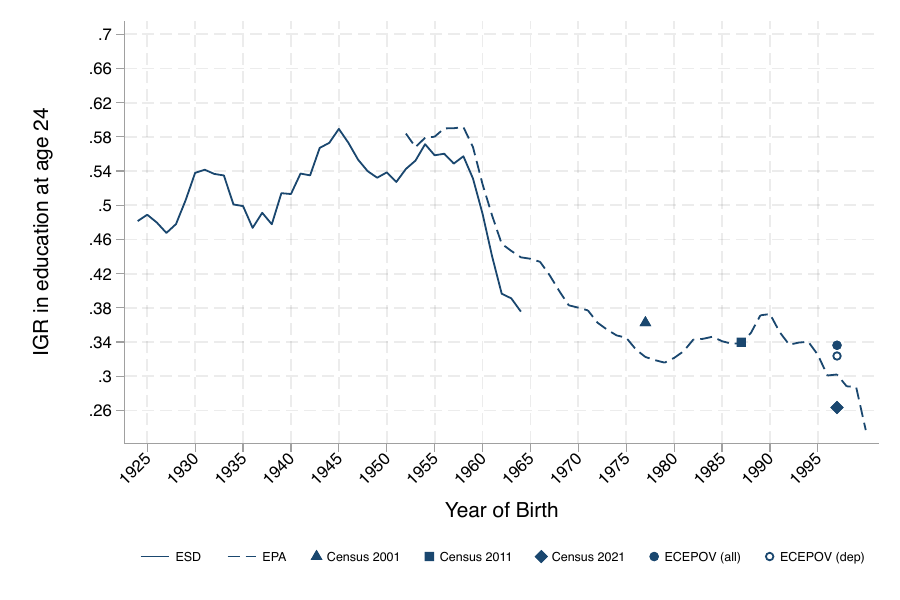}
    \subcaption{IGR in education at age 24}
\end{subfigure}
\begin{subfigure}{0.7\textwidth}
    \centering
    \includegraphics[width=.995\linewidth]{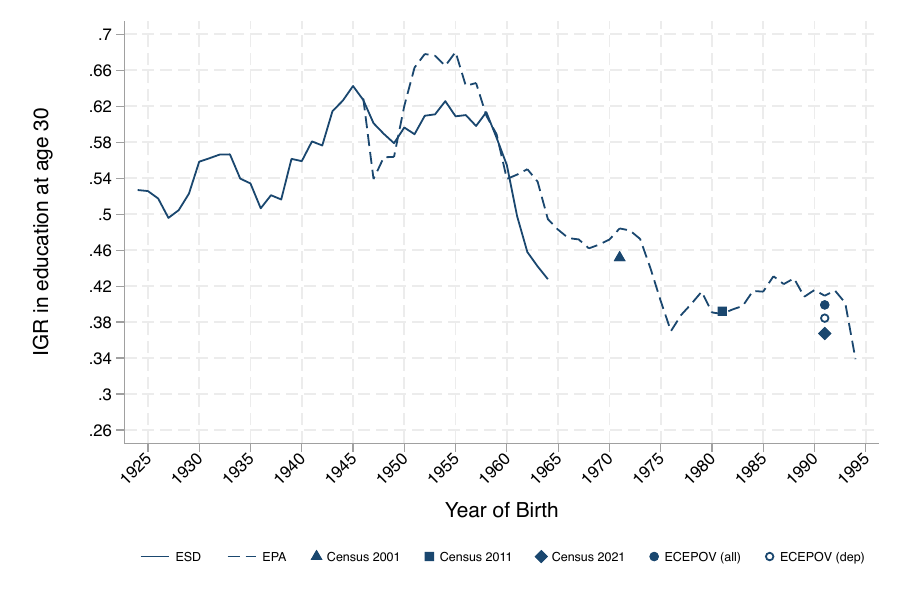}
    \subcaption{IGR in education at age 30}
\end{subfigure}
\caption{Trends in the intergenerational regression coefficient at different ages}
\label{fig:national_IGR_age}
\floatfoot{Notes: 3-year moving average of the intergenerational regression coefficient (IGR) in years of schooling, measured at the children's age of 24 (Panel a) and 30 (Panel b). Years of schooling are harmonized across six data sources: ESD, EPA, ECEPOV, Census 2001, Census 2011, and Census 2021. For ECEPOV, separate estimates are provided for dependent children residing with their parents (\textit{dep}) and the full sample (\textit{all}). See Appendix \ref{apvardefs1} for data harmonization details.}
\end{figure}

\begin{figure}[ht]
\begin{subfigure}{0.7\textwidth}
    \centering
    \includegraphics[width=.995\linewidth]{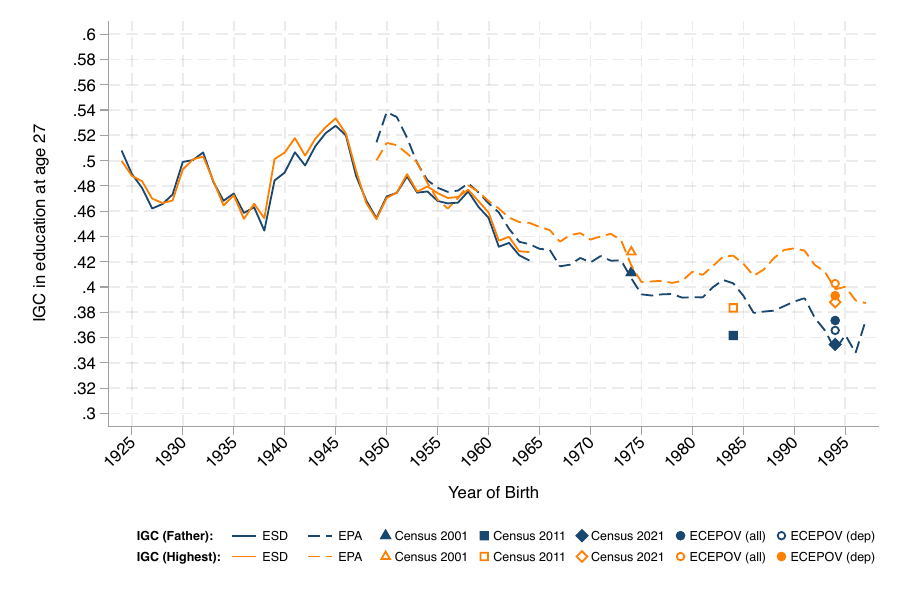}
    \subcaption{IGC using fathers' or parents' maximum education}
    \label{fig:robustness_IGC_parmax}
\end{subfigure}
\begin{subfigure}{0.7\textwidth}
    \centering
    \includegraphics[width=.995\linewidth]{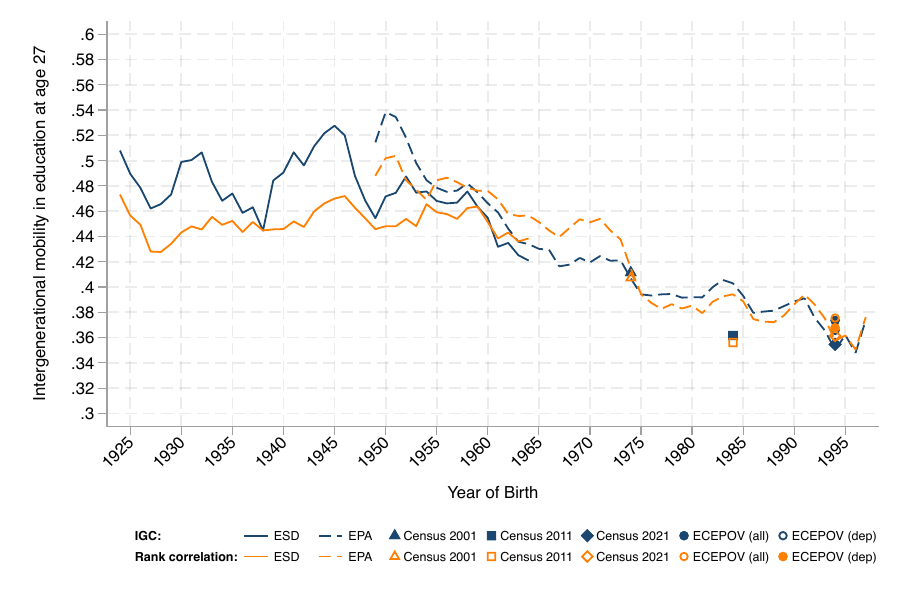}
    \subcaption{IGC and rank correlation in education}
    \label{fig:robustness_IGC_rank}
\end{subfigure}
\caption{Comparing different measures of intergenerational mobility in education}
\label{fig:robustness_IGC_parmax_rank}
\floatfoot{Notes: The figure reports 3-year moving averages by birth cohort. Panel (a) plots the intergenerational correlation (IGC) in years of schooling, measured at the children's age of 27 and using either fathers' or the maximum of parents' years of schooling. Panel (b) compares the IGC with the rank correlation between fathers' and children's schooling. ESD, EPA, ECEPOV, Census 2001, Census 2011, and Census 2021. For ECEPOV, separate estimates are provided for dependent children residing with their parents (\textit{dep}) and the full sample (\textit{all}). See Appendix \ref{apvardefs1} for data harmonization details.}
\end{figure}

\begin{figure}[ht]
\centering
\includegraphics[width=.7\linewidth]{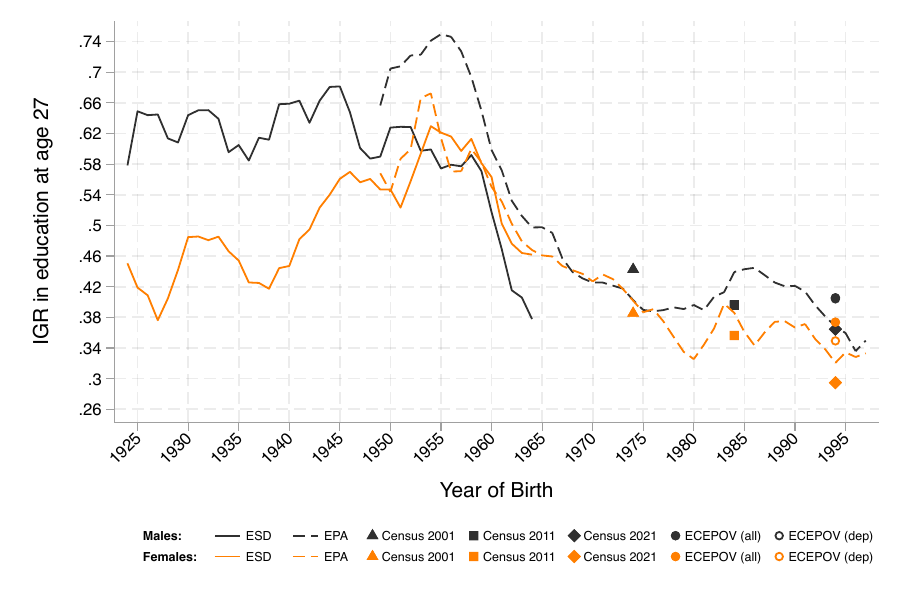}
\caption{Trends in the intergenerational regression coefficient in Spain by gender}
\label{fig:national_IGR_gender}
\floatfoot{Notes: 3-year moving average of the intergenerational regression coefficient (IGR) in years of schooling, disaggregated by gender. Black lines represent estimates for male children, while orange lines represent estimates for female children. Years of schooling are harmonized across six data sources: ESD, EPA, ECEPOV, Census 2001, Census 2011, and Census 2021. For ECEPOV, separate estimates are provided for dependent children residing with their parents (\textit{dep}) and the full sample (\textit{all}). See Appendix \ref{apvardefs1} for data harmonization details.}
\end{figure}

\begin{figure}[ht]
\begin{subfigure}{0.7\textwidth}
    \centering
    \includegraphics[width=.995\linewidth]{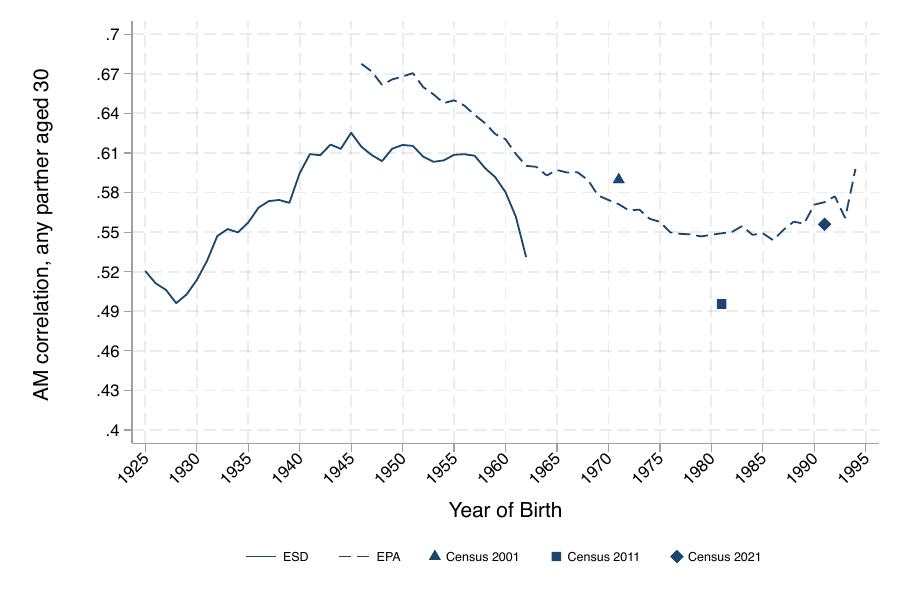}
    \caption{AM in education at age 30}
\end{subfigure}
\begin{subfigure}{0.7\textwidth}
    \centering
    \includegraphics[width=.995\linewidth]{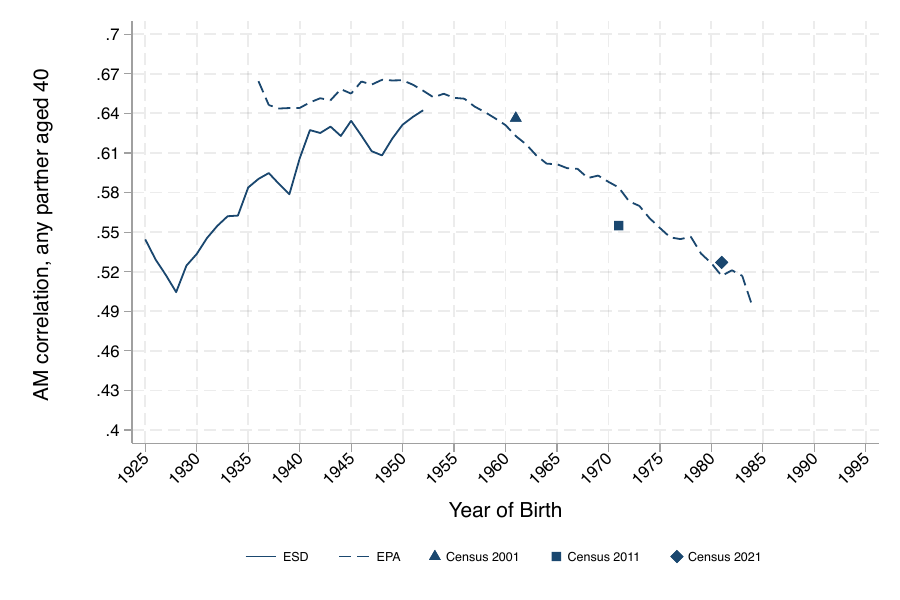}
    \caption{AM in education at age 40}
\end{subfigure}
\caption{Trends in assortative mating in education at different ages}
\label{fig:national_AM_age}
\floatfoot{Notes: 3-year moving average of the spousal correlation (AM) in years of schooling, measured when either partner is aged 30 (Panel a) or aged 40 (Panel b). Years of schooling are harmonized across five data sources: ESD, EPA, Census 2001, Census 2011, and Census 2021. See Appendix \ref{apvardefs1} for data harmonization details.}
\end{figure}

\begin{figure}
\begin{subfigure}{.495\linewidth}
    \centering
    \includegraphics[width=.99\linewidth]{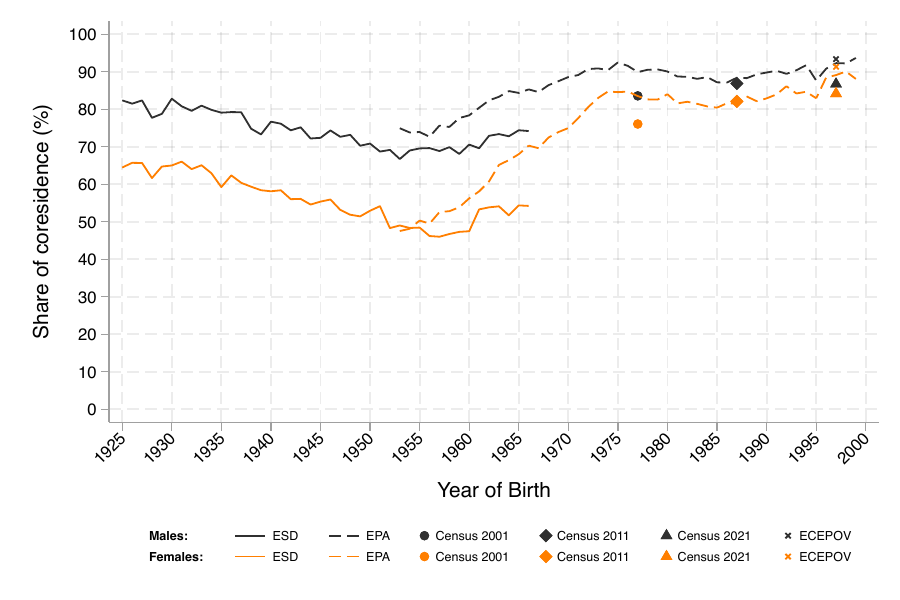}
    \caption{Coresidence share at age 24}
    \label{fig:national_sharedependence_gendera}
\end{subfigure}%
\begin{subfigure}{.495\linewidth}
    \centering
    \includegraphics[width=.99\linewidth]{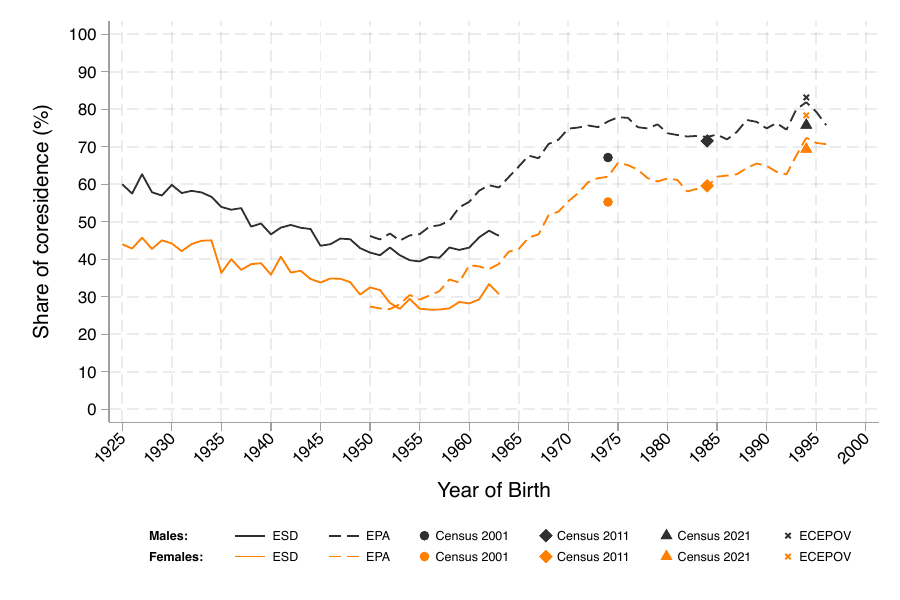}
    \caption{Coresidence share at age 27}
    \label{fig:national_sharedependence_genderb}
\end{subfigure}\\
\centering
\begin{subfigure}{.495\linewidth}
    \centering
    \includegraphics[width=.99\linewidth]{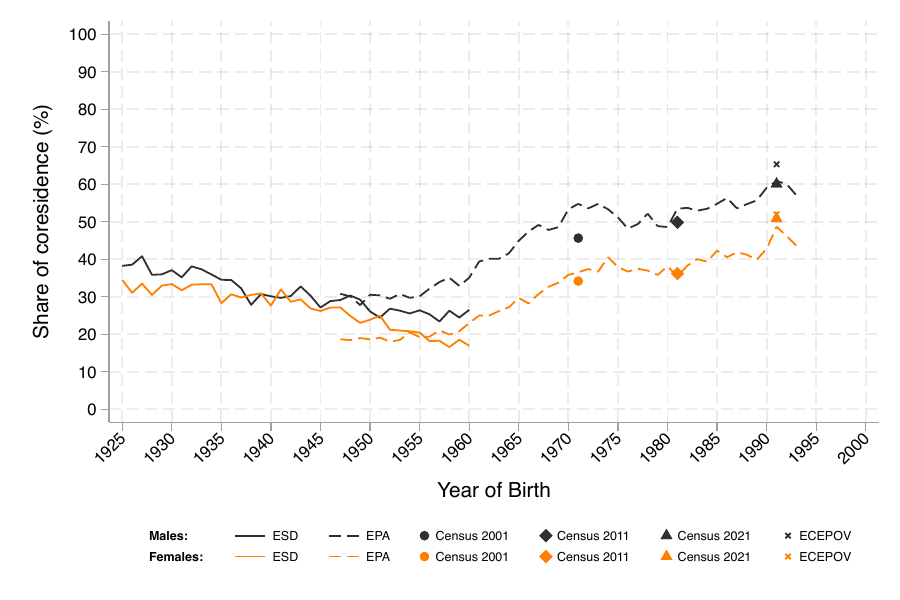}
    \caption{Coresidence share at age 30}
    \label{fig:national_sharedependence_genderc}
\end{subfigure}
\caption{Coresidence share by gender at ages 24, 27, and 30}
\label{fig:national_sharedependence_gender}
\floatfoot{Notes: The percentage of children residing in their parental household (\textit{coresident/dependent children}) is presented by gender and age at measurement across cohorts. Panels (a), (b), and (c) correspond to measurement ages of 24, 27, and 30, respectively. Black lines represent dependency shares for male children, while orange lines represent females. For the ESD dataset, coresidence shares are estimated using retrospective self-reported ages of leaving the parental household. For the EPA, Censuses, and ECEPOV samples, the analysis is restricted to individuals observed within the relevant cohort and age at the time of measurement.}
\end{figure}

\begin{figure}[h]
 \includegraphics[scale=1, clip=true]{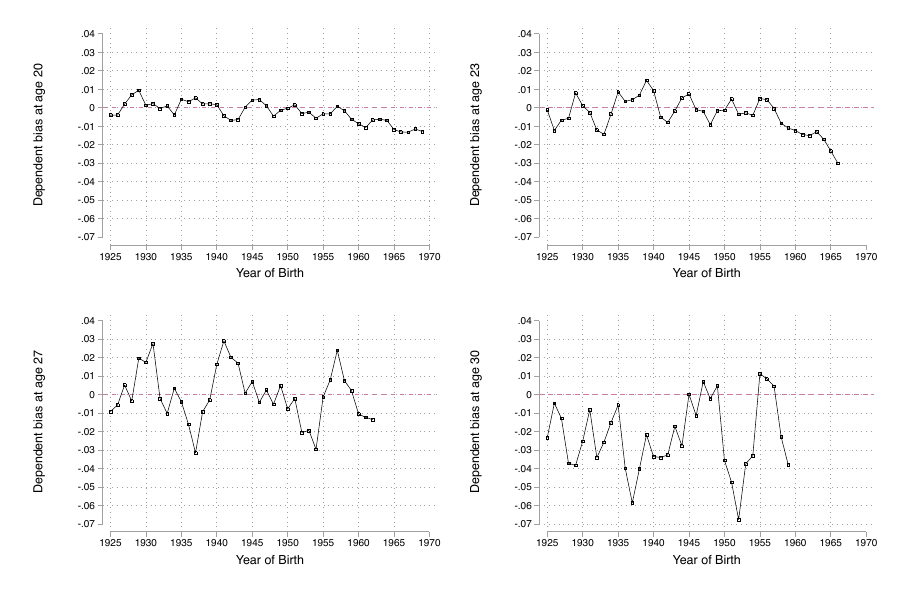}
 \caption{Dependent bias: Difference between coresident and benchmark IGR estimates}
 \label{fig:hilger_bias_obs}  
  \floatfoot{Notes: The figure shows the dependent bias across cohorts, measured as the difference between the IGR estimate based on \textit{coresident/dependent} only samples and the benchmark (\textit{dependent + indepedent}) IGR estimates, at the specified age of measurements of 20 (top-left panel), 23 (top-right panel), 27 (bottom-left panel), and 30 (bottom-right panel). A value of zero indicates no bias, since both IGR estimates are equal. Each cohort represents a moving average that incorporates data from adjacent cohorts.}
\end{figure}

\begin{figure}[h]
 \includegraphics[scale=1, clip=true]{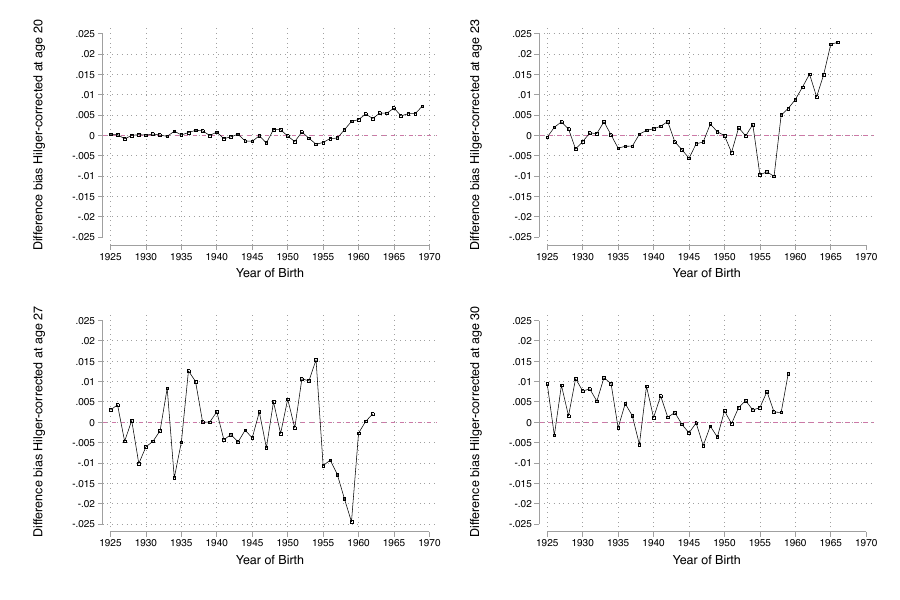}
 \caption{Bias correction: Comparing coresident and Hilger-corrected to the benchmark IGR estimates}
 \label{fig:hilger_bias_correction}  
  \floatfoot{Notes: The figure displays the difference in absolute bias achieved by the Hilger correction method. The bias reduction (y-axis) is calculated as the absolute difference between the coresident IGR estimate and the benchmark minus the absolute difference between the Hilger-corrected IGR estimate and the benchmark. A value of zero indicates that the coresident and Hilger-corrected IGR estimates are equally distant from the benchmark. A positive (negative) value indicates that the Hilger-corrected IGR estimate is closer to (further from) the benchmark than the coresident IGR estimate. Each cohort represents a moving average that incorporates data from adjacent cohorts.}
\end{figure}

\begin{figure}
\centering
 \includegraphics[scale=1]{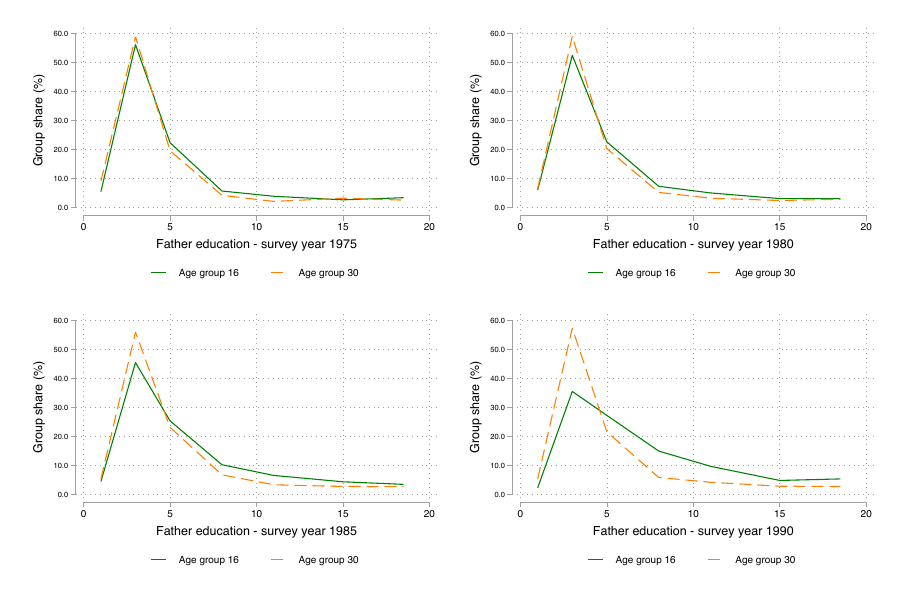}
 \caption{Assessing the smooth cohort assumption}
 \label{fig:IGR_hilger_smooth}  
  \floatfoot{Notes: The figure tests the smooth cohort assumption by comparing parental education group shares at ages 16 and 30 within the same ``fictitious'' survey year. For recent cohorts (bottom-right panel), the smooth cohort assumption breaks down due to rapid increases in parental educational attainment for post-1950 cohorts (see Figure \ref{USfigure}).}
\end{figure}

\clearpage
\section{Institutional Background: Spanish Education System in the 20th Century}\label{apinstitutional}
\setcounter{figure}{0}
\renewcommand{\thefigure}{D\arabic{figure}}

Spain's educational system historically lagged behind that of other European countries. In response to high illiteracy and the Church's strong influence over schooling, the 1857 \textit{Public Instruction Law} established compulsory, tuition-free primary education for children aged 6 to 9, regulated private primary education, and implemented a more science-oriented curriculum in secondary education. Subsequent reforms introduced a two-track school system in 1901, with students selected at age 10 into secondary education or continued primary schooling, and raised the statutory compulsory schooling age from 9 to 12 through a series of measures in 1901, 1909, and 1913, before being extended to age 14 in 1923.\footnote{\, The \href{https://www.boe.es/diario_gazeta/comun/pdf.php?p=1901/10/30/pdfs/GMD-1901-303.pdf}{Royal Decree of October 26, 1901} increased compulsory schooling to age 12 but had inferior legal rank to the Public Instruction Law. The compulsory age was raised to 12 again on June 23, 1909, by amending Articles 7 and 8 of the Public Instruction Law (\href{https://www.boe.es/diario_gazeta/comun/pdf.php?p=1909/06/25/pdfs/GMD-1909-176.pdf}{Gaceta de Madrid No.\ 176, 1909}), but established only six months of compulsory attendance from age 10 to 11 and three months from age 11 to 12. The \href{https://www.boe.es/diario_gazeta/comun/pdf.php?p=1913/07/20/pdfs/GMD-1913-201.pdf}{Royal Decree of July 18, 1913} increased voluntary primary schooling by one grade but also had an inferior legal rank to the Public Instruction Law. The \emph{Estatuto General del Magisterio de Primera Ense\~nanza} (\href{https://www.boe.es/gazeta/dias/1923/05/19/pdfs/GMD-1923-139.pdf}{Gaceta de Madrid No.\ 139, 1923}) also had Royal Decree rank.} In practice, however, these mandates were only partially enforced, and compliance remained low, especially in rural areas.

During the early Second Spanish Republic (1931--1939), education policy shifted toward a broad modernization agenda aimed at reducing rural illiteracy, limiting Church influence, and moving toward a more unified system. Between 1931 and 1933, the Republic undertook a major expansion of primary schooling that increased the teacher workforce \citep{cuno2013reforma} and improved compliance with compulsory attendance, particularly in rural areas. However, the implementation of several ambitious proposals was hindered by political turmoil and budgetary constraints caused by the international economic crisis \citep{molero1975segunda}. Following the Spanish Civil War (1936--1939), the Franco regime (1939--1975) carried out an ideological purge of the entire civil sector workforce, dismissing about 10\% of primary school teachers, and temporarily reversing previous gains \citep{GrebolNavarro2025}. Dismissals were disproportionately concentrated among more competent teachers and were even more severe at higher levels of the educational system (e.g., 23\% among full professors), contributing to a deterioration in educational quality. The subsequent consolidation and reorganization of the system delayed progress in educational attainment for nearly a decade and rolled back earlier advances in pedagogical training.

A new comprehensive primary education law was enacted in 1945.\footnote{\, See \textit{Ley de 17 de julio de 1945 sobre Educaci\'on Primaria}, \href{https://www.boe.es/diario_gazeta/comun/pdf.php?p=1945/07/18/pdfs/BOE-1945-199.pdf}{BOE No.\ 199, 1945} (in Spanish).} Consistent with the Francoist Regime's ideology, the reform reinforced Catholic instruction, Spanish nationalism and language, gender segregation, and the two-track system. Although continued rural expansion and stricter enforcement raised average educational attainment levels, substantial educational inequality persisted, reflecting early tracking and the limited availability of secondary schools outside provincial capitals and major cities.

Two decades later, compulsory schooling was extended to age 14.\footnote{\, See \textit{Ley 27/1964, de 29 de abril, sobre ampliaci\'on del periodo de escolaridad obligatoria hasta los catorce a\~nos}, \href{https://www.boe.es/boe/dias/1964/05/04/pdfs/A05696-05696.pdf}{BOE No.\ 107, 1964} (in Spanish).} Full compliance, however, was not achieved until the implementation of the 1970 General Education Law (\textit{Ley General de Educaci\'on}, LGE). The LGE was introduced as part of a broader comprehensive-school movement across Europe in the 1960s, motivated by concerns about equal opportunity and human-capital accumulation (see \citealp{meghir2005educational} for Sweden, \citealp{pekkarinen2009school} for Finland, and \citealp{aakvik2010measuring} for Norway). The religious and national-ideological aspects emphasized during the early Francoist dictatorship were relaxed in favor of a more pedagogical, practical approach. The previous two-track school system was replaced by a uniform eight-year comprehensive school, with student selection into vocational or academic tracks postponed from age 10 to 14. The reform was introduced in 1970 and gradually implemented over four years, and full compliance with the compulsory schooling age of 14 was reached by 1974.

Finally, the 1990 \textit{Ley Org\'anica General del Sistema Educativo} (LOGSE) extended compulsory education to age 16 and established a unified structure for compulsory lower secondary schooling (\textit{Educaci\'on Secundaria Obligatoria}, ESO).\footnote{\, Additional comparatively minor reforms were enacted during the early democratic period after 1978.} After completing ESO, students could pursue an academic track (\textit{Bachillerato}) leading to university or a vocational track (\textit{Formaci\'on Profesional}, FP) providing professional qualifications.

\end{document}